\documentclass[aps, prd, twocolumn, amsmath, floats,floatfix, superscriptaddress, nofootinbib,
showpacs]{revtex4}

\usepackage{amssymb}
\usepackage{amsmath}
\usepackage{verbatim}
\usepackage{mathrsfs}
\usepackage{amsfonts}
\usepackage{latexsym}
\usepackage{epsfig}
\usepackage{color}
\usepackage{graphicx,subfigure}
\usepackage{units}
\usepackage{slashbox}
\usepackage{natbib}

\begin{document}

\title{Quasistationary solutions of self-gravitating scalar fields around collapsing stars} 

\author{Nicolas Sanchis-Gual}
\affiliation{Departamento de
  Astronom\'{\i}a y Astrof\'{\i}sica, Universitat de Val\`encia,
  Dr. Moliner 50, 46100, Burjassot (Val\`encia), Spain}

\author{Juan Carlos Degollado} 
\affiliation{Departamento de Ciencias Computacionales,
Centro Universitario de Ciencias Exactas e Ingenier\'ia, Universidad de Guadalajara\\
Av. Revoluci\'on 1500, Colonia Ol\'impica C.P. 44430, Guadalajara, Jalisco, Mexico}

\author{Pedro J. Montero} 
\affiliation{Max-Planck-Institute f{\"u}r Astrophysik, Karl-Schwarzschild-Str. 1, 85748, Garching 
bei M{\"u}nchen, Germany}

\author{Jos\'e A. Font}
\affiliation{Departamento de
  Astronom\'{\i}a y Astrof\'{\i}sica, Universitat de Val\`encia,
  Dr. Moliner 50, 46100, Burjassot (Val\`encia), Spain}
\affiliation{Observatori Astron\`omic, Universitat de Val\`encia, C/ Catedr\'atico 
  Jos\'e Beltr\'an 2, 46980, Paterna (Val\`encia), Spain}
  
\author{Vassilios Mewes}
\affiliation{Departamento de
  Astronom\'{\i}a y Astrof\'{\i}sica, Universitat de Val\`encia,
  Dr. Moliner 50, 46100, Burjassot (Val\`encia), Spain}


\date{July 2015}


\begin{abstract}  

Recent work has shown that scalar fields around black holes can form long-lived, 
quasistationary configurations surviving for cosmological timescales. With this 
requirement, scalar fields cannot be discarded as viable candidates for 
dark matter halo models in galaxies around central supermassive black 
holes (SMBH). One hypothesis for the formation of most SMBHs at high
redshift is the gravitational collapse of supermassive  stars (SMS) with masses of
$\sim10^5 \rm {M_{\odot}}$. Therefore, a constraint for the existence of quasi-bound states of scalar 
fields is their survival to such dynamic events. To answer this question we
present in this paper the results of a series of 
numerical relativity simulations of gravitationally collapsing, spherically 
symmetric stars surrounded by self-gravitating scalar fields. We use an ideal fluid 
equation of state with adiabatic index $\Gamma=4/3$ which is adequate to simulate
radiation-dominated isentropic SMSs. Our results confirm the 
existence of oscillating, long-lived, self-gravitating scalar field 
configurations around non-rotating black holes after the collapse of the stars.
\end{abstract}




\maketitle

\section{Introduction}\label{sec:introduction}

The formation of galactic black holes is a fundamental issue
in the study of galaxy formation. Most galaxies host
black holes in their nuclei. Very luminous quasars at redshifts
higher than 6 are believed to be powered by supermassive black holes (SMBH) with masses of
$\sim 10^{9} \rm {M_{\odot}}$~\cite{RIS_0,2041-8205-801-1-L11}. However, it
is still unknown how such SMBH can form and grow to reach the current
values when the Universe was less than $10^{9}$ years old.

One possible scenario for 
the formation of SMBH in the center of galaxies is that they were formed through 
accretion and merging of the first stellar remnants~\cite{Bromm:2002hb,Haiman01}.
Another scenario assumes that central black holes formed via
direct collapse (see~\cite{Begelman:2006db,Begelman10,Mayer10}). This route assumes that
fragmentation, which depends on efficient cooling, is suppressed. Depending on the rate and efficiency of the inflowing
mass, there may be different outcomes. A low rate of mass
accumulation would favor the formation of isentropic SMSs, with masses $\geq 5 \times 10^{4} \rm {M_{\odot}}$, which then would evolve as
equilibrium configurations dominated by radiation
pressure before collapsing to form a SMBH seed~\cite{Shibata02,MonteroJM12}.

Scalar fields have been considered in cosmology to provide 
inflationary solutions in the early universe~\cite{Lidsey:1995np} and as an 
alternative explanation for dark energy~\cite{Padmanabhan:2002ji}. In addition, they 
have also been proposed as strong candidates for dark 
matter~\cite{Sahni:1999qe,Hu:2000ke,Matos:2000ng,Matos:2000ss}. 
In these models one assumes that dark matter is composed by 
bosonic particles which may condensate into macroscopic 
objects forming boson stars or forming halos around black 
holes. The fact that most galaxies host a SMBH at their centers has 
raised many questions regarding the dynamics of the whole system. 

The issue of whether scalar field halos are able to survive 
around black holes for cosmological timescales
has been addressed in several 
occasions (see e.g.~\cite{Burt:2011pv} and references therein). It has been shown in 
the test-field limit in the background spacetime of a Schwarzschild black 
hole, that scalar field configurations do remain around 
SMBHs for cosmological timescales, 
provided the mass of the scalar particle is of the order of 
$10^{-22} - 10^{-24}$eV. Further studies considered also the 
backreaction of spacetime~\cite{Sanchis-Gual:2015, Witek:2012tr,Barranco:2015},
and confirmed the existence of such long-lived states.
These gravitationally trapped, scalar field 
configurations are states characterized 
by precise complex frequencies. It has been proved that 
these states are generic in the sense that they appear after 
some time in the evolution of the scalar field even when  
arbitrary initial data is used~\cite{Barranco:2013rua}.
This observation seems to be enough to guarantee (as we show below) that any 
choice of initial data will create a scalar field environment around the black hole. 

The main objective of this work is to assess the role of black 
hole formation in the presence of scalar field configurations  
which resemble the quasi-bound states found in previous investigations
where the black hole is assumed to exist from the start.
To accomplish this goal we perform numerical relativity 
simulations which track the gravitational collapse of a 
polytropic star in the presence of a scalar field environment. 
The numerical techniques and the code used are those already described 
in~\cite{Sanchis-Gual:2015}. We shall refer to that work as 
Paper 1. In this work we also take into account the evolution of the
relativistic hydrodynamic equations. In order to simplify our 
model, we consider that the scalar field and the matter of the 
star are coupled only through gravity.

The paper is organized as follows: In Section~\ref{sec:formalism} 
we present the basic equations and discuss the initial data used in 
our simulations. Section~\ref{sec:numerics} briefly describes our numerical
approach.
In Section~\ref{sec:num_results} we discuss our 
findings and describe some properties of the solutions. Finally in 
Section~\ref{sec:conclusions} we sum up our concluding remarks. Throughout the paper Greek 
indices run over spacetime indices (0 to 3), while 
Latin indices run over space indices only (1 to 3). We use units in which $c=G=M_{\odot}=1$.

\section{Basic equations}\label{sec:formalism}

We investigate the dynamics of a self-gravitating scalar field
configuration around a polytropic star collapsing to a black hole by 
solving numerically the coupled Einstein-Klein-Gordon system and the
general relativistic hydrodynamics equations. The Einstein 
equations in covariant form read 
\begin{equation}
 R_{\alpha\beta}-\frac{1}{2}g_{\alpha\beta}R=8\pi T_{\alpha\beta} \ ,
\label{eq:Einstein}
\end{equation}
where $R_{\alpha\beta}$ is the Ricci tensor of the 4-dimensional
spacetime, $g_{\alpha\beta}$ is the spacetime metric,
$R$ is the Ricci scalar, and the matter content is given for our system by a combined stress-energy tensor
\begin{equation}
 T_{\alpha\beta} = T_{\alpha\beta}^{\text{F}}+ T_{\alpha\beta}^{\text{SF}},
\end{equation}
where $T_{\alpha\beta}^{\text{F}}$ is the stress-energy tensor of the (perfect) fluid defined by
\begin{equation}
T_{\alpha\beta}^{\text{F}}=\rho h u_{\alpha}u_{\beta} + p g_{\alpha\beta}\,,
\label{matter}
\end{equation}
and $T_{\alpha\beta}^{\text{SF}}$  is the part corresponding to the scalar field, given by
\begin{equation}
T_{\alpha\beta}^{\text{SF}}=\partial_{\alpha}\Phi 
\partial_{\beta}\Phi-\frac{1}{2}g_{\alpha\beta}\left(\partial^{\sigma}\Phi\partial_{\sigma}\Phi+
\mu^2\Phi^2 \right)\ .
\label{eq:tmunu}
\end{equation}
In Eq.~(\ref{matter}) $\rho$ is the rest-mass density, 
$p$ is the pressure, $u_{\alpha}$ is the fluid four velocity and $h = 1 + \varepsilon + p/\rho$ is 
the relativistic specific enthalpy, 
with $\varepsilon$ being the specific internal energy. 
We will assume that each type of matter obeys their own conservation law, in other words, 
that the coupling between them comes purely from  gravity. 
The equations of motion for the  scalar field are 
obtained from the 
Klein-Gordon equation
\begin{equation}
 \Box \Phi-\mu^2\Phi=0 \ ,
\label{eq:KG}
\end{equation}
where the D'Alambertian operator is defined by $\Box:=
(1/\sqrt{-g})\partial_{\alpha}(\sqrt{-g}g^{\alpha\beta}\partial_{\beta})$. We
follow the convention that $\Phi$ is dimensionless and $\mu$ has
dimensions of (length)$^{-1}$. 

In the following we present the different systems of evolution (and constraint) equations
we solve to carry out our study. While we mainly include this information to make the paper 
self-contained, we keep these sections as concise as possible, and refer the interested reader to our previous papers~\cite{Montero:2012yr,Sanchis-Gual:2014,Sanchis-Gual:2015} for further details. 

\subsection{Einstein's equations}

Under the assumption of spherical symmetry the 1+1 line element may be written as
\begin{equation}
 ds^2 = e^{4\chi } (a(t,r)dt^2+ r^2\,b(t,r)  d\Omega^2) \ ,
\end{equation}
with $d\Omega^2 = \sin^2\theta d\varphi^2+d\theta^2$ being the solid angle element
and $a(t,r)$ and $b(t,r)$ two non-vanishing metric functions. Moreover $\chi$ is related to
the conformal factor $\psi$ as $\psi = e^\chi = (\gamma/\hat \gamma)^{1/12}$, with $\gamma$
and $\hat\gamma$ being the determinants of the physical and conformal 3-metrics, respectively. They 
are conformally related by $\gamma_{ij} = e^{4 \chi} \hat
\gamma_{ij}$.

In particular we adopt the BSSN formalism
of Einstein's equations~\cite{Baumgarte98,Shibata95},  where the evolved
fields are the conformally related 3-dimensional metric, the conformal exponent $\chi$, the trace of the extrinsic
curvature $K$, the independent component of the traceless part of the conformal extrinsic curvature,
and the radial component of the conformal connection functions (we refer to
\cite{Alcubierre:2010is,Montero:2012yr} for further details).

The matter fields appearing on the right-hand-side of the evolution
equations for the gravitational fields (see~\cite{Alcubierre:2010is,Montero:2012yr}), $\mathcal{E}$,
$j_{i}$, $S_i$, and $S_{ij}$ include the contribution of both the fluid and the scalar field, 
i.e.~$\mathcal{E}=\mathcal{E}^{\rm F}+\mathcal{E}^{\rm{SF}}$, etc. The matter source terms for the 
fluid given by the stress-energy tensor of Eq.~\eqref{matter}  read
\begin{eqnarray}
\mathcal{E}^{\rm F}&\equiv& n^{\alpha}n^{\beta}T^{\rm F}_{\alpha\beta},\label{eq17}\\
j_{r}^{\rm F}&\equiv&-\gamma^{\alpha}_{r}n^{\beta}T^{\rm F}_{\alpha\beta},\label{eq18}\\
S_{a}^{\rm F}&\equiv&(T^{\rm F})^{r}_{r},\label{eq19a}\\
S_{b}^{\rm F}&\equiv&(T^{\rm F})^{\theta}_{\theta}\label{eq19b}.
\end{eqnarray}

The Hamiltonian and momentum constraints are given by the following two equations that we compute 
to monitor the accuracy of the numerical evolutions: 
\begin{eqnarray}
\mathcal{H}&\equiv& R-(A^{2}_{a}+2A_{b}^{2})+\frac{2}{3}K^{2}-16\pi 
\mathcal{E}=0,\label{hamiltonian}\\
\mathcal{M}_{r}&\equiv&\partial_{r}A_{a}-\frac{2}{3}\partial_{r}K+6A_{a}\partial_{r}\chi\nonumber\\
&+&(A_{a}-A_{b})\biggl(\frac{2}{r}+\frac{\partial_{r}b}{b}\biggl)-8\pi j_{r}=0,\label{momentum}
\end{eqnarray}

where $A_{a}$ and $A_{b}$ are the contraction of the traceless part of the conformal extrinsic curvature, 
$A_{a}\equiv\hat A^{r}_{r},\, A_{b}\equiv\hat A^{\theta}_{\theta}$.
In addition to the evolution fields, there are two more variables left 
undetermined, the lapse function $\alpha$, and the shift vector $\beta^{i}$. Our code 
can handle arbitrary gauge conditions and for the simulations reported in this paper we 
use the same conditions of Paper 1, namely
 the so-called {``non-advective 1+log''} condition~\cite{Bona:1997prd} for the 
lapse, and a variation of the {``Gamma-driver''} condition for the shift vector 
\cite{Alcubierre:2003ab,Alcubierre:2010is}. 

\subsection{Klein-Gordon equation}

To solve the Klein-Gordon equation we introduce two first-order variables defined as:
\begin{eqnarray}
\Pi &:=& 
n^{\alpha}\partial_{\alpha}\Phi=\frac{1}{\alpha}(\partial_{t}\Phi-\beta^{r}\partial_{r}\Phi) \ ,\\
\Psi&:=&\partial_{r}\Phi \ .
\end{eqnarray}
Therefore, using Eq.~(\ref{eq:KG}) we obtain the following system of first-order equations: 
\begin{eqnarray}
\partial_{t}\Phi&=&\beta^{r}\partial_{r}\Phi+\alpha\Pi \ ,\\
\partial_{t}\Psi&=&\beta^{r}\partial_{r}\Psi+\Psi\partial_{r}\beta^{r}+\partial_{r}(\alpha\Pi) \ ,\\
\partial_{t}\Pi&=&\beta^{r}\partial_{r}\Pi+\frac{\alpha}{ae^{4\chi}}[\partial_{r}\Psi\nonumber\\
&+&\Psi\biggl(\frac{2}{r}-\frac{\partial_{r}a}{2a}+\frac{\partial{r}b}{b}+2\partial_{r}
\chi\biggl)\biggl]\nonumber\\
&+&\frac{\Psi}{ae^{4\chi}}\partial_{r}\alpha+\alpha K\Pi - \alpha \mu^{2}\Phi \ .
\label{eq:sist-KG}
\end{eqnarray}

The matter source terms  for the scalar field read
\begin{eqnarray}
\mathcal{E}^{\rm{SF}}&\equiv&n^{\alpha}n^{\beta}T^{\rm{SF}}_{\alpha\beta}=\frac{1}{2}\biggl(\Pi^{2}
+\frac{\Psi^{2}}{ae^{4\chi}}\biggl)+\frac{1}{2}\mu^{2}\Phi^{2} \label{eq:rho}\ ,\\
j^{\rm{SF}}_{r}&\equiv&-\gamma^{\alpha}_{r}n^{\beta}T^{\rm{SF}}_{\alpha\beta}=-g_{rr}\Pi\Psi ,\\
S_{a}^{\rm{SF}}&\equiv&(T^{\rm{SF}})^{r}_{r}=\frac{1}{2}\biggl(\Pi^{2}+\frac{\Psi^{2}}{ae^{4\chi}}
\biggl)-\frac{1}{2}\mu^{2}\Phi^{2} \ ,\\
S_{b}^{\rm{SF}}&\equiv&(T^{\rm{SF}})^{\theta}_{\theta}=\frac{1}{2}\biggl(\Pi^{2}-\frac{\Psi^{2}}{ae^
{4\chi}}\biggl)-\frac{1}{2}\mu^{2}\Phi^{2} \ .
\end{eqnarray}
%

\subsection{Hydrodynamics equations}

The general relativistic hydrodynamics equations, expressed through the 
conservation laws of the stress-energy tensor $T^{\mu\nu}$ and the 
continuity equation, are
\begin{equation}
\label{hydro eqs}
	\nabla_\alpha T^{\alpha\beta} = 0\;,\;\;\;\;\;\;
\nabla_\alpha \left(\rho u^{\alpha}\right) = 0\,.
\end{equation}
Following \cite{Banyuls:97c}, we write the equations of general relativistic hydrodynamics
as a first-order, flux-conservative system.  We define the fluid 3-velocity as seen by a normal 
(Eulerian) observer as
\begin{equation}
	v^{r}\equiv \frac{u^{r}}{\alpha u^{t}}+\frac{\beta^{r}}{\alpha},
\end{equation}
and the Lorentz factor between the fluid and the normal observer as $W\equiv \alpha u^{t}$.
We also define the relativistic densities of mass, momentum, and energy for the normal observer as
\begin{align}
	D & = \rho W\,,\\
	S_r & = \rho h W^2v_r\,,\\
	\tau & = \rho h W^2 - p - D\,.
\end{align}
We then assemble these variables into a vector
${\bf{U}}$ of conserved fluid variables 
\begin{equation}
	{\bf{U}}=\sqrt{\gamma}(D,S_{r},\tau).
\end{equation}
By defining the corresponding vector of fluxes ${\bf{F}}^{r}$ and
the vector of source terms ${\bf{S}}$ we can cast Eqs.~(\ref{hydro eqs}) in flux-conservative form
\begin{equation}
\label{cylcon}
	\partial_{t}{\bf{U}}+\partial_{r}{\bf{F}}^{r}={\bf{S}}.
\end{equation}
We refer to \cite{Montero:2012yr} where the explicit form of vectors 
${\bf{F}}^{r}$ and ${\bf{S}}$ is given. To close the system of hydrodynamics equations, we choose a Gamma-law equation of state
(EOS) 
\begin{equation}
\label{EOS1}
	P=\left(\Gamma -1\right)\rho\epsilon ,
\end{equation}
where $\Gamma$ is the adiabatic index.

\subsection{Scalar field initial data}

As in Paper I our choice of initial data for the scalar field is a Gaussian distribution of the form
\begin{equation}\label{eq:pulse}
 \Phi=A_0e^{-(r-r_0)^2/\lambda^2} \ ,
\end{equation}
where $A_0$ is the initial amplitude of the pulse, $r_0$ is the center of the Gaussian, and
$\lambda$ is its width. The auxiliary first-order quantities are
initialized as follows
\begin{eqnarray}
 \Pi(t=0,r)&=&0 \ , \\ 
 \Psi(t=0,r) &=& -2\frac{(r-r_0)}{\lambda^2}A_0e^{-(r-r_0)^2/\lambda^2} \ .
 \label{eq:iderivatives}
\end{eqnarray}

Likewise, we choose a conformally flat metric with $a=b=1$ together with a time symmetry condition 
$K_{ij}=0$.
With these choices and $\Pi(t=0,r) = 0$ the momentum constraint is satisfied trivially and the 
Hamiltonian 
constraint, Eq.~\eqref{hamiltonian}, yields the following equation for the conformal factor 
$\psi=e^{\chi}$,
\begin{equation}
 \partial_{rr}\psi + \frac{2}{r}\partial_{r}\psi +2\pi\psi^5 \mathcal{E}=0 \ .
\end{equation}
In order to solve this equation for our collapsing stars, we assume that the conformal factor can be 
written as 
\begin{equation}
 \psi = \psi^{\text{TOV}}+ u(r) \ , 
\end{equation}
where $\psi^{\text{TOV}}$ is the part of the conformal factor associated with the stellar 
Tolman-Oppenheimer-Volkoff (TOV)
model. By substituting this ansatz in the Hamiltonian constraint we obtain 
\begin{equation}
 \partial_{rr}u(r) +\frac{2}{r}\partial_{r}u(r)+2\pi\psi^5 \mathcal{E}^{\text{SF}} 
+2\pi(\psi^{5}-(\psi^{\text{TOV}})^{5})\mathcal{E}^{\text{TOV}}= 0 \ . 
\end{equation}
Given a distribution of the scalar field density $\mathcal{E}^{\text{SF}}$, we solve this ordinary 
differential equation for $u$ using a standard fourth-order Runge-Kutta integrator, assuming that 
$u\rightarrow 0$ as
$r\rightarrow \infty$ and regularity at the origin.

\begin{table}
\caption{Stellar TOV models: $N=3$ ($\Gamma=4/3$) polytropes with $\kappa=0.48773$. From left to 
right the columns indicate the model number,  the central rest-mass density, $\rho_c$,  the central 
energy density, $e_c$,  the radius of the star, $R$, the gravitational mass of the star, $M$, and 
the compactness, $M/R$.}\label{tab:table1}
\begin{ruledtabular}
\begin{tabular}{cccccc}
Model&$\rho_c$&$e_c$&$R$&$M$&$M/R$\\
\hline
1&3.11e-6&3.17e-6&185.8&1.43&0.008\\
2&3.11e-5&3.25e-5&86.3&1.31&0.015\\
3&3.11e-4&3.41e-4&39.9&1.01&0.025\\
4&3.11e-3&3.77e-3&19.2&0.80&0.042
\end{tabular}
\end{ruledtabular}
\end{table}

\begin{figure}
\begin{minipage}{1\linewidth}
\includegraphics[width=1.08\textwidth, height=0.3\textheight]{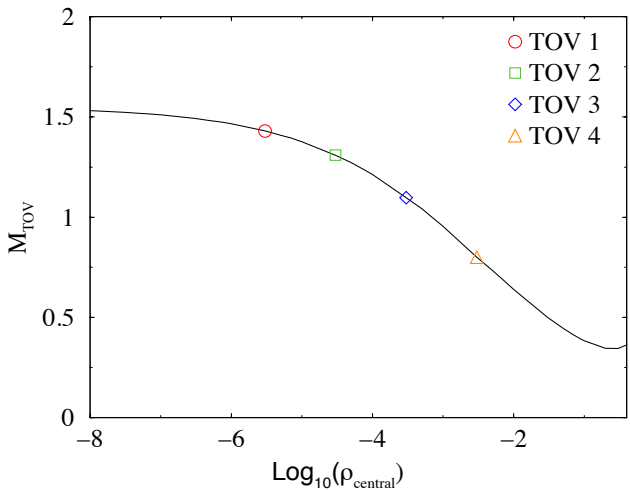} 
\caption{Mass of the polytropic TOV stars as function of their initial central rest-mass density. 
All models are in the unstable branch of the diagram.}
\label{fg:rhovsmass}
\end{minipage}
\end{figure}

\section{Numerics} 
\label{sec:numerics}

The time update of the different systems of evolution equations we have to solve in our code 
(Einstein, Klein-Gordon, and Euler) is done using the same type of techniques we have extensively 
used in our previous work (see in particular~\cite{Montero:2012yr,Sanchis-Gual:2015}). We address 
the interested reader to those references for full details on the particular numerical techniques 
implemented in the code. Here, we simply mention that the evolution equations are integrated using 
the second-order PIRK method developed by \cite{Isabel:2012arx,Casas:2014}.  This 
method allows to handle the singular terms that appear in the evolution equations due to our choice 
of curvilinear coordinates. The derivatives in the spacetime evolution are computed using a 
fourth-order centered finite difference approximation on a uniform grid except 
for advection terms for which we adopt a fourth-order upwind scheme.  We also use fourth-order 
Kreiss-Oliger dissipation to avoid high frequency noise appearing near the outer boundary. The difference with our 
previous work \cite{Sanchis-Gual:2015} is that we are also evolving the hydrodynamics conserved quantities explicitly.
Correspondingly, the equations of hydrodynamics are solved using the HLLE approximate Riemann solver in tandem with 
the second-order MC reconstruction scheme~\cite{Montero:2012yr}.

\section{Results} 
\label{sec:num_results}

\subsection{Stellar models and code convergence}

We construct spherically-symmetric stellar models by solving the Tolman-Oppenheimer-Volkoff (TOV)
equations for a polytropic EOS,
\begin{equation}
p = \kappa \, \rho^{1+1/N},
\end{equation}
where $\kappa$ is the polytropic constant and $N$ the polytropic index. These data are evolved in 
our code with 
the Gamma-law EOS, Eq.~(\ref{EOS1}), with $\Gamma = 1 + 1/N$. We adopt $N=3$ ($\Gamma=4/3$), which 
is the appropriate value for a SMS. Moreover, we select the value of $\kappa=0.48773$ in our code units. The polytropic 
constant is set to this value to keep the mass of the star close to one, namely in the range 
$M=\lbrace0.8,1.43\rbrace$ for our four different TOV models. These
models are gravitationally unstable spherical relativistic stars that collapse to black holes in dynamical timescales by artificially decreasing 
the pressure by a small amount (typically 5\%) at $t=0$. The main characteristics of our sample of 
stellar models are reported in Table~\ref{tab:table1}. Their corresponding location in the mass vs 
central density diagram is displayed in Fig.~\ref{fg:rhovsmass}. The models are constructed using 
the RNS code \cite{Stergioulas:1995b}.

The collapsing polytropic stars are surrounded by a scalar field cloud 
whose functional form is given by Eq.~(\ref{eq:pulse}). We investigate 
the behaviour of the scalar field in a highly dynamical evolution in 
which the stars undergo gravitational collapse. The end result of the 
collapse is the formation of a Schwarzschild black hole whose mass is 
the sum of the mass of the corresponding original star and a portion of 
the scalar field mass, the part that is accreted. The aim of this 
investigation is to find out whether in such a dynamical scenario, part 
of the scalar field can still survive in the form of long-lived 
quasi-bound states.

\begin{figure}
\begin{minipage}{1\linewidth}
\hspace{-0.5cm}\includegraphics[width=1.055\textwidth, height=0.3\textheight]{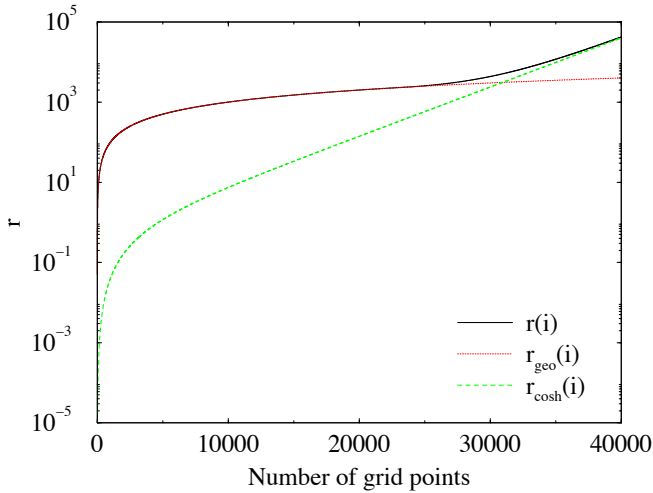} 
\caption{Extent of the radial grid as a function of the number of grid points. The black solid line represents the radial coordinate profile of the composite grid, the red dotted line the geometric progression patch and the green dashed line the hyperbolic cosine patch.}
\label{fg:grid}
\end{minipage}
\end{figure}

To carry out our study we set up and evolve 80 different initial scalar field models. These models 
are characterized by different values of the initial energy of the scalar field, namely 
$E_{0}=\lbrace3.2\times10^{-5},0.50,2.85,4.3\rbrace$ and different values of the scalar field mass 
$\mu=\lbrace0.05,0.08,0.1,0.15,0.2\rbrace$. The choice of the values for the energy is made to span 
all regimes of interest, from the test-field regime to that for which the self-gravity of the scalar 
field becomes important. For all models, the distribution 
of the scalar field pulse is placed at the same initial location $r_{0}=100$ and  they all share the 
same initial width $\lambda=50$. 

We have used a logarithmic radial grid in our simulations setting to $\Delta r=0.1$ the minimum radial resolution at the origin. This is different from the grid we used in Paper 1. The reason behind this choice is to reduce the number of radial points without loosing accuracy in the inner region and placing the outer boundary sufficiently far away to avoid unwanted reflections from the scalar field. Our grid is composed of two patches, a geometrical progression in the interior and a hyperbolic cosine outside. Using the inner grid alone requires too many grid points to place the outer boundary sufficiently far from the origin, while the hyperbolic cosine patch produces very small grid spacings in the inner region of the grid, leading to prohibitively small timesteps due to the CFL condition. The two patches are defined at each radial point as
\begin{eqnarray}
r_{\rm{geo}}(i)&=&i\,\Delta r\,g_{\rm{prog}}^{i}-r_{\rm{min}}\\
r_{\rm{cosh}}(i)&=&\rm{cosh}[i\,(-\rm{log}((1+r_{\rm{max}})\nonumber\\
&+&\sqrt{2r_{\rm{max}}+r^{2}_{\rm{max}}})/(N_{r}))]-1,
\end{eqnarray} 
where $\Delta r$ is the minimum resolution, $r_{\rm{min}}$ and $r_{\rm{max}}$ the inner and outer boundary respectively, and $N_{r}$ the total number of radial points. The outer boundary of the computational domain is placed at 
$r_{\rm{max}}=4\times10^4$, far enough as to not affect the dynamics in the inner region during the entire 
simulation. For the matching of the two patches, we compute the grid spacings $\Delta r_i$ for the two patches $r_{\rm{geo}}(i)$ and $r_{\rm{cosh}}(i)$ at each grid point and when the difference between the grid spacings is at a minimum, we change from the interior grid to the hyperbolic cosine grid from the corresponding radial point onwards. In Fig. \ref{fg:grid} we plot the two patches and the resulting grid as a function of the number of grid points. The time step is given by 
$\Delta t=0.5\Delta r$ in order to obtain long-term stable simulations. The final time of the 
numerical evolutions is $3.5\times10^4$.

\begin{figure}
\begin{minipage}{1\linewidth}
\includegraphics[width=1.055\textwidth, height=0.3\textheight]{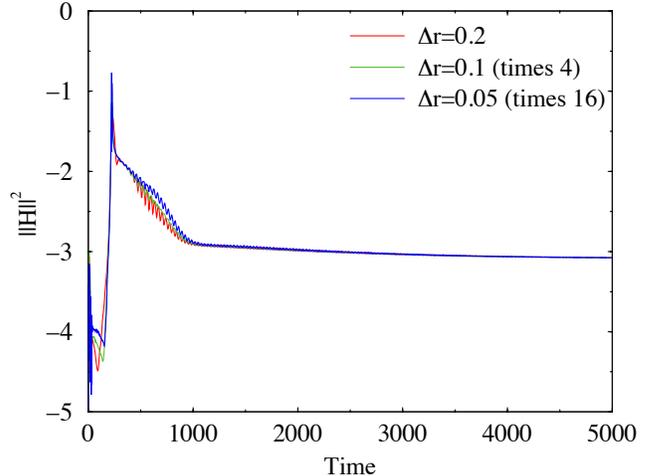} 
\caption{Time evolution of the L2 norm of the Hamiltonian constraint for an initial scalar field 
pulse with 
$E_0=5\times10^{-3}$ and mass $M\mu=0.1$ for the TOV model 4. The plot shows results for three 
different resolutions, 
rescaled by the factors corresponding to second-order convergence.}
\label{fg:Converg}
\end{minipage}
\end{figure}

\begin{table*}
\vspace{-1cm}
\caption{Initial parameters and most relevant quantities for the scalar field and collapsing TOV 
model 1 ($M=1.43$). From left to right the columns report: the name of the model,  the initial 
amplitude of the pulse, $A_0$, the scalar field mass, $\mu$, the real part of the angular frequency 
$\omega$ for the fundamental mode of oscillation and the first overtone, the time of the collapse of 
the TOV star and the formation of the black hole, $t_{\rm{col}}$, the final mass of the apparent 
horizon of the black hole, $M_{\rm{AH}}$, the initial ADM mass of the system, $M_{\rm{ADM}}$, and 
the initial and final energy of the scalar field, $E_{0}$ and $E_{\rm{final}}$ respectively. The 
initial Gaussian pulse is located at $r_0=100$ with half-width $\lambda=50$.}
\label{tab:mod1}
\begin{ruledtabular}
\begin{tabular}{cccccccccc}
Model&$A_{0}$&$\mu$&\multicolumn{2}{c}{$ 
\,\,\omega$}&$t_{\text{col}}$&$M_{\rm{AH}}$&$M_{\rm{ADM}}$&$E_{0}$&$E_{\text{final}}$\\
\cline{4-5}
 &&&1&2 \\

\hline
1\_1a&5.00E-5&0.05&0.04990&...&2494&1.40&1.43&3.2E-5&3.1E-5\\
1\_1b&3.30E-5&0.08&0.07935&0.07988&2494&1.40&1.43&3.2E-5&2.0E-5\\
1\_1c&2.65E-5&0.10&0.09874&0.09964&2494&1.40&1.43&3.2E-5&1.2E-5\\
1\_1d&1.80E-5&0.15&0.14901&0.14954&2494&1.40&1.43&3.2E-5&1.2E-5\\
1\_1e&1.35E-5&0.20&0.19783&0.19927&2494&1.40&1.43&3.2E-5&5.0E-6\\
\hline
1\_2a&6.30E-3&0.05&0.05009&...&1706&1.40&1.92&0.50&0.49\\
1\_2b&4.12E-3&0.08&0.07953&0.08007&1582&1.67&1.92&0.50&0.22\\
1\_2c&3.33E-3&0.10&0.09859&0.09999&1508&1.76&1.92&0.50&0.15\\
1\_2d&2.25E-3&0.15&0.15008&...&1376&1.77&1.92&0.50&0.14\\
1\_2e&1.70E-3&0.20&0.20017&...&1306&1.90&1.92&0.50&0.02\\
\hline
1\_3a&1.50E-2&0.05&0.05099&...&774&1.49&4.20&2.85&2.71\\
1\_3b&9.82E-3&0.08&0.07909&0.08115&748&3.38&4.19&2.84&0.81\\
1\_3c&7.95E-3&0.10&0.10197&...&738&3.47&4.19&2.84&0.70\\
1\_3d&5.36E-3&0.15&0.15314&...&728&4.10&4.19&2.84&0.07\\
1\_3e&4.04E-3&0.20&0.20412&...&730&4.13&4.20&2.84&0.02\\
\hline
1\_4a&1.85E-2&0.05&0.05153&...&606&1.83&5.72&4.31&3.81\\
1\_4b&1.22E-2&0.08&0.08276&...&608&4.49&5.77&4.36&1.18\\
1\_4c&9.85E-3&0.10&0.10341&...&612&5.06&5.75&4.34&0.54\\
1\_4d&6.65E-3&0.15&0.15512&...&626&5.53&5.76&4.35&0.07\\
1\_4e&5.00E-3&0.20&0.20665&...&642&5.60&5.73&4.33&0.007\\

\end{tabular}
\end{ruledtabular}
\end{table*}

\begin{table*}
\caption{Same as Table~\ref{tab:mod1} but for TOV model 2.}\label{tab:mod2}
\begin{ruledtabular}
\begin{tabular}{cccccccccc}
Model&$A_{0}$&$\mu$&\multicolumn{2}{c}{$ 
\,\,\omega$}&$t_{\text{col}}$&$M_{\rm{AH}}$&$M_{\rm{ADM}}$&$E_{0}$&$E_{\text{final}}$\\
\cline{4-5}
 &&&1&2 \\

\hline
2\_1a&5.00E-5&0.05&0.04991&...&750&1.28&1.31&3.2E-5&3.1E-5\\
2\_1b&3.30E-5&0.08&0.07953&...&750&1.28&1.31&3.2E-5&2.1E-5\\
2\_1c&2.65E-5&0.10&0.09909&0.09981&750&1.28&1.31&3.2E-5&1.0E-5\\
2\_1d&1.80E-5&0.15&0.14650&0.14918&750&1.28&1.31&3.2E-5&9.9E-6\\
2\_1e&1.35E-5&0.20&0.19802&0.19927&750&1.28&1.31&3.2E-5&2.7E-6\\
\hline
2\_2a&6.30E-3&0.05&0.05009&...&640&1.28&1.80&0.50&0.49\\
2\_2b&4.12E-3&0.08&0.07971&0.08025&650&1.54&1.80&0.50&0.23\\
2\_2c&3.33E-3&0.10&0.09859&0.10008&654&1.64&1.80&0.50&0.16\\
2\_2d&2.25E-3&0.15&0.14936&0.15009&660&1.67&1.80&0.50&0.10\\
2\_2e&1.70E-3&0.20&0.20017&...&666&1.76&1.80&0.50&0.03\\
\hline
2\_3a&1.50E-2&0.05&0.05117&...&438&1.36&4.07&2.84&2.71\\
2\_3b&9.82E-3&0.08&0.07918&0.08170&472&3.20&4.07&2.83&0.90\\
2\_3c&7.95E-3&0.10&0.10197&...&488&3.39&4.07&2.83&0.65\\
2\_3d&5.36E-3&0.15&0.15314&...&516&3.98&4.07&2.83&0.09\\
2\_3e&4.04E-3&0.20&0.20412&...&536&4.01&4.07&2.83&0.02\\
\hline
2\_4a&1.85E-2&0.05&0.05170&...&386&1.58&5.60&4.30&3.92\\
2\_4b&1.22E-2&0.08&0.08276&...&426&4.32&5.65&4.35&1.18\\
2\_4c&9.85E-3&0.10&0.10341&...&444&5.05&5.63&4.33&0.44\\
2\_4d&6.65E-3&0.15&0.15512&...&478&5.40&5.63&4.33&0.08\\
2\_4e&5.00E-3&0.20&0.20678&...&500&5.48&5.62&4.32&0.008\\

\end{tabular}
\end{ruledtabular}
\end{table*}

\begin{table*}
\caption{Same as Table~\ref{tab:mod1} but for TOV model 3.}\label{tab:mod3}
\begin{ruledtabular}
\begin{tabular}{cccccccccc}
Model&$A_{0}$&$\mu$&\multicolumn{2}{c}{$ 
\,\,\omega$}&$t_{\text{col}}$&$M_{\rm{AH}}$&$M_{\rm{ADM}}$&$E_{0}$&$E_{\text{final}}$\\
\cline{4-5}
 &&&1&2 \\

\hline
3\_1a&5.00E-5&0.05&0.04991&...&228&1.09&1.10&3.2E-5&3.1E-5\\
3\_1b&3.30E-5&0.08&0.07971&...&228&1.09&1.10&3.2E-5&2.6E-5\\
3\_1c&2.65E-5&0.10&0.09928&0.09982&228&1.09&1.10&3.2E-5&1.4E-5\\
3\_1d&1.80E-5&0.15&0.14758&0.14936&228&1.09&1.10&3.2E-5&1.1E-5\\
3\_1e&1.35E-5&0.20&0.19855&0.19945&228&1.09&1.10&3.2E-5&3.1E-6\\
\hline
3\_2a&6.30E-3&0.05&0.05009&...&226&1.10&1.60&0.50&0.49\\
3\_2b&4.12E-3&0.08&0.07989&...&228&1.22&1.60&0.50&0.36\\
3\_2c&3.33E-3&0.10&0.09928&0.10018&228&1.40&1.60&0.50&0.16\\
3\_2d&2.25E-3&0.15&0.14953&0.15026&228&1.50&1.60&0.50&0.09\\
3\_2e&1.70E-3&0.20&0.19827&0.20017&228&1.53&1.60&0.50&0.04\\
\hline
3\_3a&1.50E-2&0.05&0.05117&...&220&1.13&3.91&2.82&2.71\\
3\_3b&9.82E-3&0.08&0.07953&0.08133&226&2.94&3.91&2.82&0.91\\
3\_3c&7.95E-3&0.10&0.10197&...&226&3.22&3.91&2.82&0.61\\
3\_3d&5.36E-3&0.15&0.15314&...&228&3.72&3.90&2.81&0.11\\
3\_3e&4.04E-3&0.20&0.20412&...&228&3.82&3.91&2.82&0.03\\
\hline
3\_4a&1.85E-2&0.05&0.05170&...&218&1.23&5.35&4.28&4.02\\
3\_4b&1.22E-2&0.08&0.08276&...&226&4.15&5.39&4.32&1.15\\
3\_4c&9.85E-3&0.10&0.10341&...&226&4.82&5.35&4.30&0.43\\
3\_4d&6.65E-3&0.15&0.15512&...&228&5.17&5.38&4.31&0.08\\
3\_4e&5.00E-3&0.20&0.20682&...&228&5.29&5.37&4.29&0.01\\

\end{tabular}
\end{ruledtabular}
\end{table*}

\begin{table*}
\caption{Same as Table~\ref{tab:mod1} but for TOV model 4.}\label{tab:mod4}
\begin{ruledtabular}
\begin{tabular}{cccccccccc}
Model&$A_{0}$&$\mu$&\multicolumn{2}{c}{$ 
\,\,\omega$}&$t_{\text{col}}$&$M_{\rm{AH}}$&$M_{\rm{ADM}}$&$E_{0}$&$E_{\text{final}}$\\
\cline{4-5}
 &&&1&2\\

\hline
4\_1a&5.00E-5&0.05&0.04991&...&72&0.75&0.80&3.2E-5&3.1E-5\\
4\_1b&3.30E-5&0.08&0.07989&...&72&0.75&0.80&3.2E-5&3.1E-5\\
4\_1c&2.25E-5&0.10&0.09963&...&72&0.75&0.80&2.3E-5&1.9E-5\\
4\_1d&1.80E-5&0.15&0.14882&0.14973&72&0.75&0.80&3.2E-5&1.4E-5\\
4\_1e&1.35E-5&0.20&0.19927&0.19963&72&0.75&0.80&3.2E-5&7.5E-5\\
\hline
4\_2a&6.30E-3&0.05&0.05009&...&72&0.75&1.29&0.50&0.49\\
4\_2b&4.12E-3&0.08&0.08007&...&72&0.77&1.29&0.50&0.47\\
4\_2c&3.33E-3&0.10&0.09982&0.10035&72&0.89&1.28&0.49&0.35\\
4\_2d&2.25E-3&0.15&0.14829&0.15008&72&1.09&1.29&0.50&0.19\\
4\_2e&1.70E-3&0.20&0.19908&0.20035&72&1.24&1.29&0.50&0.04\\
\hline
4\_3a&1.50E-2&0.05&0.05117&...&72&0.84&3.60&2.80&2.72\\
4\_3b&9.82E-3&0.08&0.08078&0.08168&72&2.61&3.59&2.79&0.91\\
4\_3c&7.95E-3&0.10&0.10214&...&72&2.63&3.60&2.80&0.88\\
4\_3d&5.36E-3&0.15&0.15314&...&72&3.38&3.59&2.79&0.11\\
4\_3e&4.04E-3&0.20&0.20412&...&72&3.48&3.60&2.80&0.04\\
\hline
4\_4a&1.85E-2&0.05&0.05182&...&72&0.86&5.03&4.24&4.08\\
4\_4b&1.22E-2&0.08&0.08029&0.08276&72&3.53&5.04&4.29&1.45\\
4\_4c&9.85E-3&0.10&0.10341&...&72&4.30&5.06&4.27&0.64\\
4\_4d&6.65E-3&0.15&0.15512&...&72&4.82&5.06&4.27&0.10\\
4\_4e&5.00E-3&0.20&0.20681&...&72&4.93&5.05&4.26&0.01\\

\end{tabular}
\end{ruledtabular}
\end{table*}

\begin{figure*}
\begin{center}
\subfigure{\includegraphics[width=0.5\textwidth]{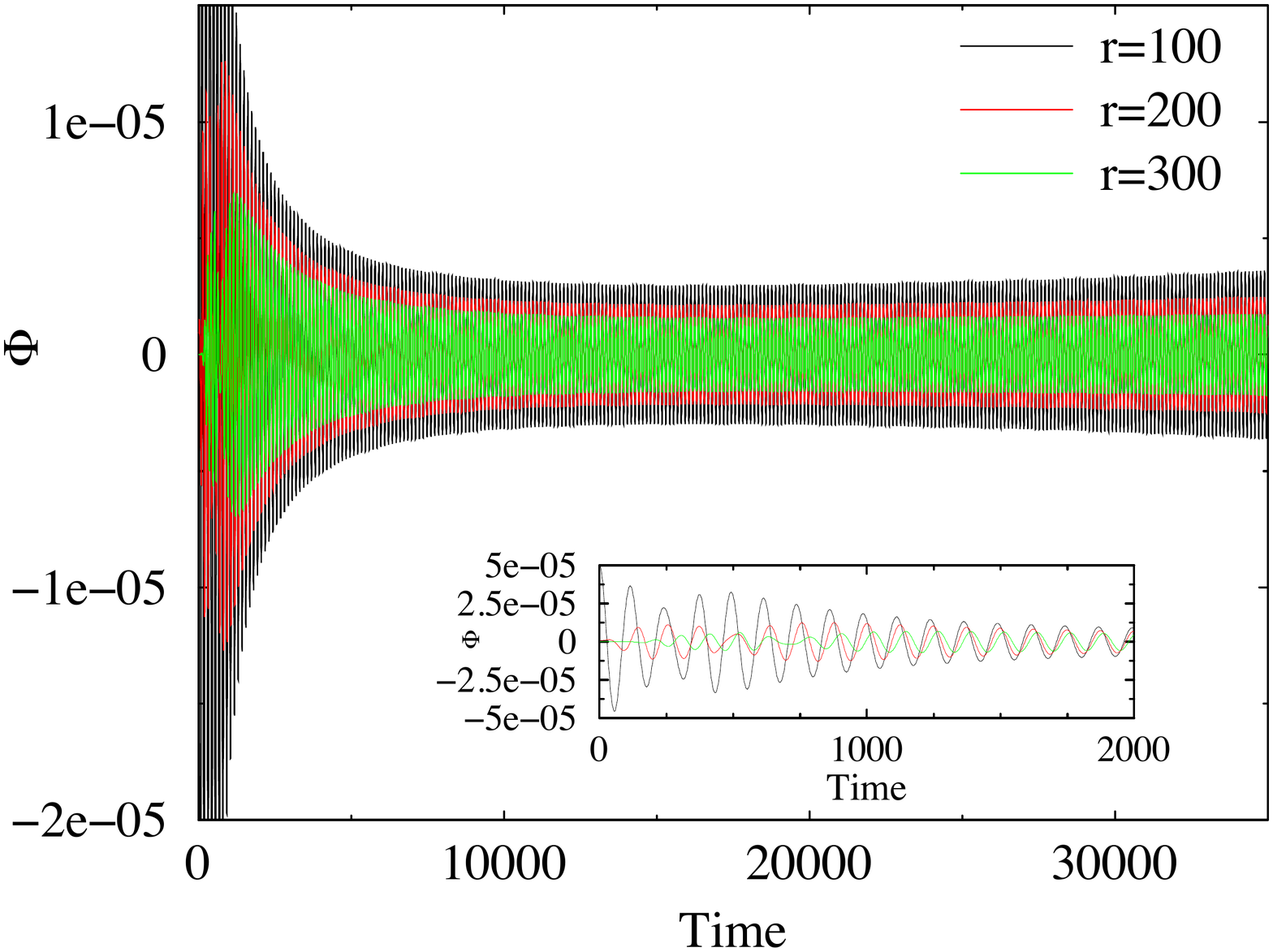}}
\hspace{-1cm}\subfigure{\includegraphics[width=0.5\textwidth]{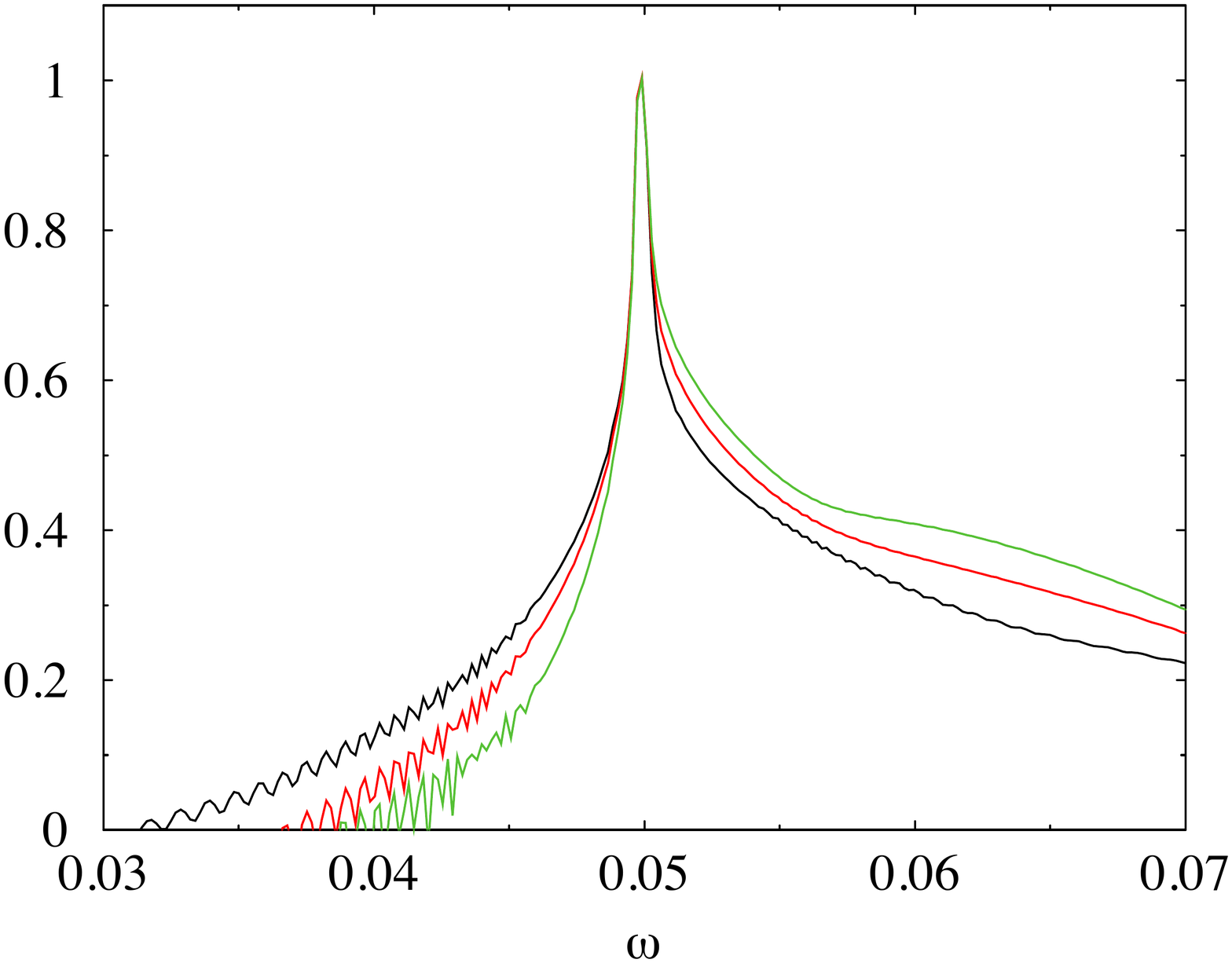}}\\
\vspace{-1cm}\subfigure{\includegraphics[width=0.5\textwidth]{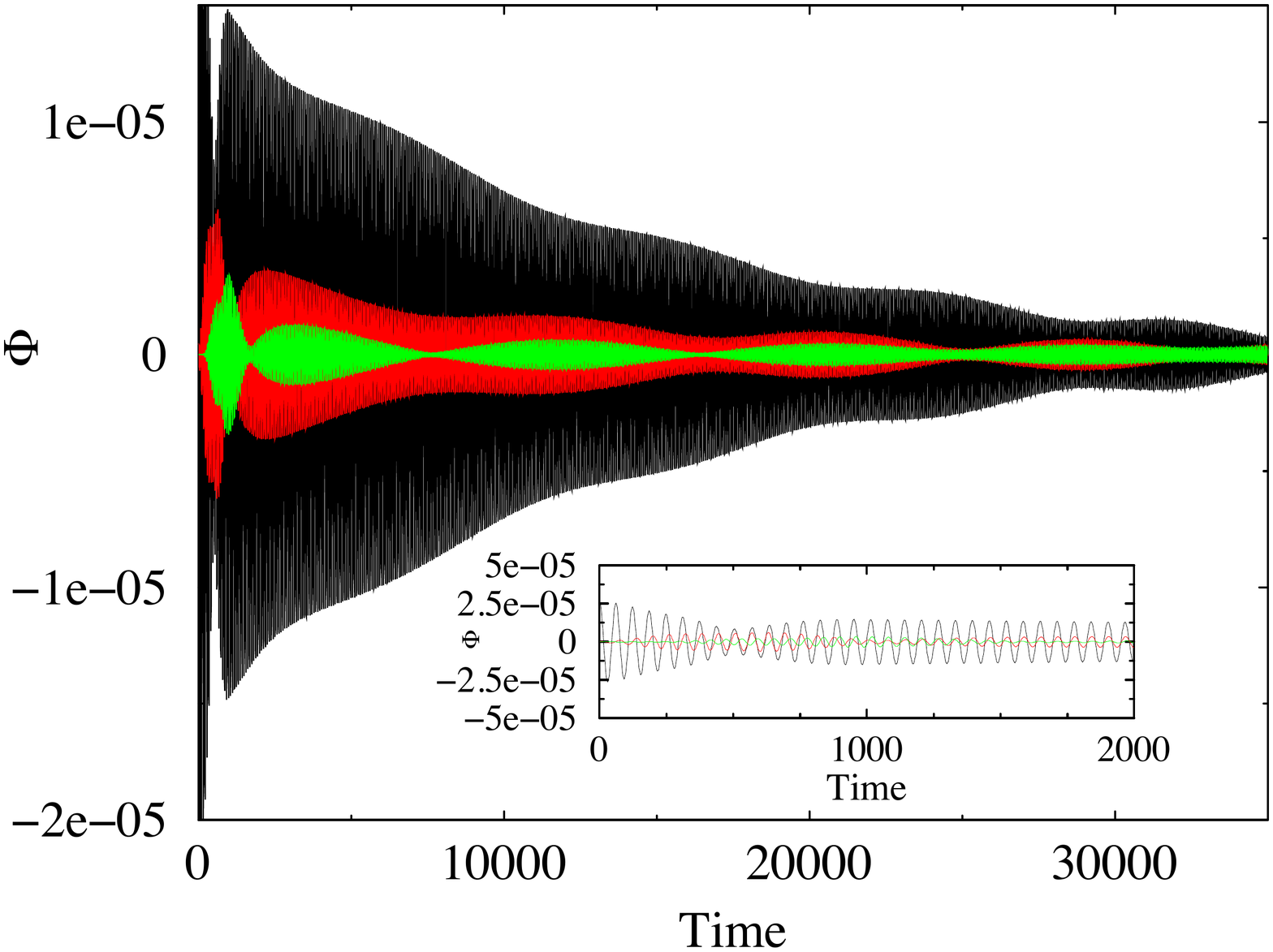}}
\hspace{-1cm}\subfigure{\includegraphics[width=0.5\textwidth]{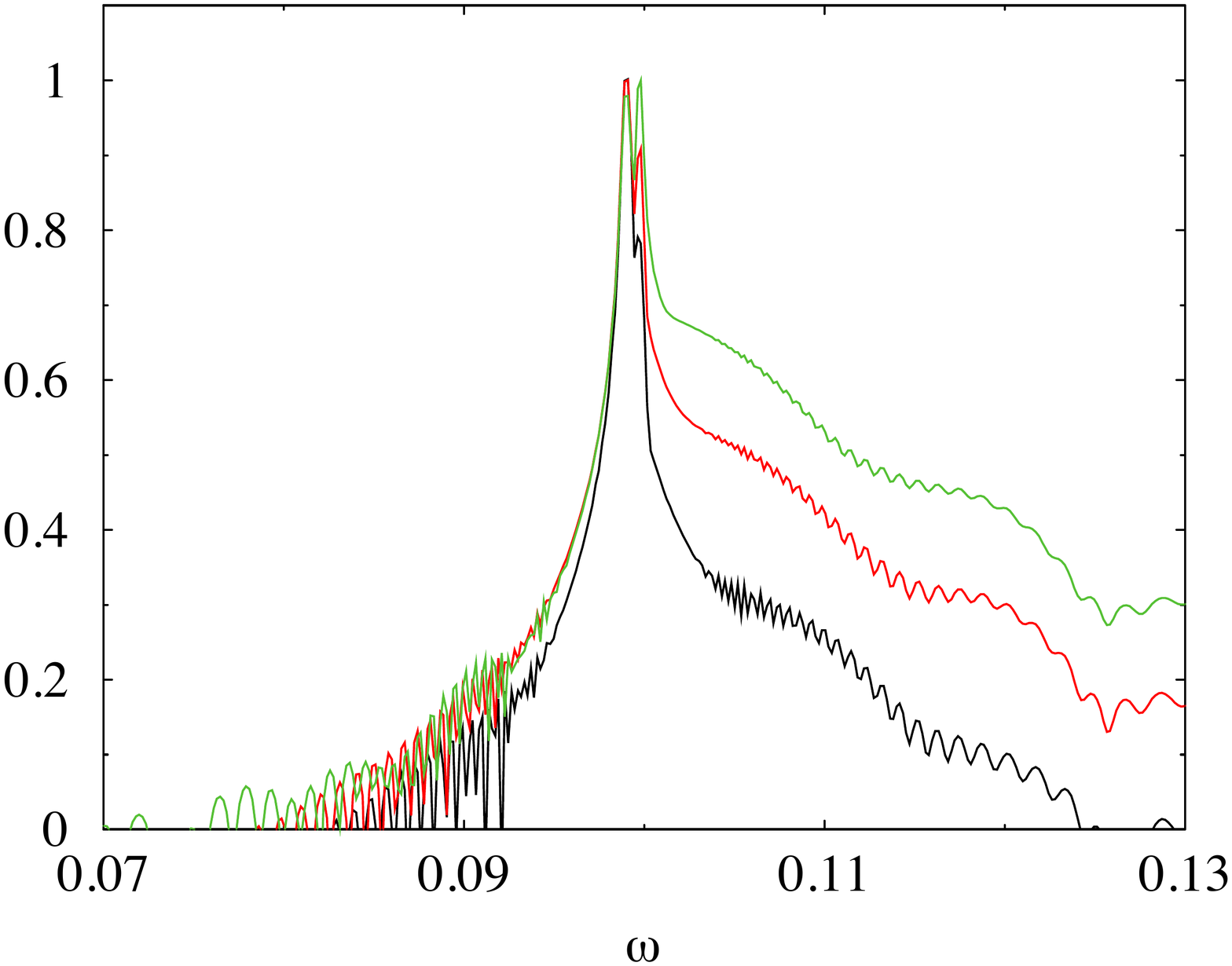}}\\
\vspace{-1cm}\subfigure{\includegraphics[width=0.5\textwidth]{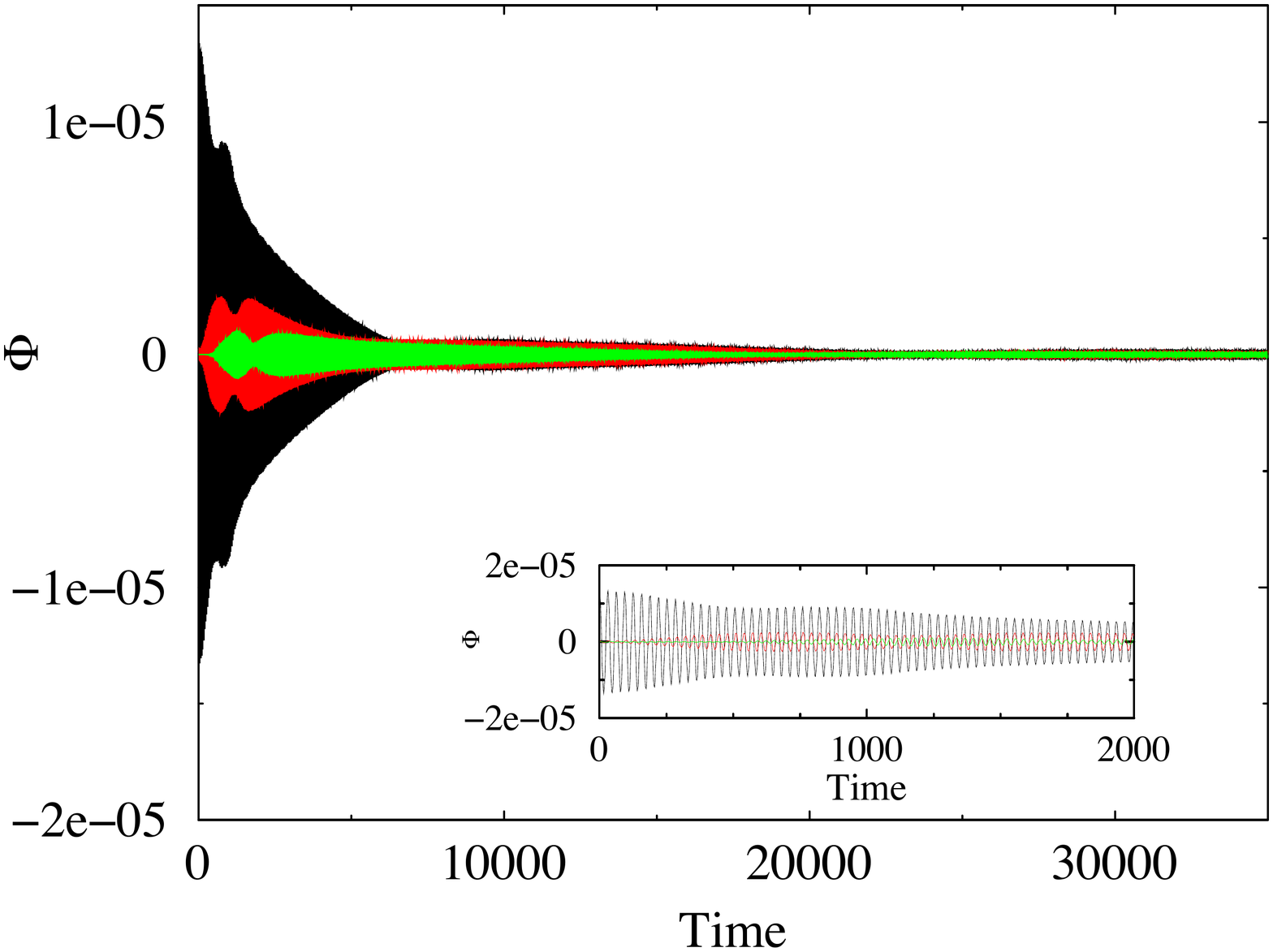}}
\hspace{-1cm}\subfigure{\includegraphics[width=0.5\textwidth]{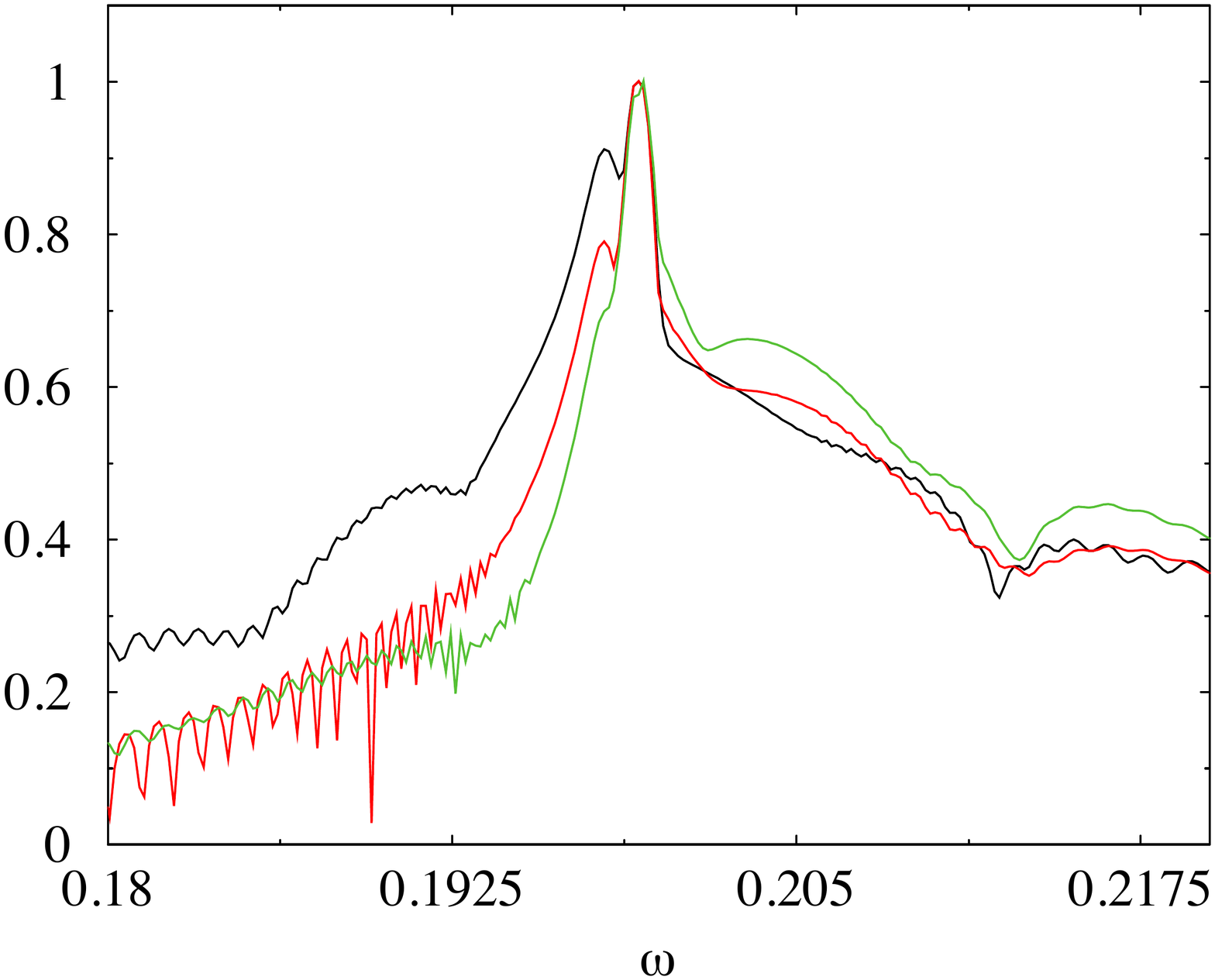}}
\caption{{\it Left column}: Time evolution of the scalar field with initial energy 
$E_{0}=3.2\times10^{-5}$ for the TOV model 2 and three different scalar field masses 
$\mu=\lbrace0.05,0.1,0.2\rbrace$ (from top to bottom). The inset shows a magnified view of the 
initial 2000 units of time in the evolution. {\it Right column}: Fourier transform of the evolution 
of the scalar field. $\omega$ refers to the real part of the frequencies. The units in the vertical 
axis are arbitrary and are normalized by the amplitude of the fundamental mode. Each curve corresponds
to a different extraction radius, as indicated in the legend of the top-left plot.}
\label{fg:SF1}
\end{center}
\end{figure*}

\begin{figure}
\begin{center}
\subfigure{\includegraphics[width=0.5\textwidth]{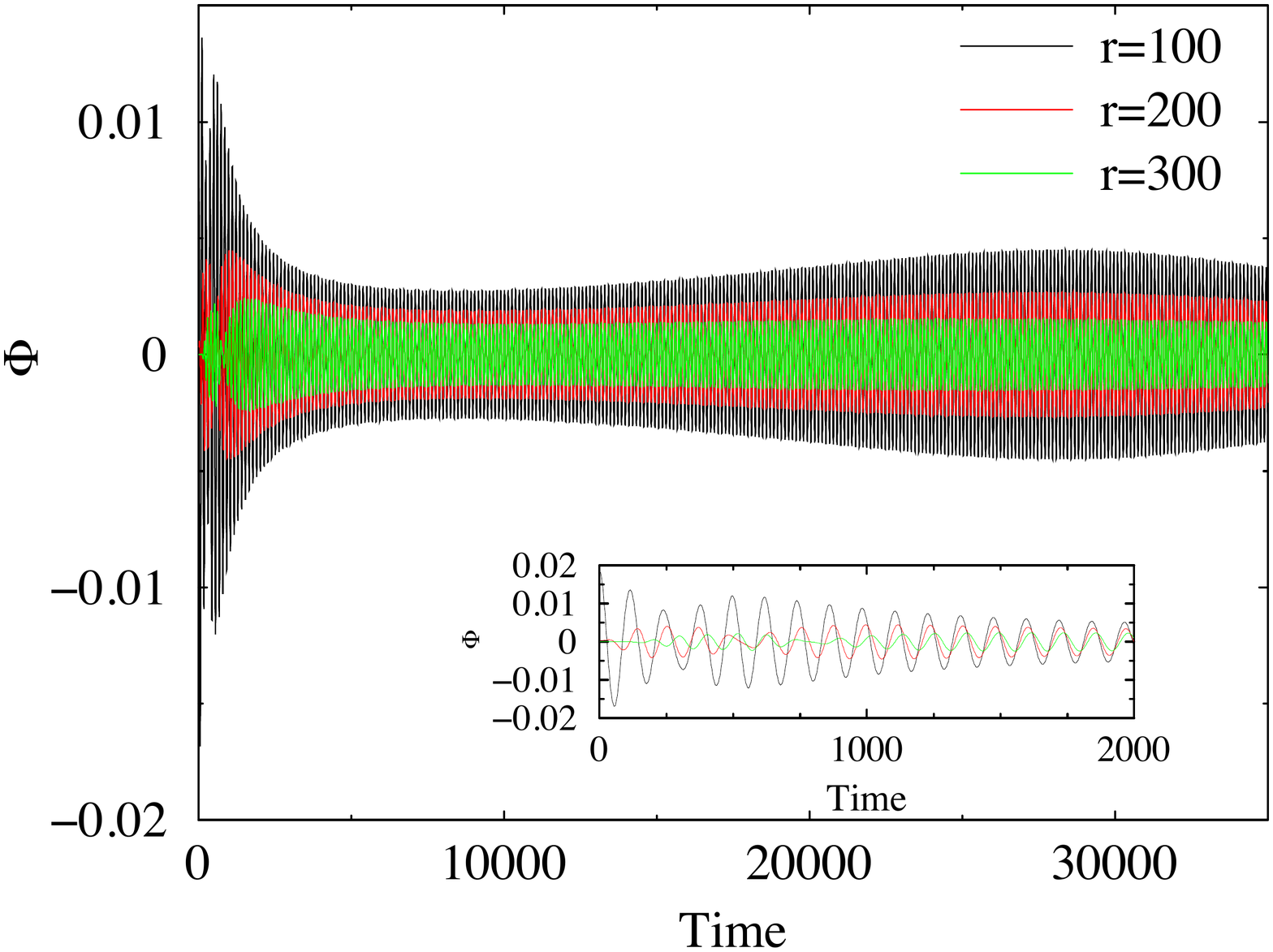}}\\
\vspace{-1.0cm}\subfigure{\includegraphics[width=0.5\textwidth]{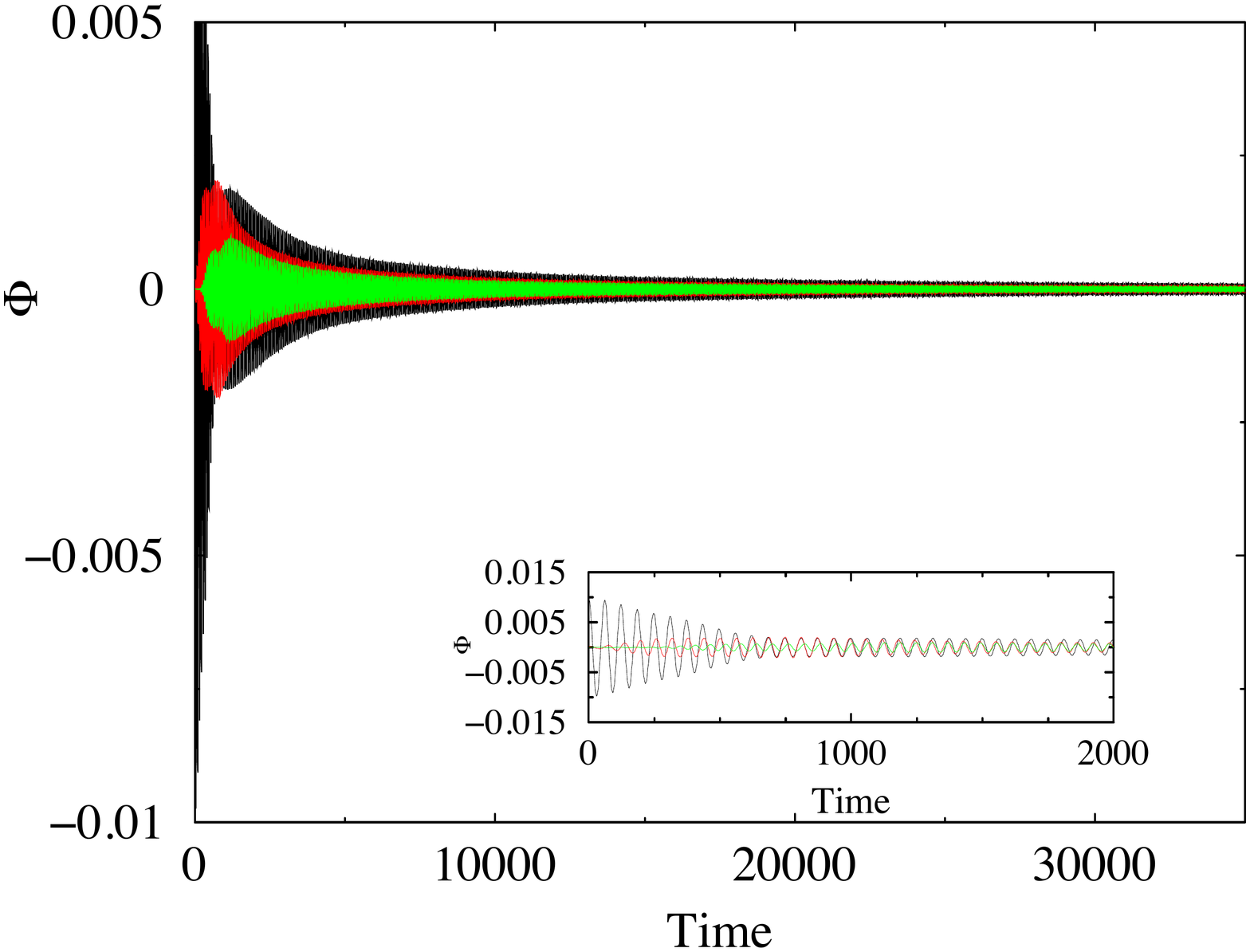}}\\
\vspace{-1.0cm}\subfigure{\includegraphics[width=0.5\textwidth]{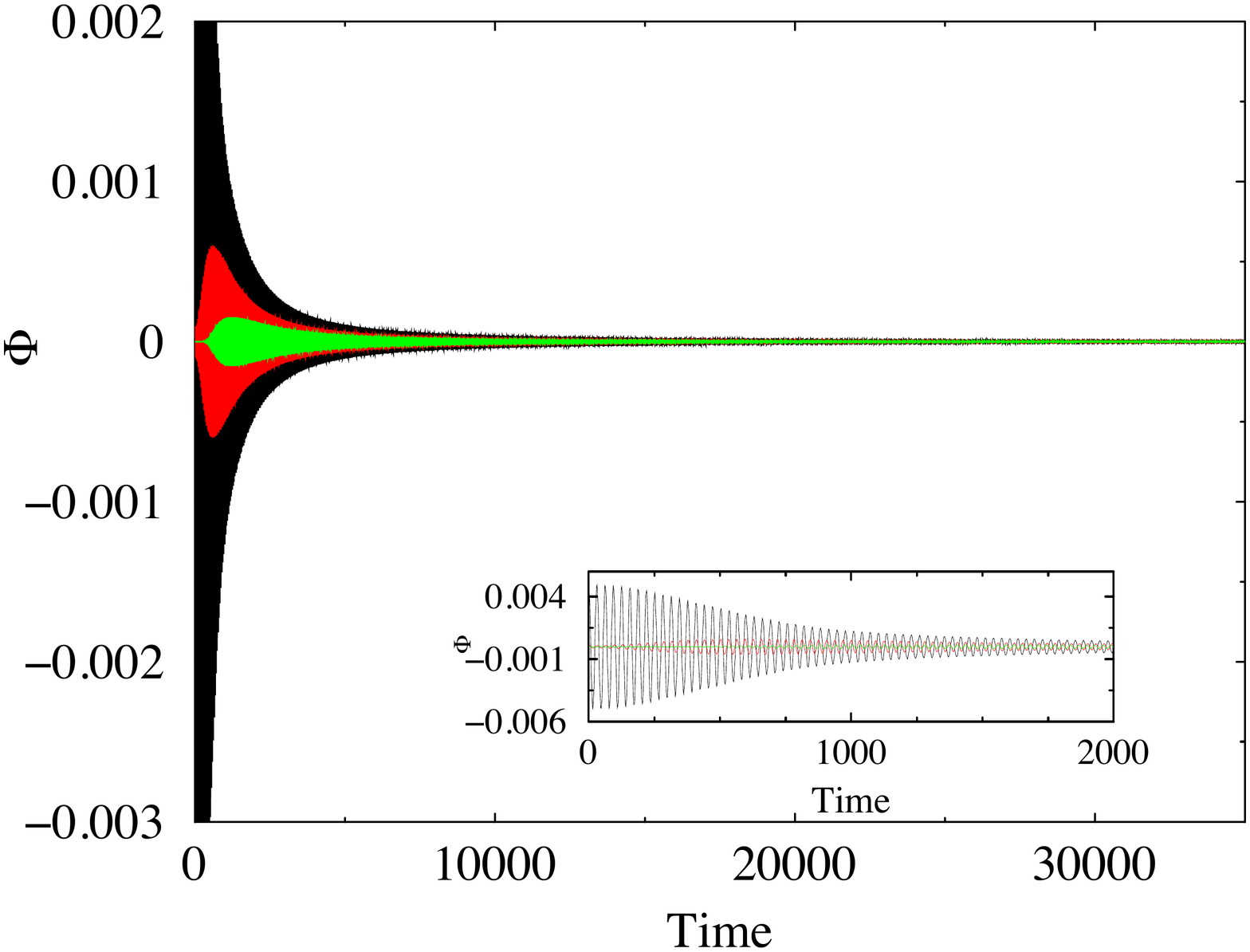}}
\caption{Time evolution of the scalar field with initial energy $E_{0}=4.3$ for the TOV model 2 and 
three different scalar field masses $\mu=\lbrace0.05,0.1,0.2\rbrace$ (from top to bottom). The inset 
shows a magnified view of the initial 2000 units of time in the evolution.}
\label{fg:SF2}
\end{center}
\end{figure}

\begin{figure}
\begin{center}
\subfigure{\includegraphics[width=0.5\textwidth]{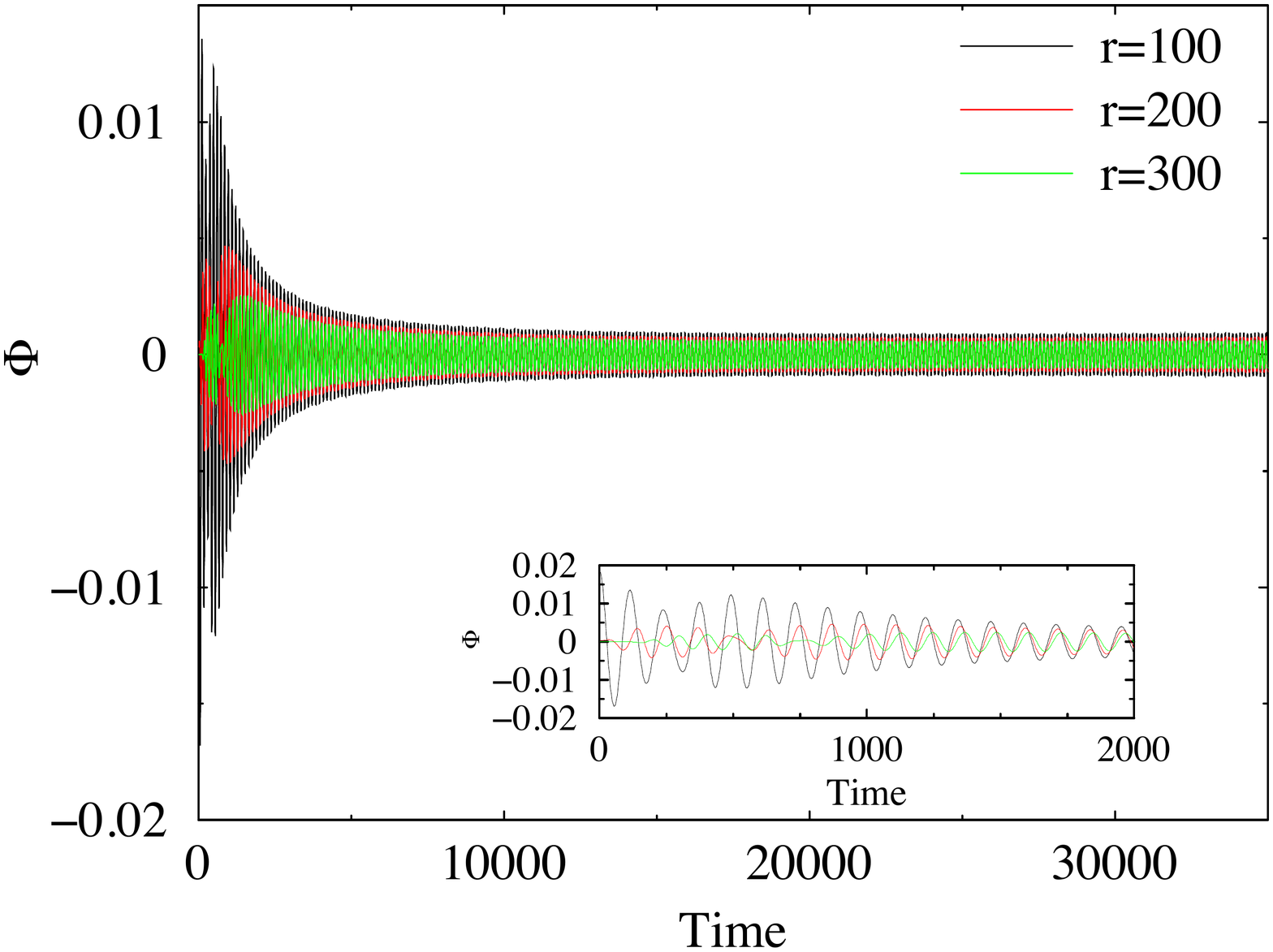}}\\
\vspace{-1.0cm}\subfigure{\includegraphics[width=0.5\textwidth]{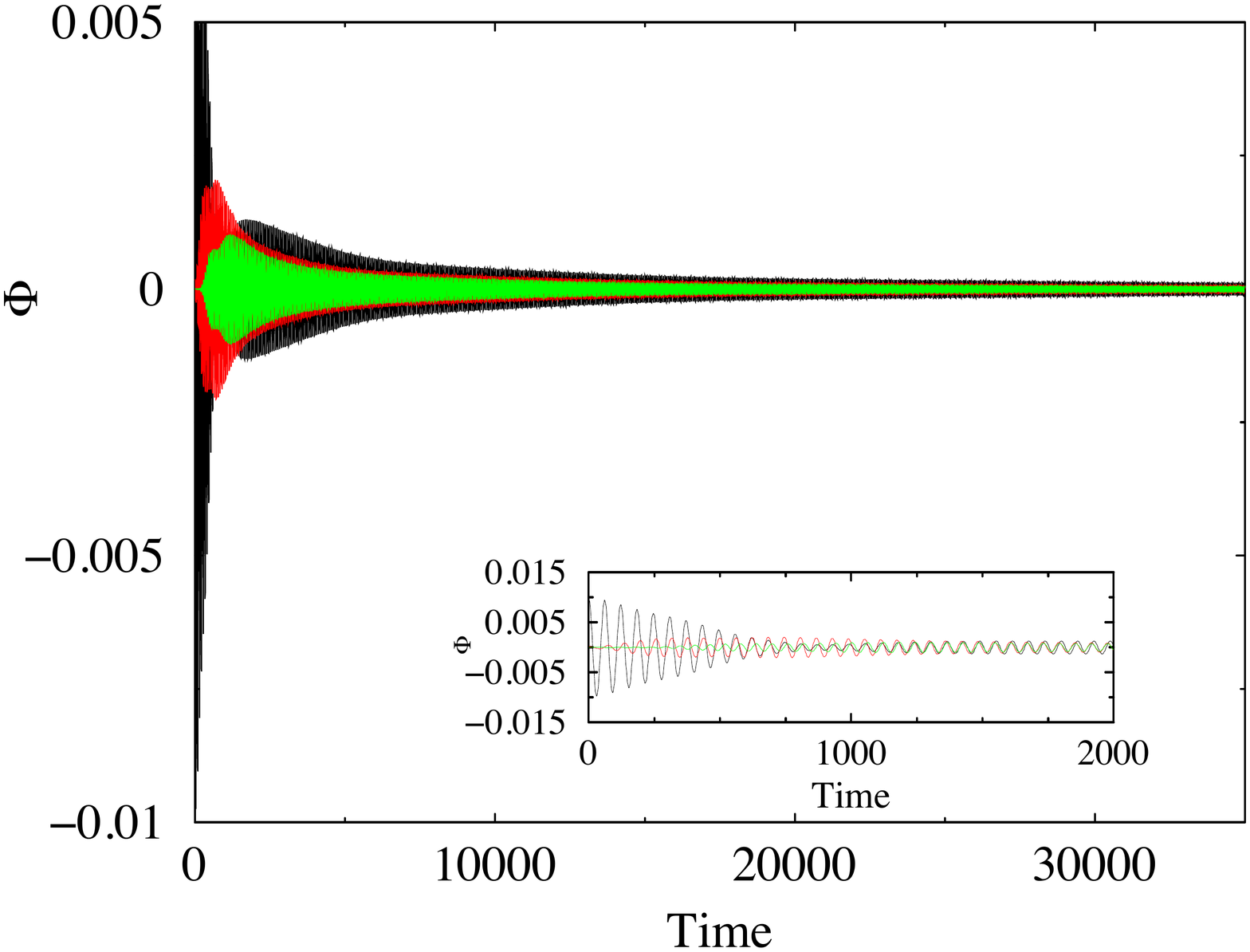}}\\
\vspace{-1.0cm}\subfigure{\includegraphics[width=0.5\textwidth]{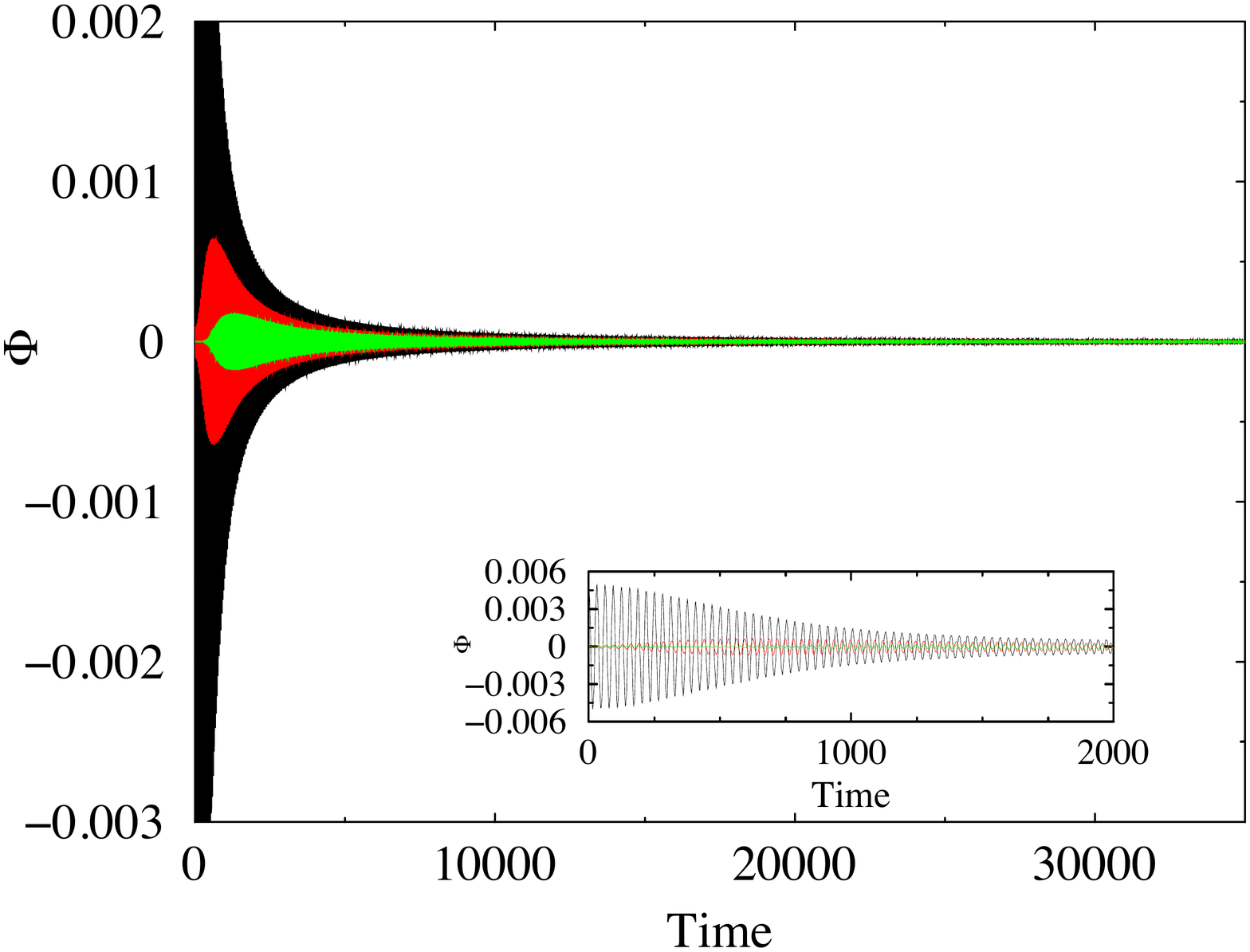}}
\caption{Time evolution of the scalar field with initial energy $E_{0}=4.3$ for the TOV model 4 and 
three different scalar field masses $\mu=\lbrace0.05,0.1,0.2\rbrace$ (from top to bottom). The inset 
shows a magnified view of the initial 2000 units of time in the evolution.}
\label{fg:SF3}
\end{center}
\end{figure}

While the convergence of the code was already satisfactorily tested in Paper 1, we nevertheless need 
to assess it again here since we now deal with an augmented system of equations than in Paper 1 due 
to the presence of the hydrodynamical terms. In order to test the convergence of the code we 
performed three simulations with different resolutions $\Delta r = \lbrace0.2,0.1,0.05\rbrace$. In 
Fig.~\ref{fg:Converg} we plot the rescaled evolution of the L2 norm of the Hamiltonian constraint 
for a particular choice of the initial energy, $E_0=4.3$, scalar field mass, $\mu=0.1$, and for 
the TOV model 4, obtaining the expected second-order convergence of our PIRK time-evolution scheme. 
We note that the same result is achieved irrespective of the combination of parameters considered 
and of the TOV model.

Comparing simulations with different resolutions, we estimate that the error in the final black hole 
mass is $\sim 2$\% while the error estimates for the scalar field energy and ADM mass at the end of 
the simulations are $\sim 2-3$\%. Those errors decrease as $(\Delta r)^2$ with increasing resolution. 
Taking into account that our simulations are performed up to a (significantly long) final time of 
$3.5\times10^{4}$, the error bars for our main quantifiable quantity, the frequencies of oscillation 
of the scalar field (see next section), are $\sim 3\times10^{-5}$.

\subsection{Non-linear quasi-bound states}

We solve the Einstein-Klein-Gordon-Euler system using the initial data 
given by Eqs.~(\ref{eq:pulse})-(\ref{eq:iderivatives}) and let the 
scalar field evolve in the rapidly changing gravitational field of a 
collapsing polytropic star. As in Paper 1 we analyze the results of 
the simulations by extracting a time series for the scalar field 
amplitude at a set of observation points located at fixed radii 
$r_{\rm{ext}}$ (typically at three radii located at $r_{\rm{ext}}=100, 
200$ and 300). To identify the frequencies at which the scalar 
field oscillates we perform a Fast Fourier transform after a given 
number of time steps and obtain the corresponding power spectrum.  
The very existence of these frequencies for the long-term simulations 
we perform is the main indicator we use to demonstrate the presence of 
quasi-bound states.

The main results of our simulations are summarized in Tables \ref{tab:mod1}-\ref{tab:mod4}. These 
four tables contain for each TOV model, the different choices for the scalar field mass $\mu$, the 
initial amplitude of the pulse $A_{0}$, the Arnowitt-Deser-Misner (ADM) mass 
$M_{\rm{ADM}}$ of the whole system, defined as
\begin{equation}\label{eq:admmass}
M_{\rm{ADM}}=-\frac{1}{2\pi}\lim_{r\rightarrow \infty}\oint_{S}\partial_{j}\psi dS^{j}\quad,
\end{equation}
and the initial energy of the scalar field $E_{0}$
\begin{equation}\label{eq:scalar}
E=\int_{2M}^{\infty}\mathcal{E}^{\text{SF}} dV  \ ,
\end{equation}
where $\mathcal{E}^{\text{SF}}$ is defined in Eq.~(\ref{eq:rho}). 
As the second-to-last column of each table shows, we vary the initial amplitude for the different 
$\mu$ in order to keep the initial scalar field energy constant for four different scenarios, from 
the test-field approximation to the non-linear regime where the scalar field is self-gravitating. 
This is presented as the four subdivisions of each table. The oscillations frequencies of the scalar 
field are reported in columns 4 and 5 for the fundamental mode and first overtone, respectively.

\begin{figure}
\begin{minipage}{1\linewidth}
\includegraphics[width=1.08\textwidth, height=0.3\textheight]{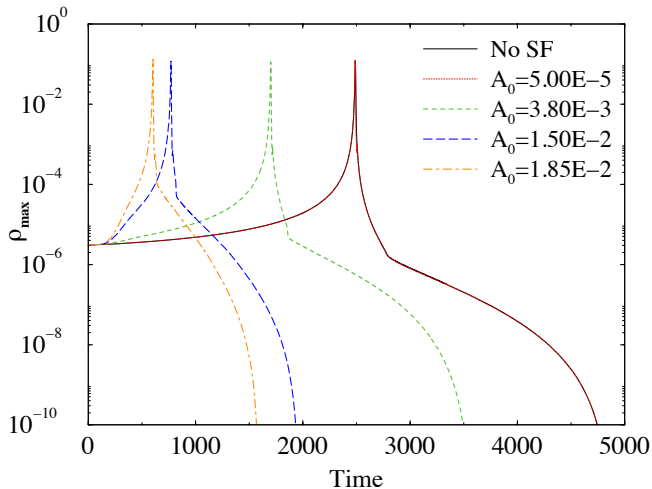} 
\caption{Time evolution of the maximum (central) rest-mass density of the TOV model 1 for different 
initial values of the amplitude of the scalar field pulse with mass $\mu=0.05$. Notice that the 
black and red solid lines overlap.}
\label{fg:maxrho}
\end{minipage}
\end{figure}

Figures \ref{fg:SF1}-\ref{fg:SF3} show the time evolution of the scalar field amplitude extracted at 
three different radii, $r_{\rm{ext}}=\lbrace100,200,300\rbrace$. In the left panels of 
Fig.~\ref{fg:SF1} we plot the evolution for the initial scalar field energy $E_{0}=3.2\times10^{-5}$ 
for the TOV model 2 and for three scalar field masses, namely $\mu=\lbrace0.05,0.1,0.2\rbrace$ from 
top to bottom. The three panels show a moderately rapid decay of the amplitude of the oscillations 
of the scalar field. However, the field amplitudes by the end of the simulations have 
not totally vanished, especially for the $\mu=0.05$ model shown in the top panel. Correspondingly, the 
right panels of Fig.~\ref{fg:SF1} present the power spectra, showing the well-defined oscillation 
frequencies of the quasi-stationary states. It is worth stressing that
the values of the frequencies do not depend on the location of the observer.

Similar time evolution plots are displayed in figure \ref{fg:SF2} but, in this case, for the larger 
initial scalar field energy of our sample, $E_{0}=4.3$ (keeping the same values of $\mu$ and TOV 
model  as in Fig.~\ref{fg:SF1}). The evolution for the same initial parameters corresponding to the 
TOV model 4 can be found in figure \ref{fg:SF3}. The comparison among these time evolution plots  
shows that for the smallest scalar field mass, $\mu=0.05$, the differences between the evolutions 
for the different TOV models are more evident. The TOV model 2 shows an increase in the amplitude of 
the oscillations at the end of the simulation due to the presence of two close frequencies giving 
rise to a beating pattern, while TOV model 4 does not show this behavior. However, for larger scalar 
field masses, the differences in the time evolutions become qualitatively less important.

The power spectra obtained from the Fourier transforms of these time 
series show a set of distinct frequencies, exemplified in 
the right panels of Fig.~\ref{fg:SF1} for a particular set of 
models. The frequency resolution is inversely proportional to the 
simulation time and hence the longer the runs the more accurate the 
frequencies.  The presence of 
these frequencies confirms the existence of long-lived spherical 
quasi-bound states of scalar fields surrounding black holes even if 
these are formed {\it dynamically} through the gravitational 
collapse of spherical stars. During the non-linear evolution of the 
system, part of the scalar field survives the gravitational collapse 
and remains in the form of trapped states. These 
results extend the validity of our previous model, put forward in 
\cite{Barranco:2012qs} and Paper 1, in the linear and non-linear 
regimes respectively, to a significantly more dynamical scenario.

In columns 4 and 5 of Tables \ref{tab:mod1}-\ref{tab:mod4} we show 
the frequencies $\omega$ of the scalar field oscillations. These 
frequencies refer only to the real part of the dominant modes and 
are labelled according to their strength on the power spectrum. For 
some models we are able to obtain several frequencies of oscillation. 
Tables \ref{tab:mod1}-\ref{tab:mod4} 
also report the mass of the apparent horizon (AH) at the end of the 
simulation $M_{\rm{AH}}$, defined as 
$M_{\rm{AH}}=\sqrt{\mathcal{A}/16\pi}$, where $\mathcal{A}$ is the 
area of the AH, the time of collapse at which the AH is first found, 
$t_{\text{col}}$,  and the scalar field energy $E_{\text{final}}$ at 
the end of the simulation. In figure \ref{fg:maxrho} we plot the 
time evolution of the central rest-mass density of the TOV model 1 
for different values of the scalar field amplitude. This figure 
shows that the time at which the TOV collapses decreases 
monotonically when the initial scalar field amplitude increases. We 
note that $t_{\text{col}}$ for the case of the test-field scalar 
field energy ($A_0=5\times 10^{-5}$) coincides perfectly with the 
corresponding time for the case with no scalar field affecting the 
density evolution of the TOV star (as both lines overlap in the 
figure). 

From Tables \ref{tab:mod1}-\ref{tab:mod4} we see that the time of 
collapse shows a clear dependence with the initial energy of the 
scalar field $E_0$ only for the less compact TOV models of our sample, 
models 1 and 2 for which the values of the compactness parameter $M/R$ 
are 0.008 and 0.015, respectively. This time becomes shorter the 
larger $E_0$. Model 3, with $M/R=0.025$, has significantly shorter 
values of $t_{\text{col}}$ and, moreover, $t_{\text{col}}$ barely 
shows any dependence on $E_0$. Finally, for the TOV model 4 with 
$M/R=0.042$ the time of collapse is the shortest of the whole sample 
and constant, $t_{\text{col}}=72$ irrespective of $E_0$. This result 
is explained by the fact that in this case the star collapses to form 
a black hole even before the scalar field pulse can reach the 
collapsing star. Therefore, the presence of the scalar field does not 
influence the collapse dynamics at all, even for the highest scalar 
field energies we consider in our sample.

The oscillation frequencies for the same scalar field parameters vary 
slightly from one TOV model to another due to the different initial 
mass of the black hole when it is formed. For most cases we find that 
the smaller the black hole mass the higher the frequencies.  This 
trend, however, does not apply for the larger energy models, for which 
the frequencies are essentially identical for all the TOV stars. This 
behaviour might be simply a numerical artifact due to the lack of 
sufficient time resolution, as with increasing evolution time  the 
frequencies could be better resolved and differences may appear. 

\begin{figure}[t]
\begin{center}
\subfigure{\includegraphics[width=0.5\textwidth]{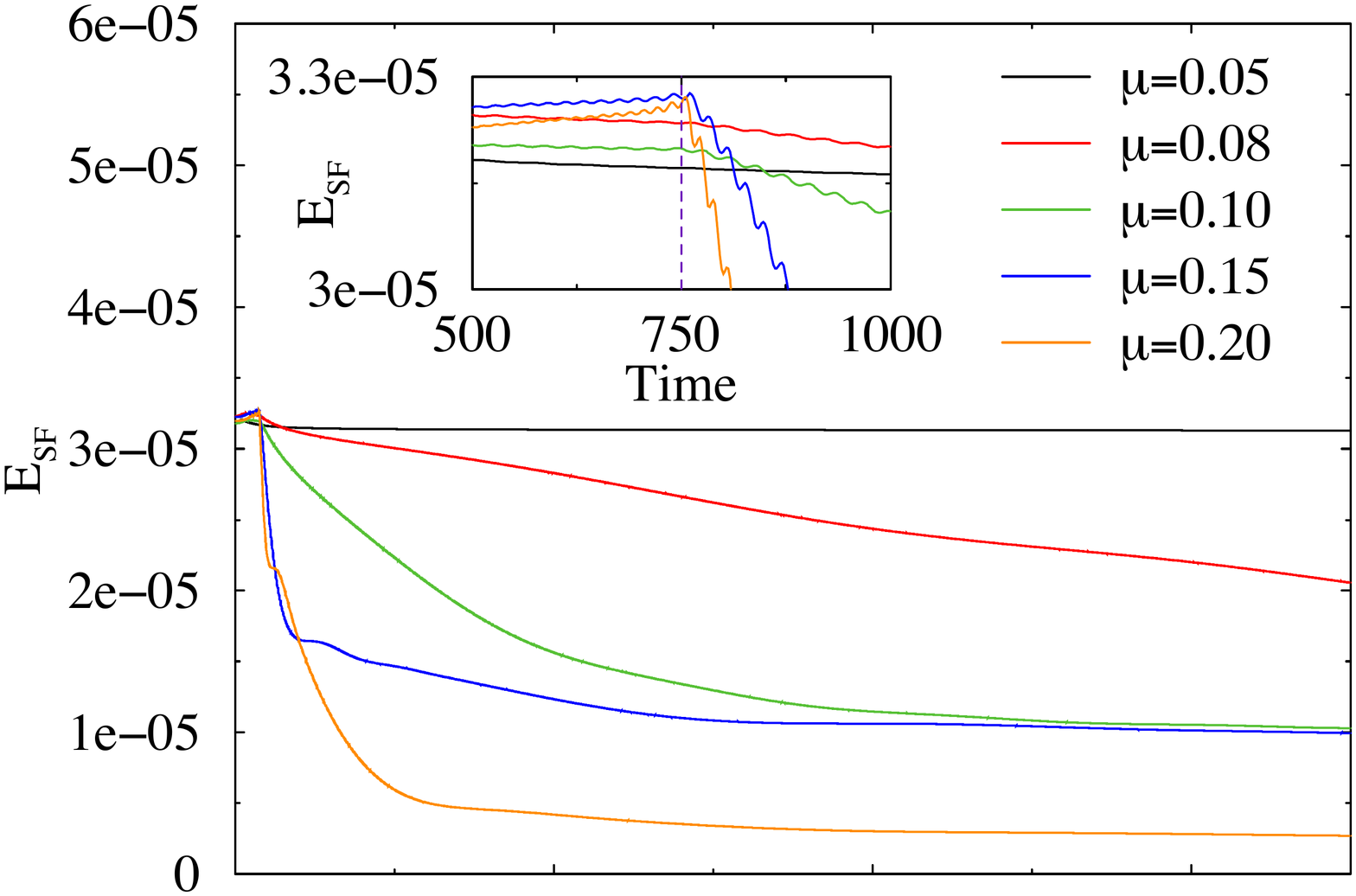}}\\
\vspace{-2.1cm}\subfigure{\includegraphics[width=0.5\textwidth]{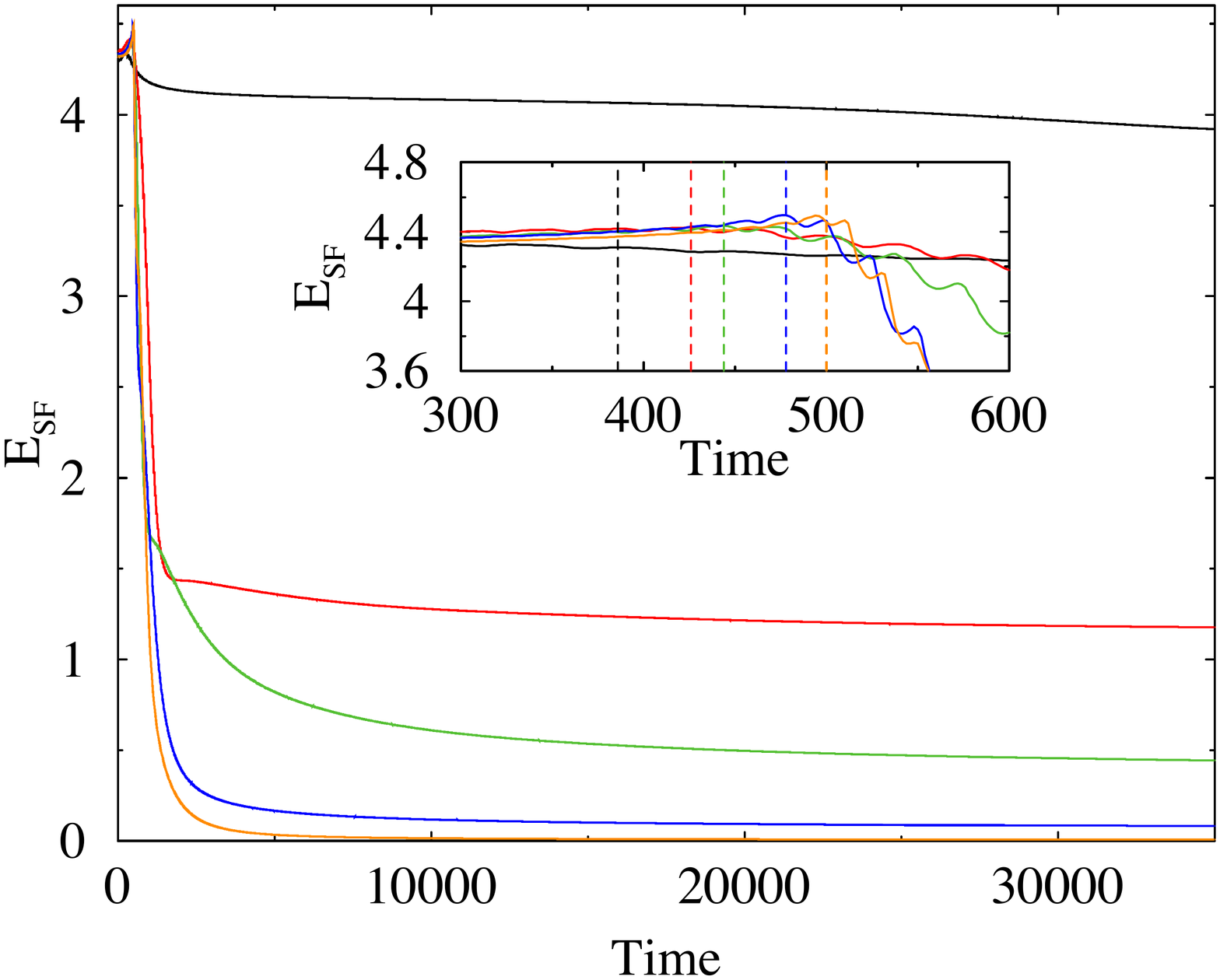}}
\caption{Time evolutions of the scalar field energy from the initial value $E_{0}=3.2\times10^{-5}$ 
({\it top panel}) and from the initial value $E_{0}\sim4.3$ ({\it bottom panel}). The evolutions are 
shown for the TOV model 2 and for all scalar field masses 
$\mu=\lbrace0.05,0.08,0.1,0.15,0.2\rbrace$. The insets show magnified views around the time of 
collapse in the corresponding evolutions. The times at which the black holes are formed are 
displayed as vertical dashed lines. The different colors are related to the different scalar field 
masses as indicated in the legend of the top panel.}
\label{fg:SF4}
\end{center}
\end{figure}

\begin{figure}[t]
\begin{center}
\subfigure{\includegraphics[width=0.5\textwidth]{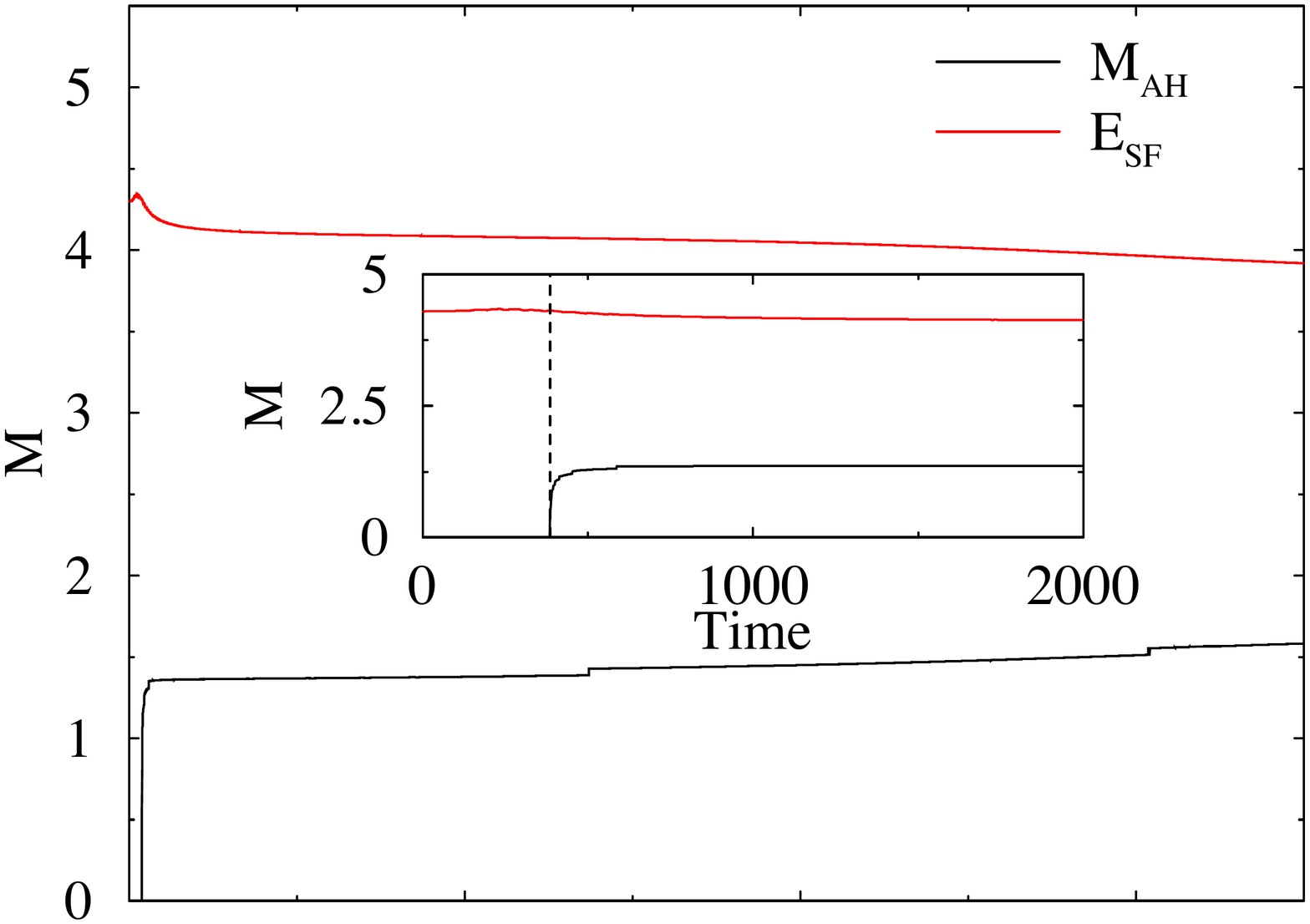}}\\
\vspace{-2.1cm}\subfigure{\includegraphics[width=0.5\textwidth]{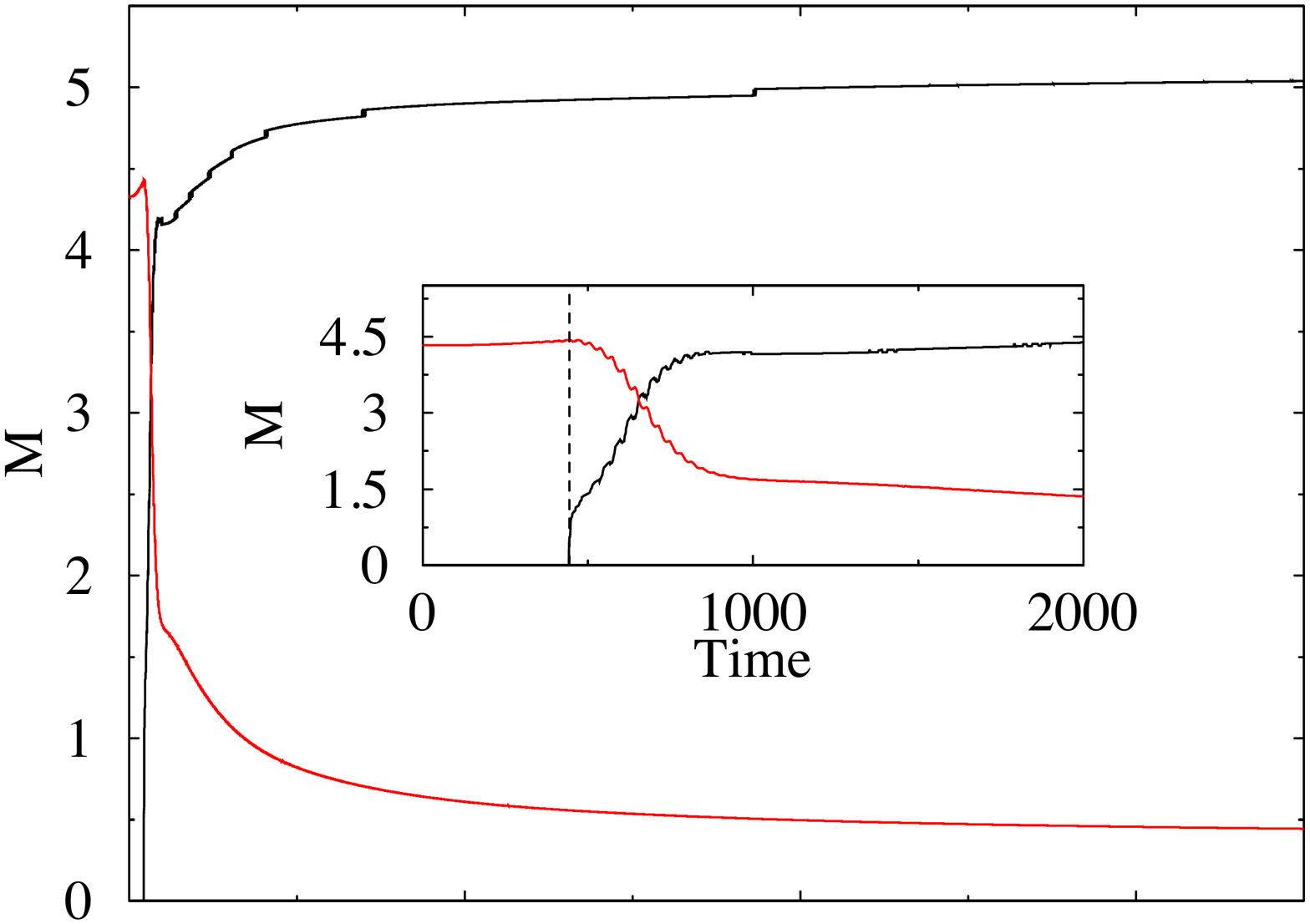}}\\
\vspace{-2.1cm}\subfigure{\includegraphics[width=0.5\textwidth]{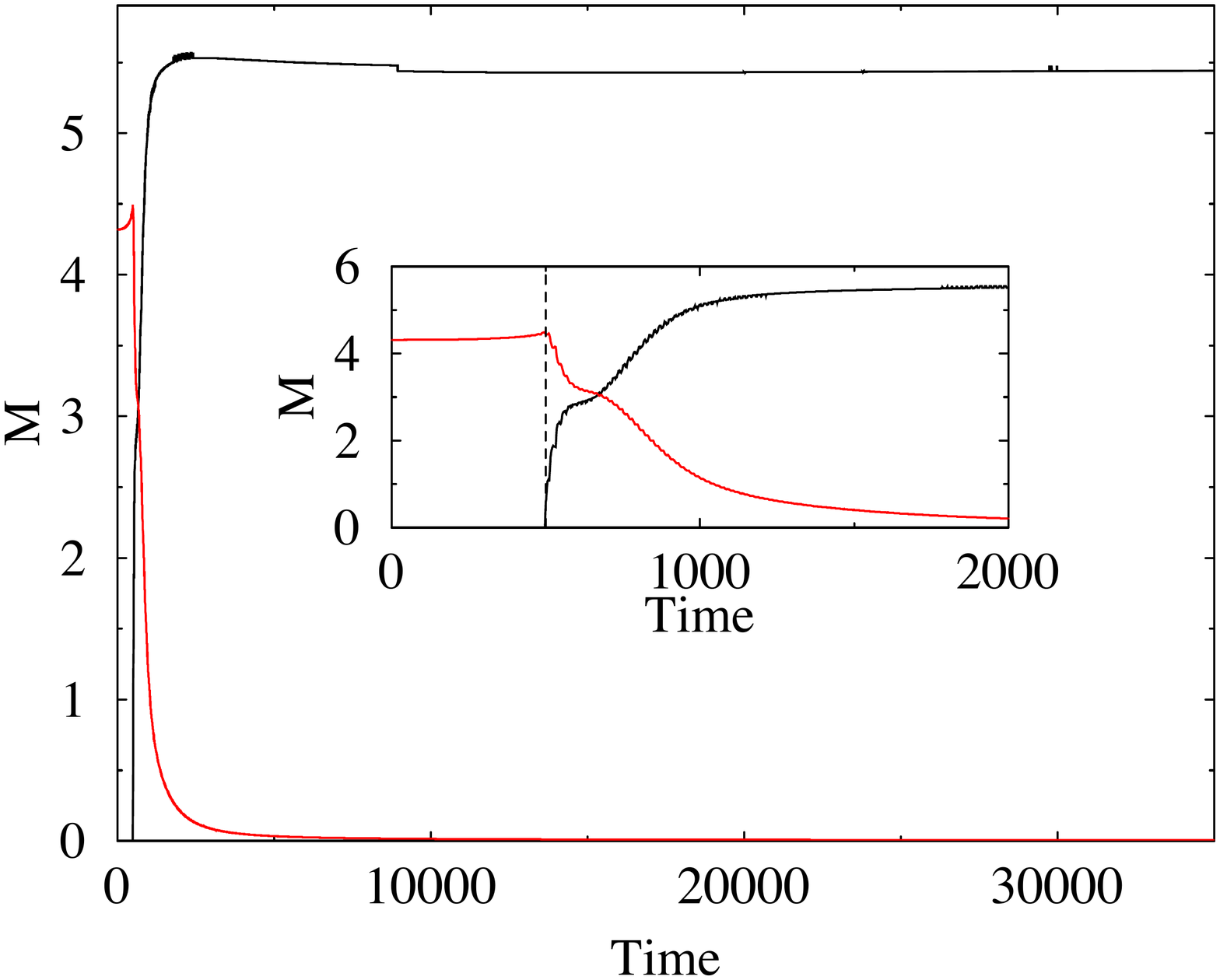}}\\
\vspace{-0.6cm}
\caption{Time evolution of the mass of the apparent horizon $M_{\rm{AH}}$ and of the scalar field 
energy for the initial $E_{0}\sim4.3$ and the scalar field masses 
$\mu=\lbrace0.05,0.1,0.2\rbrace$ from top to bottom corresponding to the TOV model 2. The inset shows a magnified view 
of initial 2000 in the evolution. The time at which the black hole is formed is displayed as a 
vertical dashed line.}
\label{fg:SF5}
\end{center}
\end{figure}

\begin{figure}
\begin{center}
\subfigure{\includegraphics[width=0.5\textwidth]{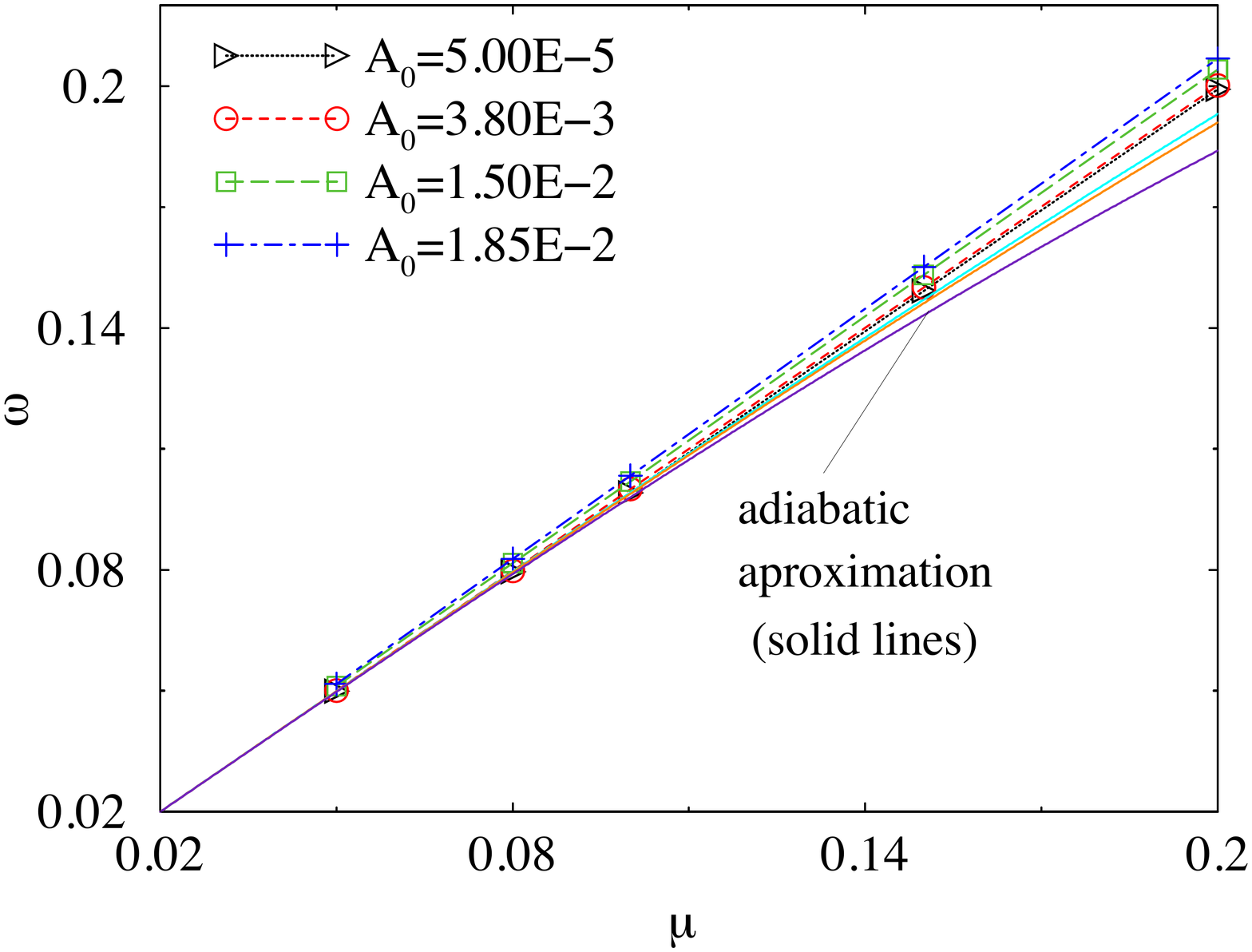}}
\vspace{-1.0cm}\\\subfigure{\includegraphics[width=0.5\textwidth]{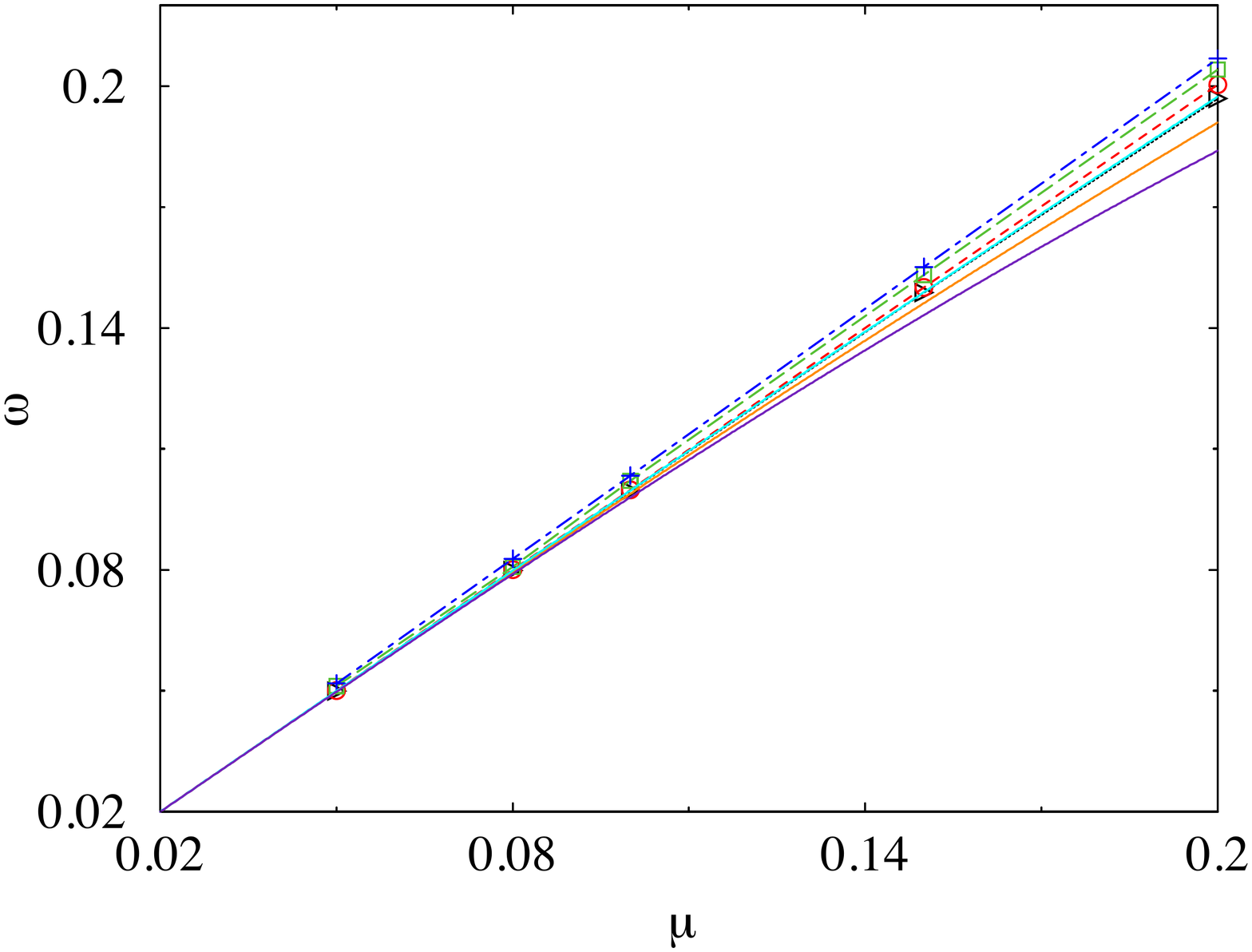}}
\caption{{\it Top panel}:  Frequency $\omega$ as a function of the scalar field mass $\mu$ for the 
models shown in Table~\ref{tab:mod2}. The semi-analytic relation given by Eq.~\eqref{eq:furuhashi} 
is also plotted for $M=\lbrace1.31,1.5,2.0\rbrace$ (solid lines). The semi-analytic values for 
$M=1.31$ corresponds to the cyan line. {\it  Bottom panel}: same as the top panel but for the models 
shown in Table~\ref{tab:mod4}. The semi-analytic relation is plotted for  
$M=\lbrace0.8,1.5,2.0\rbrace$. The semi-analytic values for $M=0.80$ corresponds to the cyan line.}
\label{fg:freq_vs_mu}
\end{center}
\end{figure}

In general our results show that when we consider large enough values of the scalar field mass 
$\mu$, a significantly large amount of the initial scalar field energy falls onto the black hole, 
and this amount increases with $\mu$ (see last columns of Tables \ref{tab:mod1}-\ref{tab:mod4}). 
This behaviour is found for all initial scalar field energies considered. On the other hand, the 
amount of scalar field absorbed by the black hole changes depending on the TOV model, that is, on 
the initial mass of the black hole when it is formed. We find that it is slightly smaller for the 
models for which the TOV mass is smaller (and the compactness of the star is correspondingly 
larger). For the models with the largest mass $\mu$ and largest amplitude $A_0$, the energies at the 
end of the simulations are just a tiny fraction of the initial values (see last row of last columns 
of Tables \ref{tab:mod1}-\ref{tab:mod4}). Even if long-lived quasi-stationary states can be found in 
such cases, their energies may be too small to question the actual survival of the scalar field. 
Nevertheless, for 
smaller values of $\mu$ this trend is reversed and only a small fraction of the initial energy of 
the scalar field is absorbed by the black hole. 

Our results cannot be directly extrapolated to the case of a typical $10^{6}M_{\odot}$ SMS 
collapsing to a black hole. However, we may still put forward an educated guess of what actually might
happen. On the one hand, SMS are less compact ($M/R\sim10^{-3}$) than our TOV models and their gravity is almost Newtonian. Taking this into account, if the quasistationary states survive to the collapse of a more compact star, it could be expected that they will also persist after the collapse of a SMS. 
On the other hand, SMS collapse occurs at $t_{\rm{col}}\sim10^{5}\,\,\rm{s}$~\cite{MonteroJM12}, which in our units is about $t_{\rm col}/M\sim1000-40000$, 
depending on the compactness of the star. Our results yield $t_{\rm col}/M\sim 90-1750$, and the simulations show that the scalar field survives at least for two to three orders of magnitude beyond the collapse timescale. In the context of a SMS, quasistationary states may then exist for long times.

The time evolution of the scalar field energy corresponding to the TOV model 2 and for all scalar 
field masses is shown in figure \ref{fg:SF4}. The top panel corresponds to $E_0=3.2\times 10^{-5}$ 
and the bottom panel to $E_0\sim 4.3$ (the precise values are given in the second to last column of 
Table \ref{tab:mod2}, models 2.4a-2.4e).
In the case of the test-field regime, the energy decay is smooth and varies monotonically with the 
increasing mass. In the non-linear regime, there is a  significant leap (better seen in the inset of 
the bottom panel) in the amount of scalar field energy due to the initial accretion onto the black 
hole when it is formed. The decay of the energy after black hole formation (signalled by the dashed 
vertical lines in the figure) is in general fairly abrupt for all scalar field masses except for 
$\mu=0.05$. After the scalar field first interacts with the black hole, the energy reaches a phase 
during which the accretion rate is importantly reduced, allowing to the survival of the scalar field 
for long timescales (but again notice the drastic reduction in the scalar field energy for $E_0\sim 
4.3$ and $\mu=0.15$ and $\mu=0.2$). This response of the system happens faster the larger the 
initial energy of the scalar field.

\begin{table*}
\caption{Same as Tables~\ref{tab:mod1}-\ref{tab:mod4} but for evolutions of scalar fields in 
spacetimes containing initially a Schwarzschild black hole. Only the fundamental frequency is 
shown.}\label{tab:mod5}
\begin{ruledtabular}
\begin{tabular}{ccccccccc}
Model&$A_{0}$&$\mu$&$ 
\,\,\omega$&$(M_{\rm{AH}})_{0}$&$(M_{\rm{AH}})_{\text{final}}$&$M_{\rm{ADM}}$&$E_{0}$&$E_{\text{
final}}$\\
\hline
1\_4a\_bh&1.85E-2&0.05&0.05116&1.43&3.70&5.73&4.31&1.88\\
1\_4c\_bh&9.85E-3&0.10&0.10341&1.43&5.32&5.75&4.34&0.31\\
1\_4e\_bh&5.00E-3&0.20&0.20681&1.43&5.62&5.75&4.33&0.007\\
\hline
2\_4a\_bh&1.85E-2&0.05&0.05170&1.31&2.67&5.59&4.30&2.88\\
2\_4c\_bh&9.85E-3&0.10&0.10341&1.31&5.18&5.62&4.33&0.34\\
2\_4e\_bh&5.00E-3&0.20&0.20681&1.31&5.44&5.61&4.32&0.008\\
\hline
3\_4a\_bh&1.85E-2&0.05&0.05170&1.10&1.34&5.36&4.28&3.93\\
3\_4c\_bh&9.85E-3&0.10&0.10341&1.10&4.87&5.39&4.30&0.41\\
3\_4e\_bh&5.00E-3&0.20&0.20681&1.10&5.26&5.38&4.29&0.01\\
\hline
4\_4a\_bh&1.85E-2&0.05&0.05188&0.80&0.86&5.04&4.24&4.08\\
4\_4c\_bh&9.85E-3&0.10&0.10341&0.80&4.32&5.06&4.27&0.64\\
4\_4e\_bh&5.00E-3&0.20&0.20681&0.80&4.93&5.05&4.26&0.01\\
\end{tabular}
\end{ruledtabular}
\end{table*}

In Fig.~\ref{fg:SF5} we plot the evolution of the scalar field energy $E_{\text{SF}}$ (red curves) 
and the mass of the apparent horizon $M_{\rm{AH}}$ (black curves) for the TOV model 2 and the 
largest initial energy ($E_0\sim 4.3$) for three scalar field masses, 
$\mu=\lbrace0.05,0.1,0.2\rbrace$ from top to bottom. The scalar field energy remains mostly constant 
until the time of collapse, when it begins to decrease (see the corresponding insets). The decrease 
is more significant and faster the larger the scalar field mass.
Similarly, the mass of the black hole starts to increase once the apparent horizon (discontinually) 
appears, as a response to the accretion of the scalar field. 
It becomes clear from this figure that the amount of the scalar field lost in the process is 
absorbed by the black hole, becoming the only contribution to its growth when the fluid mass has 
been completely accreted by the black hole. The evolutions of both quantities are almost perfect 
mirror images from one another.

By using a matching technique, Furuhashi and Nambu showed analytically in 
\cite{Furuhashi:2004jk} that in the limit $M\mu\ll1$ the real part of the frequency of quasi-bound 
states, $\omega_{\text{qbs}}$, depends on the 
mass parameter $\mu$ as
\begin{equation}
 {\rm Re}(\omega_{\text{qbs}})\approx \mu\left[1-\frac{1}{2}(M\mu)^2 \right]\ .
\label{eq:furuhashi}
\end{equation}
In Fig.~\ref{fg:freq_vs_mu} we show a plot of the frequencies of all the configurations of 
Table~\ref{tab:mod2} and \ref{tab:mod4} as a function of $\mu$ (similar behaviour is found for the 
other models). Our numerical results are plotted with non-solid lines. We also consider in this plot 
the adiabatic approximation for the frequency given by 
Eq.~\eqref{eq:furuhashi} for different values of the mass of the black hole $M$. These analytic 
results are plotted with solid lines in the figure. 

For the TOV model 4 (bottom panel), we find that for the smallest value of $A_0$, indicated in the legend 
in the top panel of the 
figure, the corresponding semi-analytic curve (black dotted line) is indistinguishable from the one of the test 
field approximation for $M=0.8$ (cyan solid line)  showing consistency with the test-field 
approximation and the limit $M\mu\ll1$.  For the TOV model 2 (top panel), while the results are similar, 
for the larger scalar field mass the adiabatic approximation is not matched even in the test-field approximation for $M=1.31$ (cyan solid line). For all the models, we find that for greater values of $A_0$, 
the results from the non-linear approach not only deviate from the adiabatic approximation but they 
also follow the opposite trend: the frequency is greater than the frequency of the test-field limit. 
This discrepancy may be due to the violation of the condition $M\mu\ll1$ used in the analytical 
approximation and due to a breakdown of the adiabatic approximation.

These results agree with our findings in Paper 1. When the initial amplitude is increased and the model is 
far from the test-field regime, the frequencies become higher. A similar trend is found when the 
scalar field mass is small, the frequencies becoming closer to those of the test-field regime. However, 
in this case we find that for the larger amplitudes the frequencies do not differ as much as in 
Paper 1 from the frequencies matching the adiabatic approximation. This may be due to the fact that 
depending on the initial profile and position of the scalar field pulse, there can be a variable 
amount of scalar field that ends up being part of the quasi-bound states. 

\subsection{Comparison with a BH-scalar field system}

In order to further understand the influence of the collapse of the stars in the 
evolution of the scalar field, we also evolve an additional set of 12 models 
corresponding to purely Schwarzschild-like black hole spacetimes. These black holes 
have initial masses equal to those of the different TOV stars considered in the 
simulations we have just discussed. Likewise, the initial distributions of the 
scalar field clouds are the same as in the previous simulations. Therefore,  the 
scalar field interacts from the beginning with an already existing black hole. 
This is the same situation we analyzed in Paper 1, only the initial black hole 
masses have now been adapted to match those of the TOV stars. We choose only 
three scalar field masses, $\mu=\lbrace0.05,0.1,0.2\rbrace$, and the largest 
value of the initial energy, $E_0\sim 4.3$. The grid resolution, time-step, and 
final evolution time remain unchanged. Table \ref{tab:mod5} summarizes the 
results corresponding to this scenario.

In this set of simulations, as we showed in Paper 1, part of the scalar field is 
immediately accreted onto the black hole while part forms the quasi-bounds states. 
The determination of the oscillation frequencies of those states reveals that, for 
$\mu=0.1$ and 0.2,  some of the frequencies are equal to those of the collapsing 
star scenario. On the other hand, the final scalar field energy is, for the 
majority of models, smaller than that for the TOV setup but the difference is 
reduced either if the scalar field mass is increased or if the initial black hole 
mass is lowered. This discrepancy, most noticeable for TOV model 2, can be due to 
the fact that for the TOV a larger fraction of the scalar field may escape before 
the black hole forms and starts accreting. There is almost no difference between 
the two setups for TOV model 4 because, as mentioned before, the time of collapse 
is short enough to form a black hole before most of scalar field reaches the star.  
We also note that for the black hole models 1 and 2 and $\mu=0.05$ the amount of 
scalar field that falls onto the black hole is larger than for the other models 
and scalar field masses. As a result,  the mass of the final black holes increases 
significantly and the oscillation frequencies change. The dynamics of the scalar 
field in the BH and TOV setups are markedly different for this particular set of 
models.

\begin{figure*}
\begin{center}
\subfigure{\includegraphics[width=0.36\textwidth]{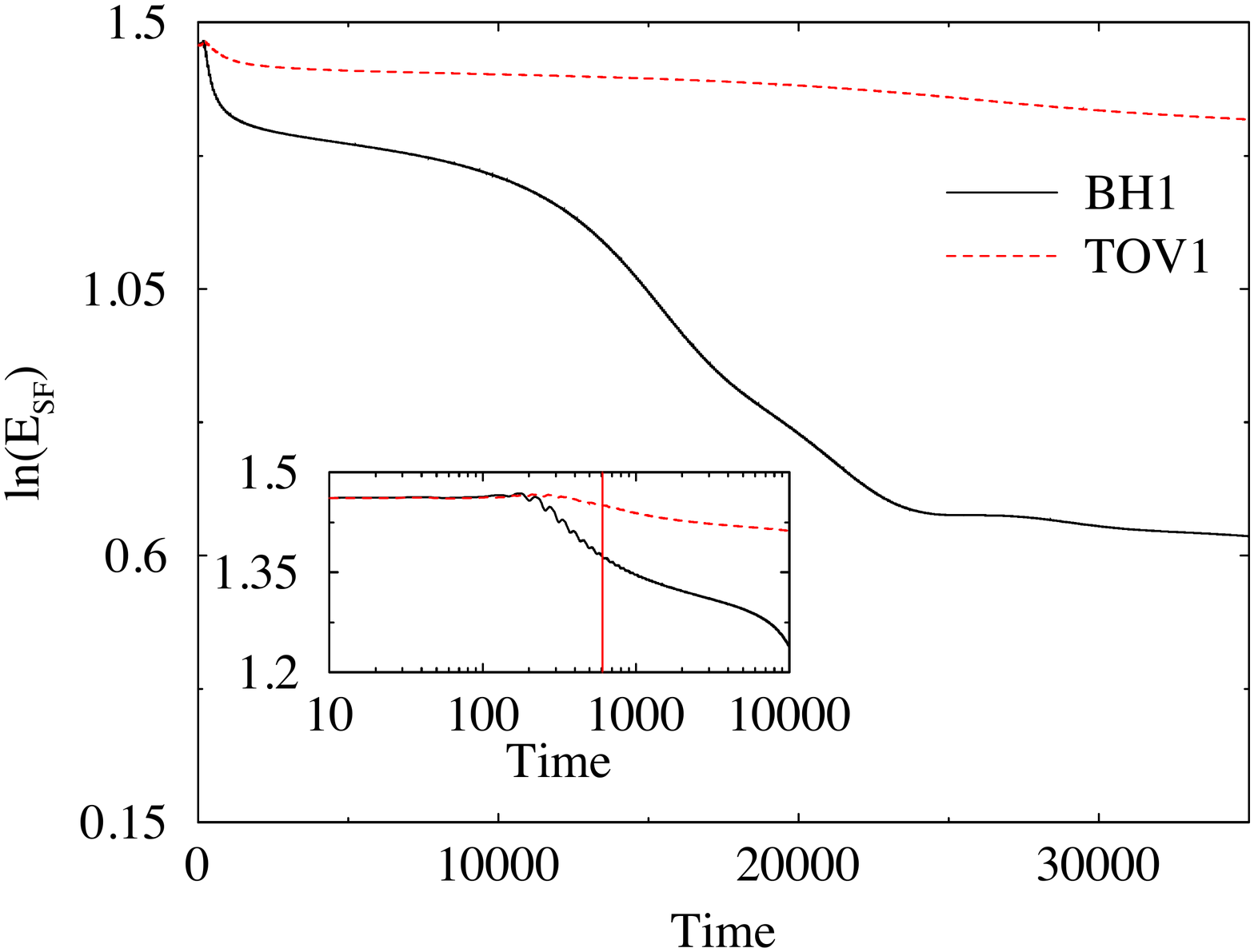}}
\hspace{-0.9cm}\subfigure{\includegraphics[width=0.36\textwidth]{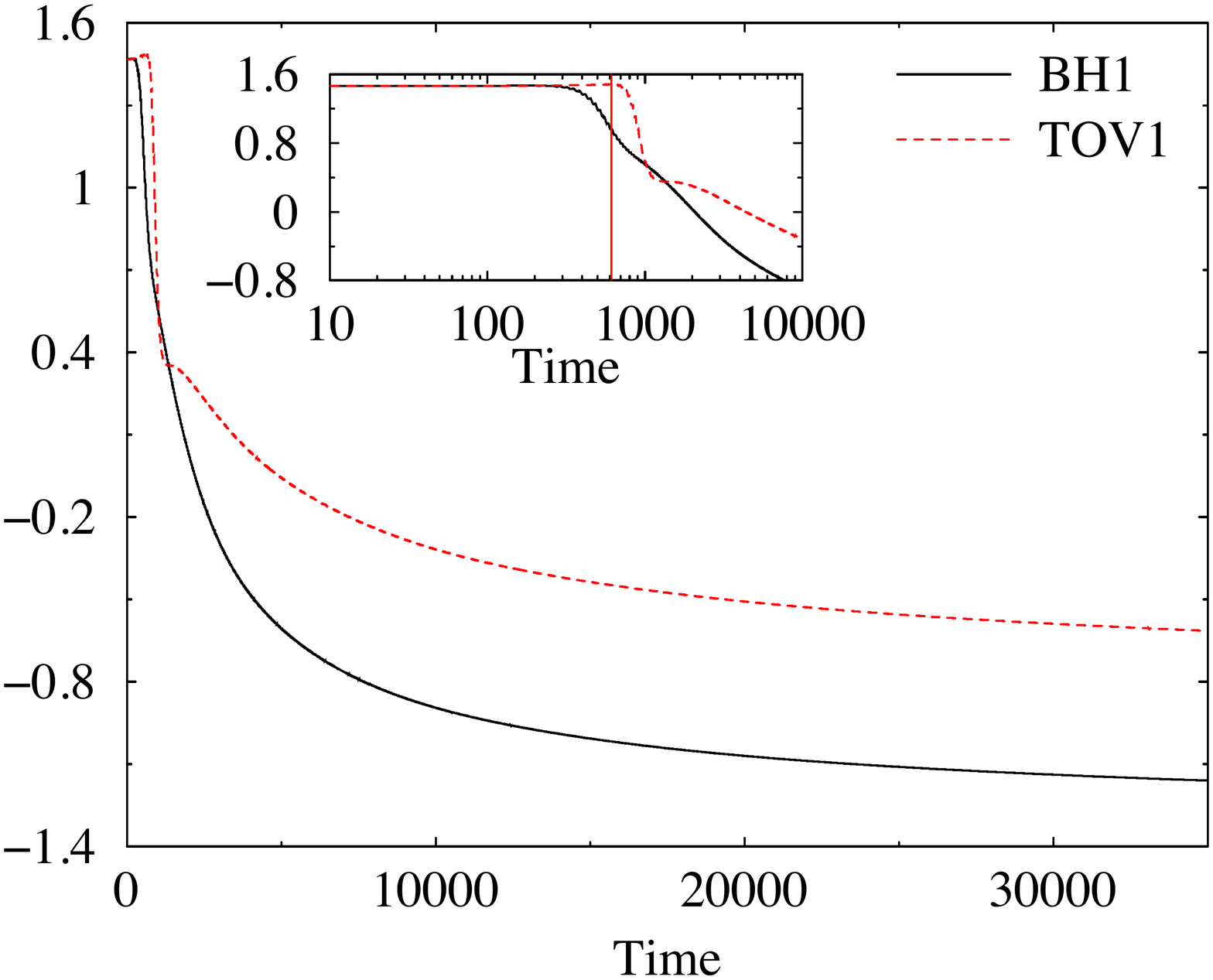}}
\hspace{-0.9cm}\subfigure{\includegraphics[width=0.36\textwidth]{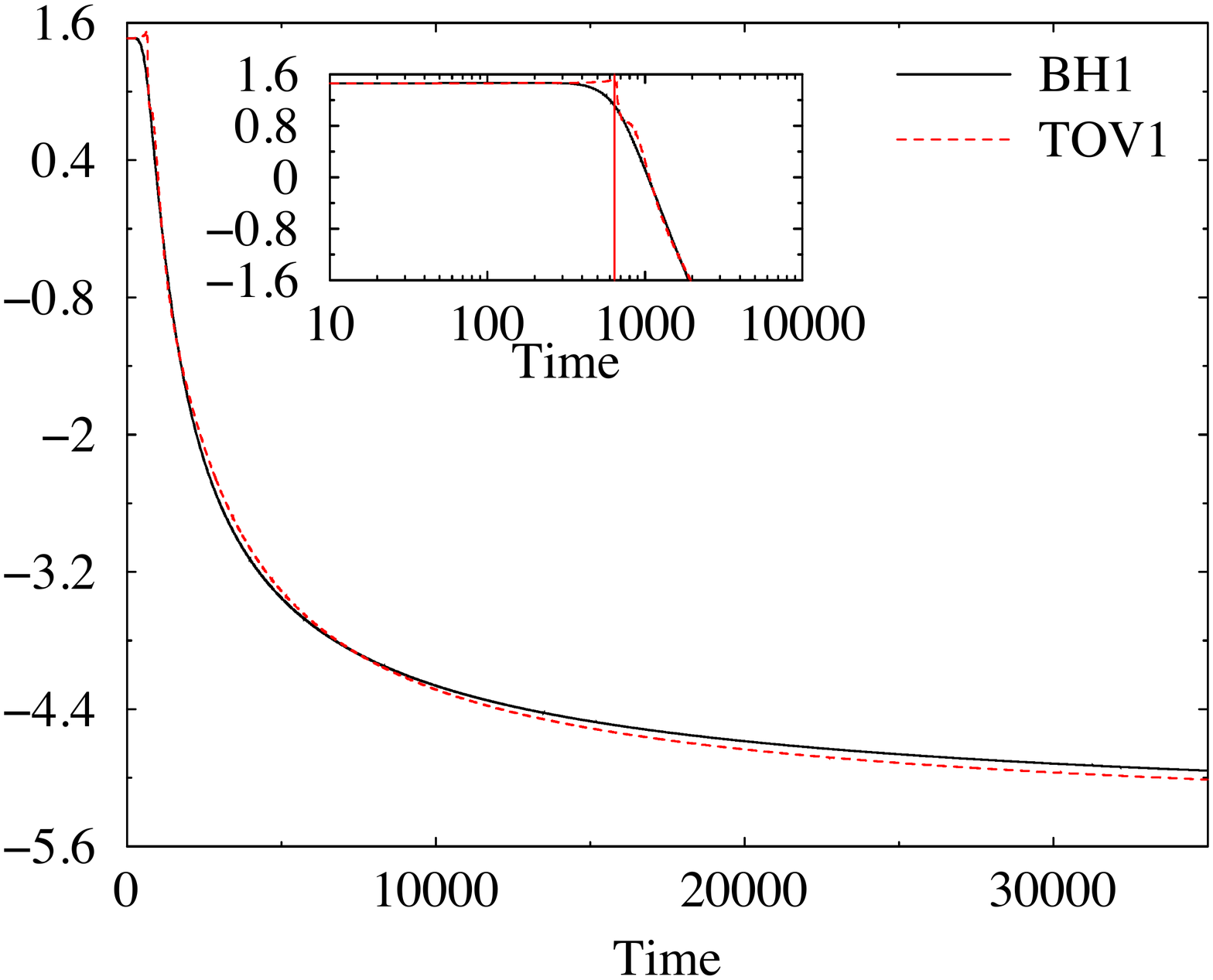}}\\
\vspace{-0.5cm}\subfigure{\includegraphics[width=0.36\textwidth]{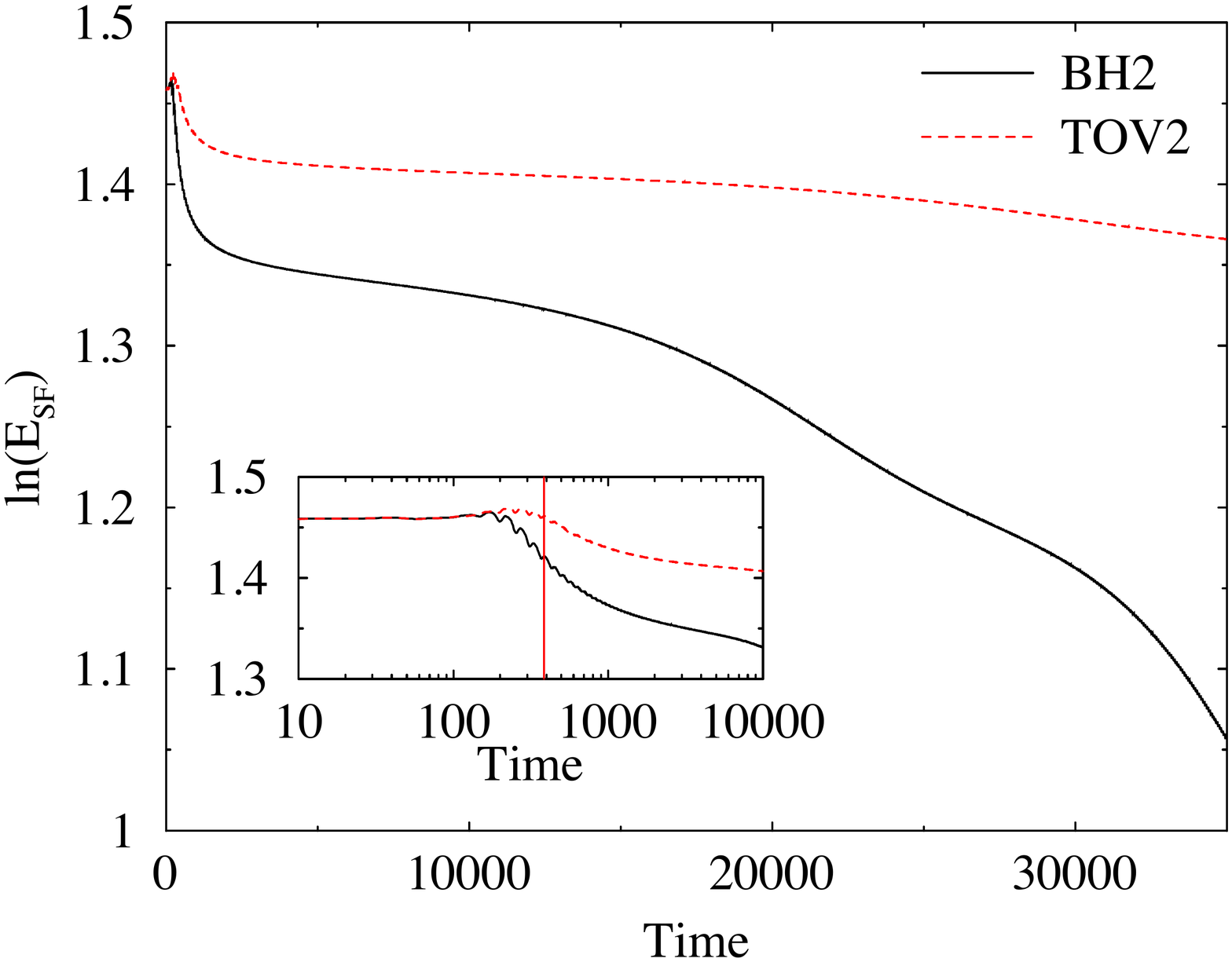}}
\hspace{-0.9cm}\subfigure{\includegraphics[width=0.36\textwidth]{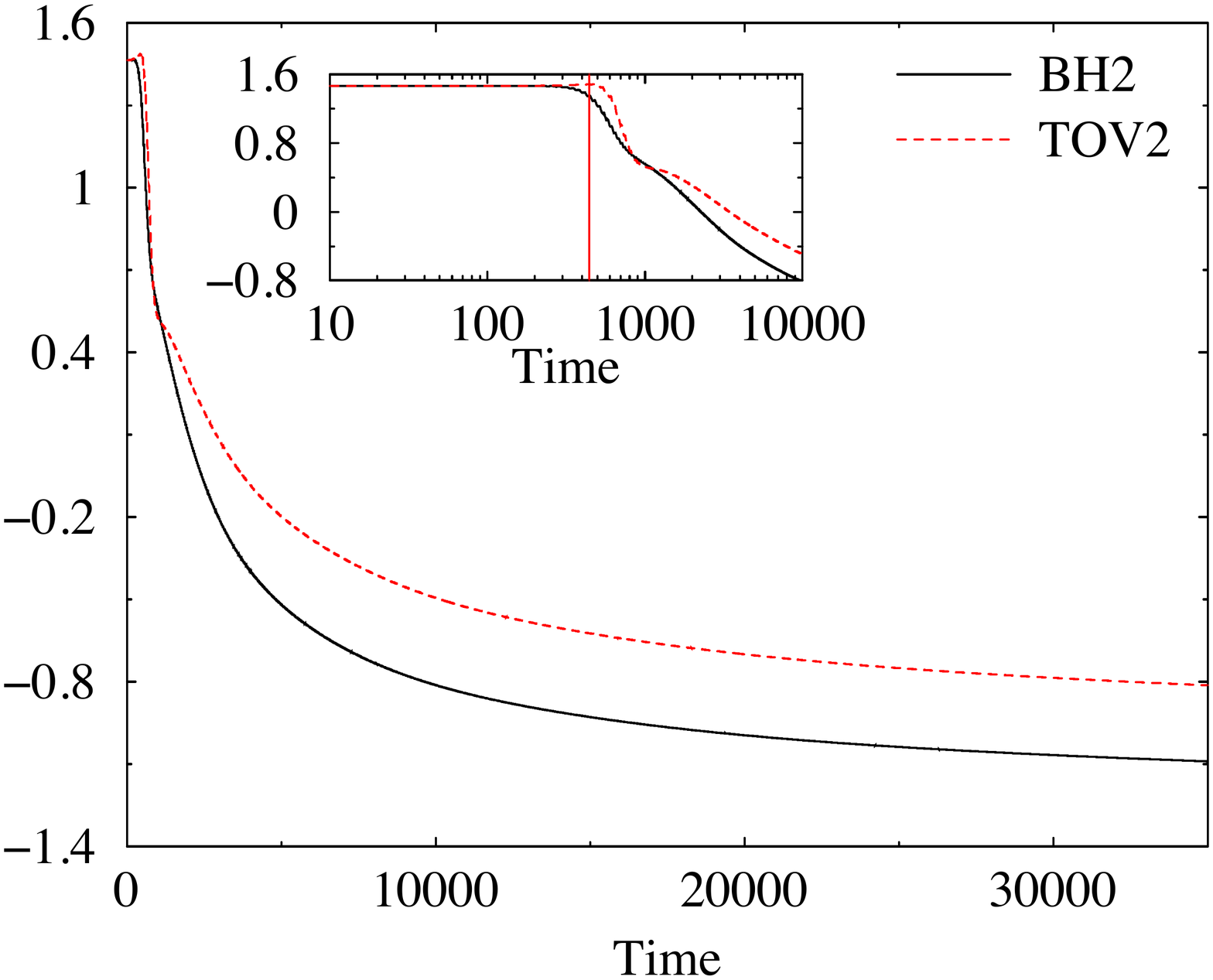}}
\hspace{-0.9cm}\subfigure{\includegraphics[width=0.36\textwidth]{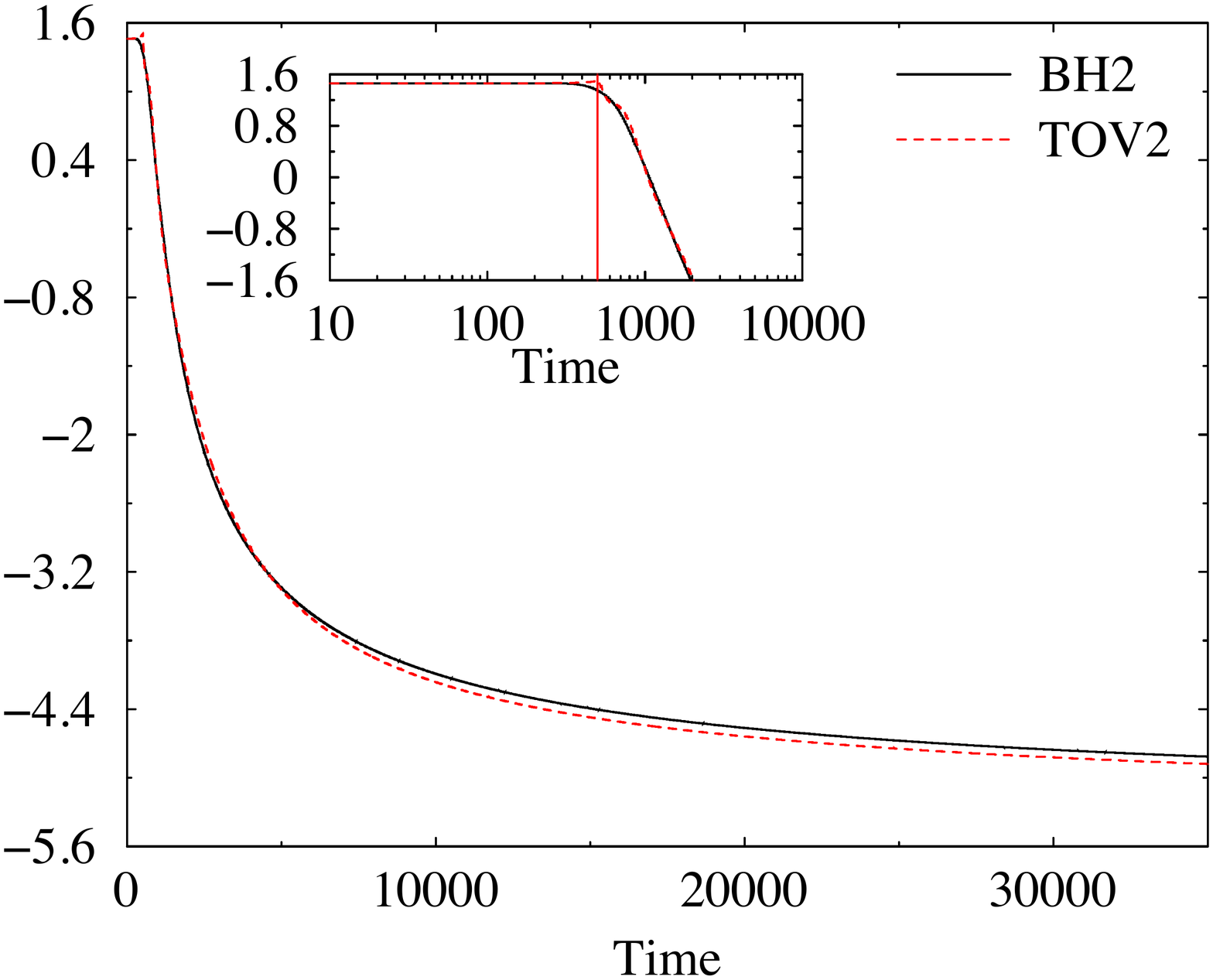}}\\
\vspace{-0.5cm}\subfigure{\includegraphics[width=0.36\textwidth]{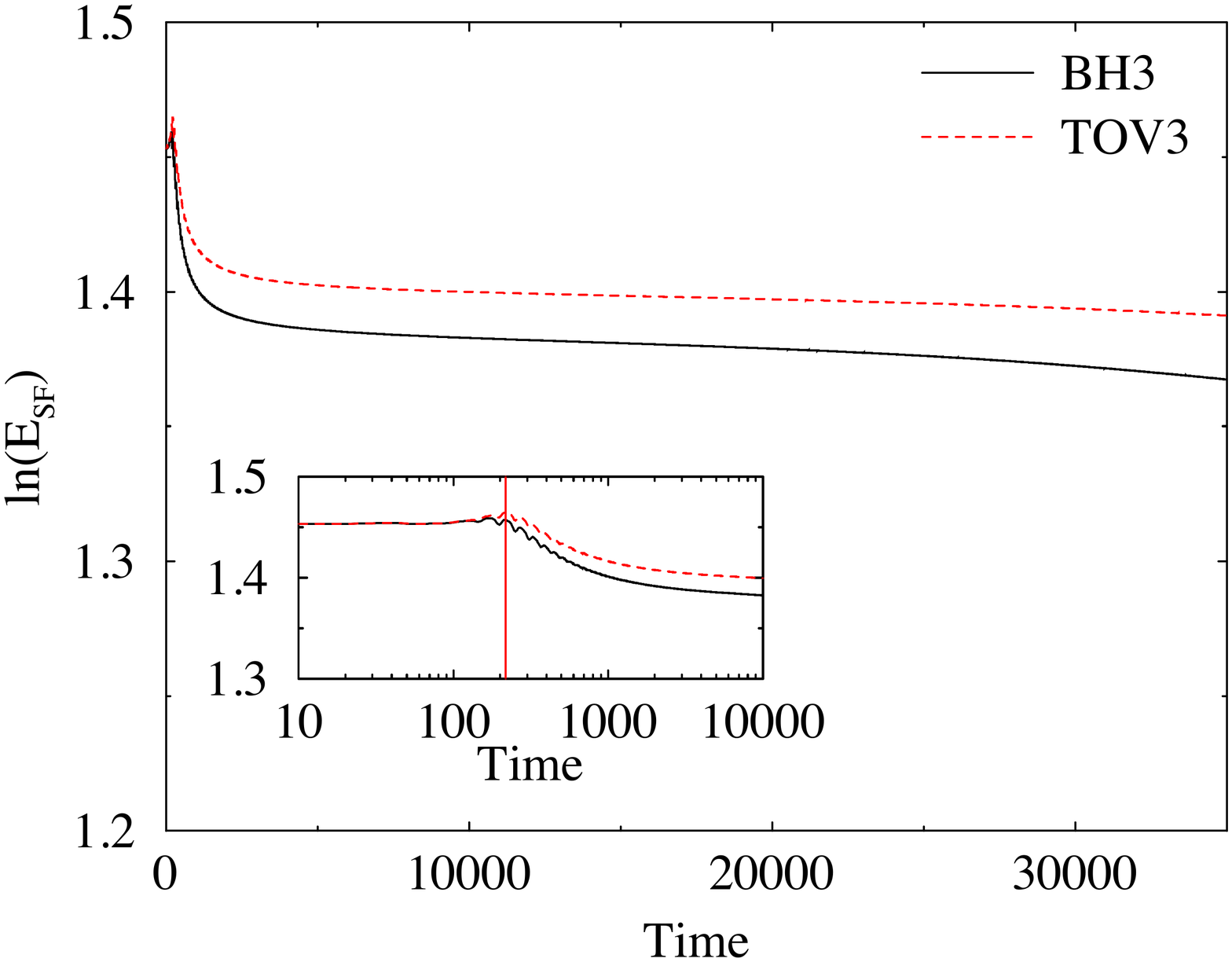}}
\hspace{-0.9cm}\subfigure{\includegraphics[width=0.36\textwidth]{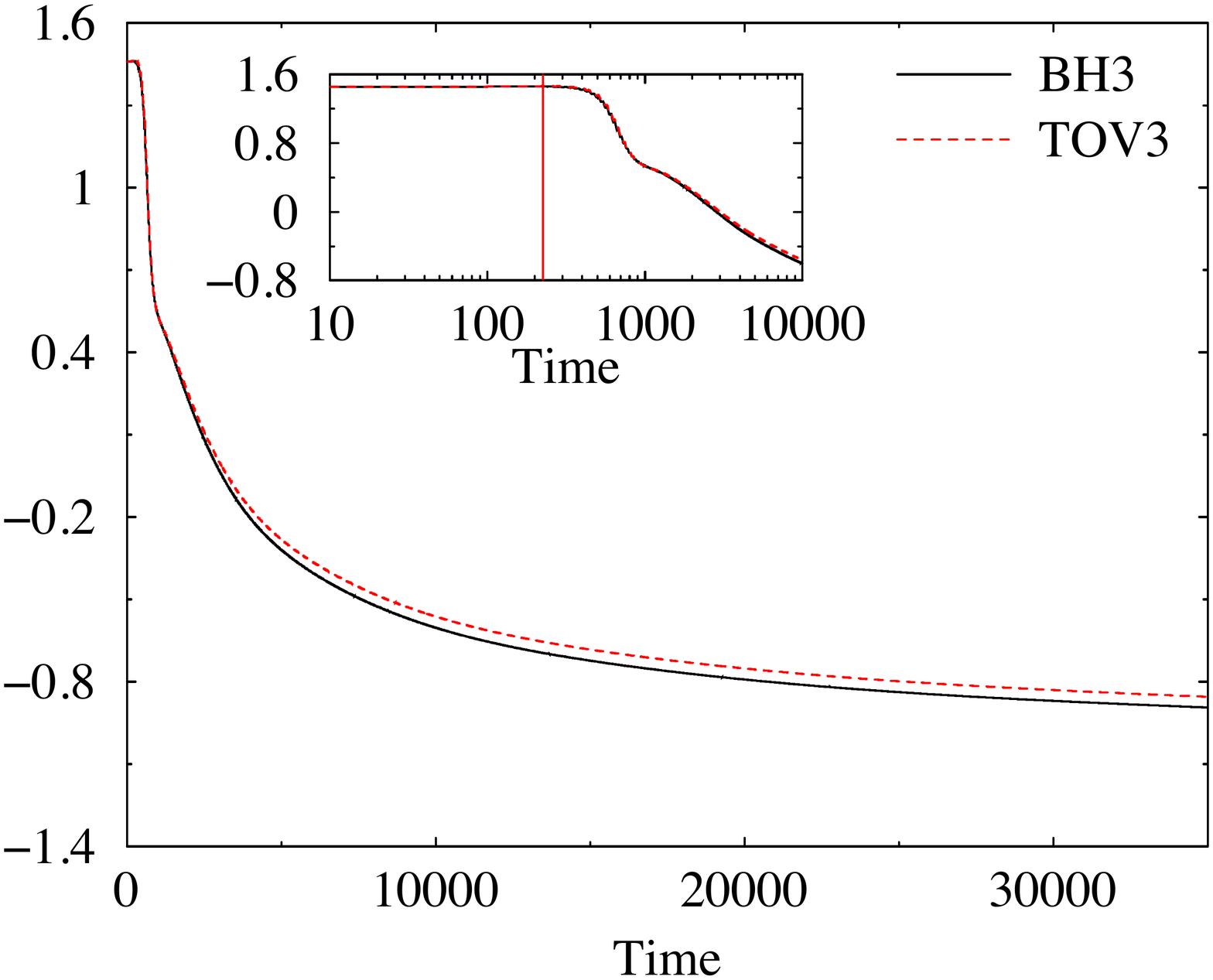}}
\hspace{-0.9cm}\subfigure{\includegraphics[width=0.36\textwidth]{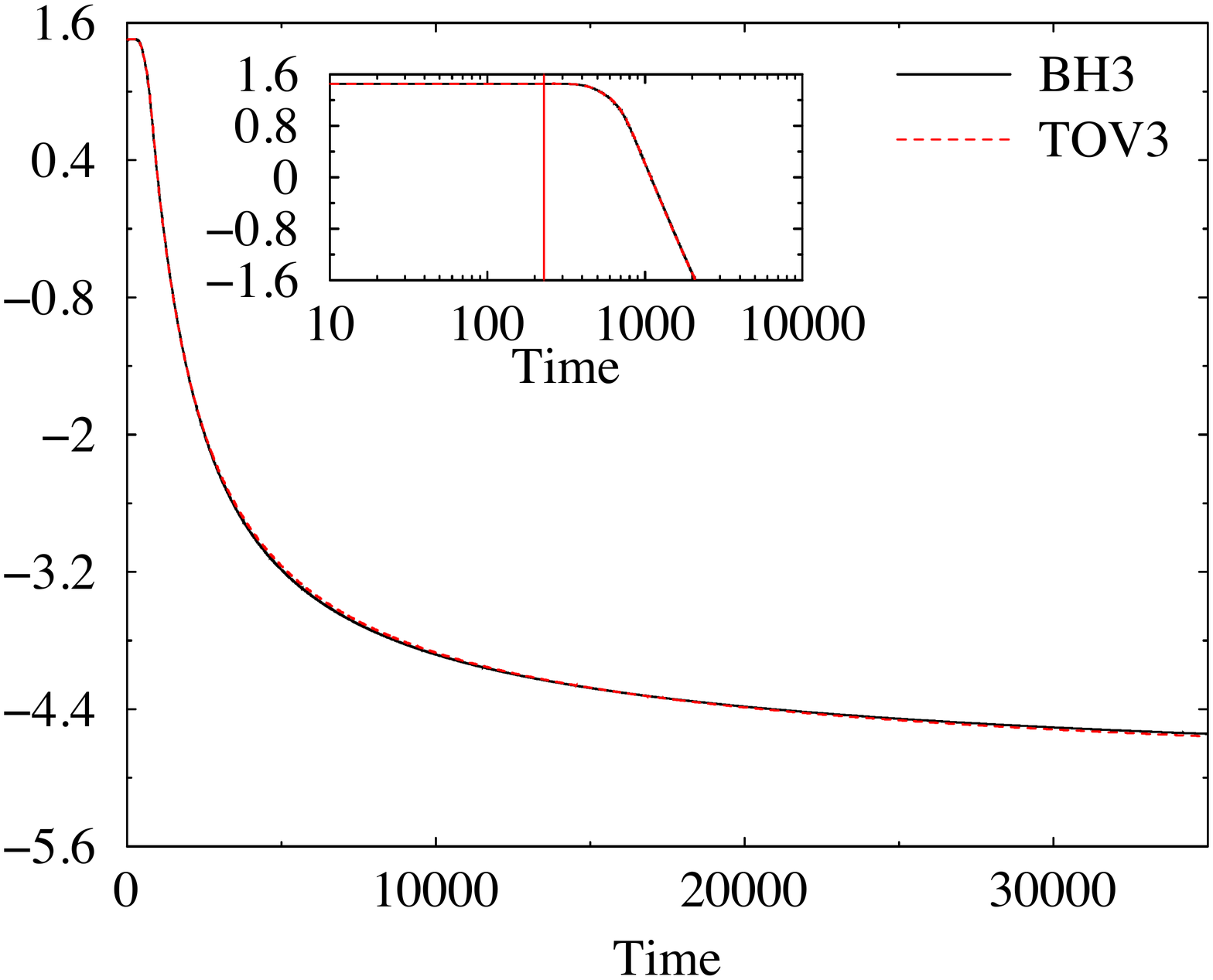}}\\
\vspace{-0.5cm}\subfigure{\includegraphics[width=0.36\textwidth]{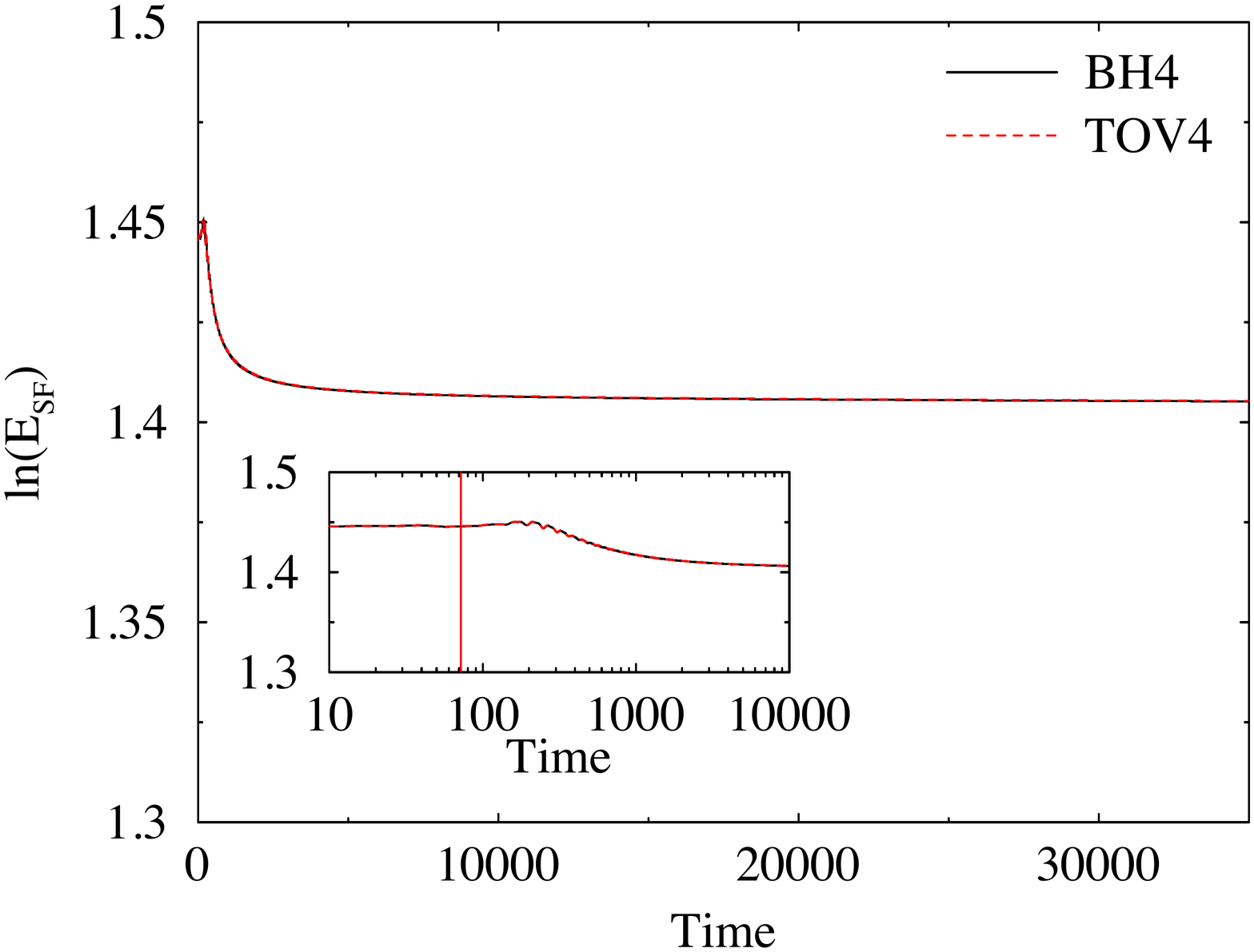}}
\hspace{-0.9cm}\subfigure{\includegraphics[width=0.36\textwidth]{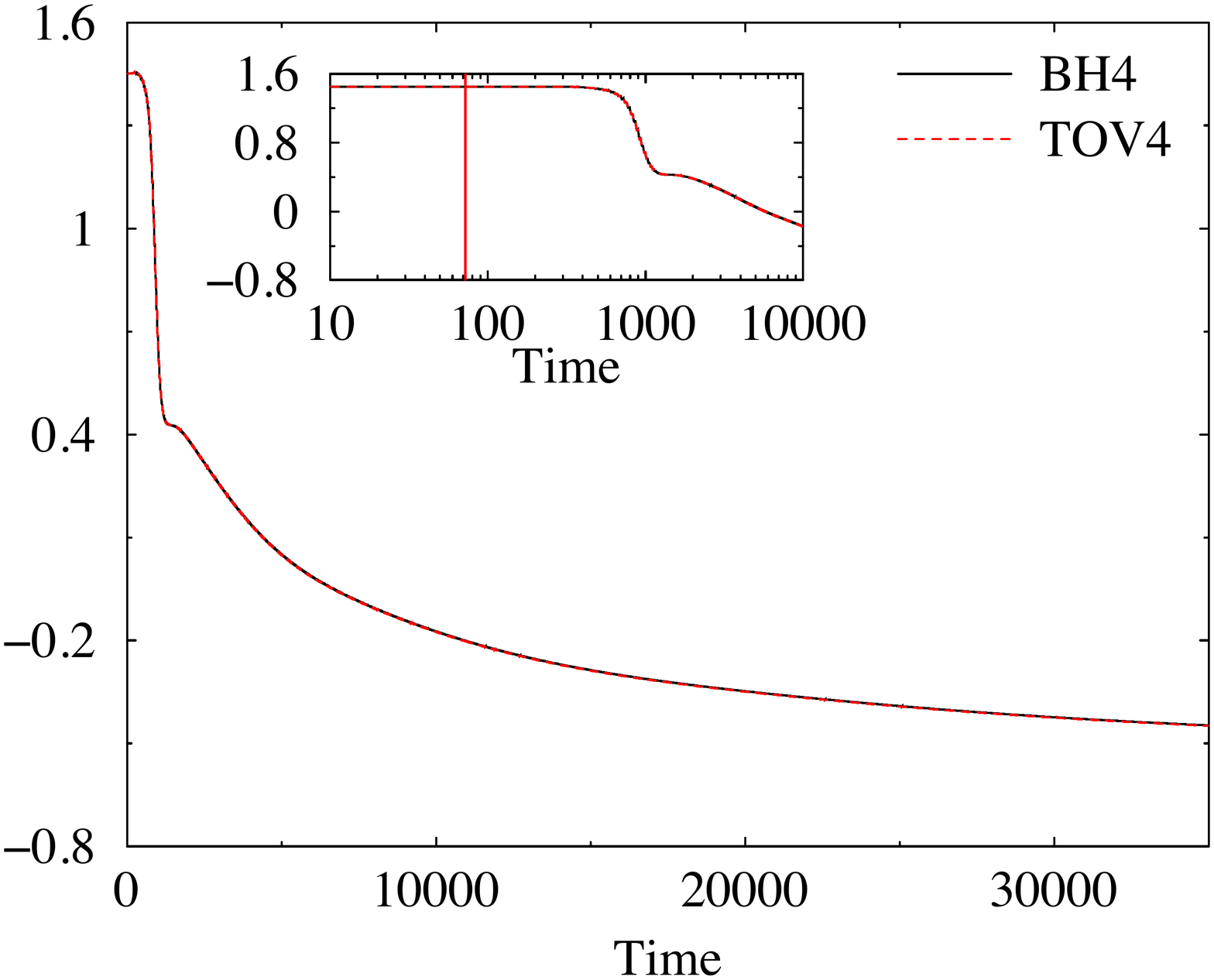}}
\hspace{-0.9cm}\subfigure{\includegraphics[width=0.36\textwidth]{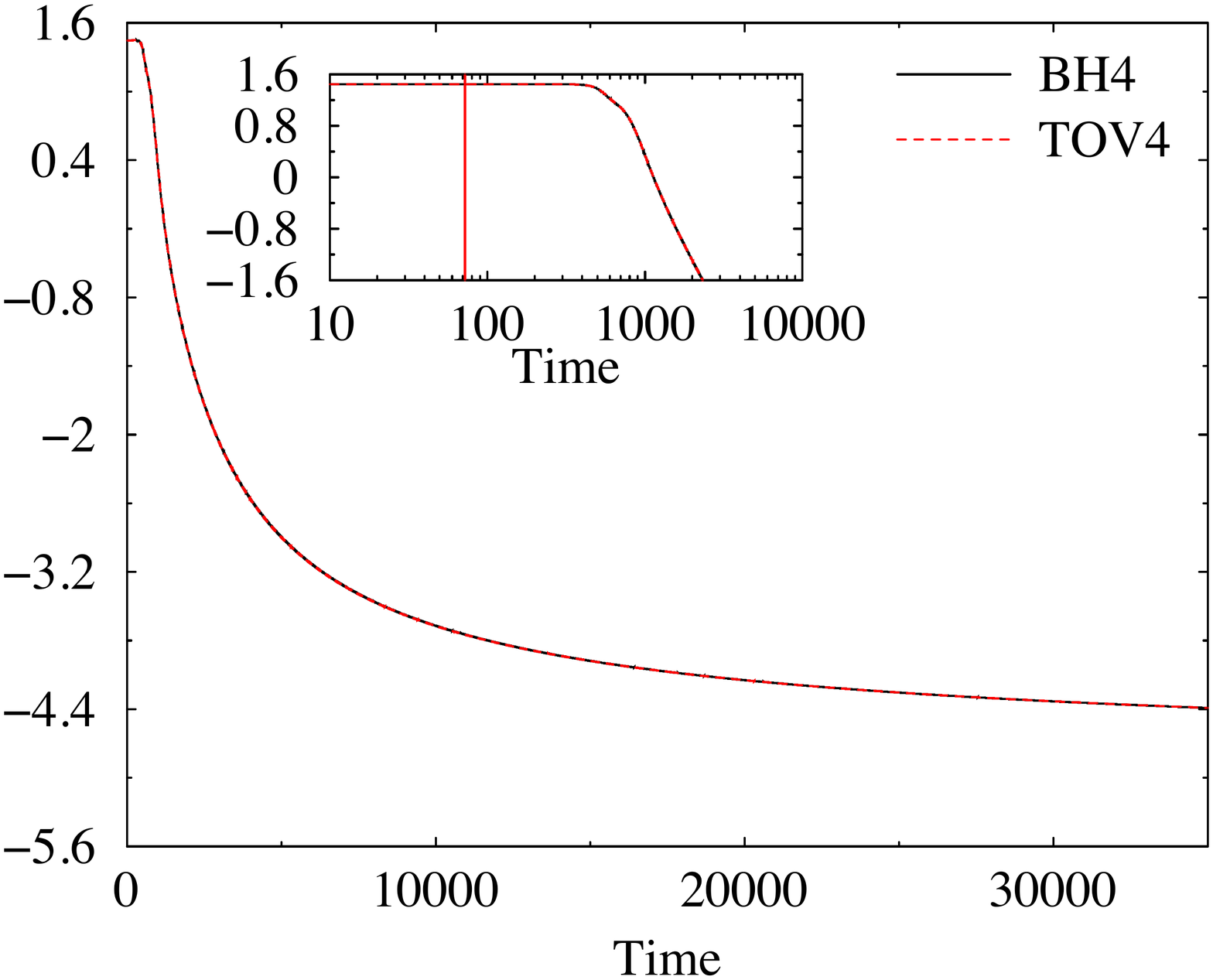}}
\caption{Comparison of the time evolution of the scalar field energy for the collapsing TOV star 
models and for Schwarzschild black holes with the same initial mass. The initial energy density is 
$E_{0}\sim4.3$. The insets show magnified views in logarithm scale of the initial 10000 units of 
time in the evolution. {\it Left column}: $\mu=0.05$. {\it Middle column}: $\mu=0.1$. {\it Right 
column}: $\mu=0.2$.}
\label{fg:SF6}
\end{center}
\end{figure*}

This is clearly visible in Fig.~\ref{fg:SF6} where we plot the evolution of the 
scalar field energy for the initial parameters described in Table \ref{tab:mod5} 
for both the TOV and BH models. As the mass of the scalar field increases from 
$\mu=0.05$ (left column of panels) to $\mu=0.2$ (right column of panels) the 
evolution of the energy becomes progressively similar, and the presence of a 
black hole from the start or at later times is unimportant for the dynamics. The 
same trend is found as the initial black hole mass or TOV mass is reduced (see 
bottom row of panels in Fig.~\ref{fg:SF6}). In this case the dynamics also 
becomes identical irrespective of the scalar field mass. The major differences 
are found in the upper left corner of the composite of panels in Fig.~\ref{fg:SF6}, 
i.e.~when $\mu$ is the smallest and the black hole and TOV initial masses are 
largest. In any event, the influence of the gravitational collapse on the dynamics 
of the scalar field and on the eventual presence of quasi-bound states is not as 
potentially threatening as we might have expected before conducting this research. 
The initial (rapid) phase of accretion by an already formed black hole seems to 
overcome any dynamical effect from the collapse of the star which might have 
affected the survival of the scalar field, even for the smaller scalar field 
masses and the larger initial black hole masses considered in our study, where 
different evolutions are observed.

\section{Conclusions} 
\label{sec:conclusions}

In this paper we have discussed the feasibility of the existence of 
quasi-stationary states of self-gravitating scalar fields in the form 
of clouds (or hairy ``wigs") around collapsing, black-hole-forming, 
stars. For our analysis we have solved numerically the 
Einstein-Klein-Gordon coupled system of equations further coupled to 
the Euler equations of general relativistic hydrodynamics. Our 
numerical approach has been limited to spherical symmetry and has been 
based on spherical coordinates and a PIRK numerical scheme for the 
time update. This research extends our previous work where the same 
issue was investigated for  black holes already existing at $t=0$ to a 
much more dynamical scenario where the black holes are formed as the 
result of the gravitational collapse of unstable (polytropic) stars.

In our non-linear study we have performed a large number of (second-order) 
accurate and long-term stable simulations of dynamical non-rotating 
collapsing stars, with an EOS corresponding to SMSs, 
surrounded by self-gravitating scalar fields. The models of our sample 
have been suitably parameterized to span a broad range of scalar field 
cases, from the test-field regime to the fully non-linear regime, along 
with different initial masses for the star. We have found that in the case 
of highly dynamical spacetimes there are states which closely resemble the 
quasi-bound states in the test-field approximation 
of~\cite{Barranco:2012qs, Barranco:2013rua}, surviving the stellar collapse 
and black hole formation. 

By performing a Fourier transform of the time series of our numerical data 
we have been able to characterize the scalar field states by their 
distinctive oscillation frequencies. It has been found that the scalar field 
oscillates with different well-defined frequencies, and that its time 
evolution produces in most of our models a characteristic beating pattern 
due to the non-linear combination of two oscillations with close frequencies. 
Such scalar field oscillations may have important imprints in a number of 
astrophysical scenarios such as gravitational wave astronomy. In 
particular~\cite{Witek:2012tr,Degollado:2014vsa} have shown recently 
that the frequency of oscillation of the scalar field leaves a distinctive
signature in the late time behaviour of the gravitational wave signal.

We have also compared the evolutions of the scalar field in 
collapsing, black-hole-forming spacetimes with spacetimes that
already contain a non-rotating black hole. This comparison has
provided a better understanding of the influence of the 
stellar collapse on the scalar field dynamics. Our results 
have shown that the differences are mainly due to the initial (or lack of) 
accretion phase onto the black hole that contributes to the growth of its 
mass and to the different dynamics, in the case of self-gravitating 
scalar fields.

Furthermore, our results indicate that for certain values of $\mu$, 
the formation of the black hole during the collapse of the star enhances 
the chances of the scalar field to remain in a quasi-bound state. 
This result may have important implications for models of black hole 
formation together with dark matter halos in the early epochs of the Universe. 
For instance, large galaxies in the scalar field model can already be formed 
at high redshift \cite{Magana:2012jc}.

The collapse of the $\sim 1 M_{\odot}$ stars considered in this work and the 
subsequent formation of the black holes modify the 
dynamics of the scalar field, but for the range of TOV masses
this type of events has very short duration. While our results cannot be directly extrapolated to the case of a typical $10^{6}M_{\odot}$ SMS 
collapsing to a black hole, it could perhaps be expected, on the basis of their smaller compactness and almost Newtonian gravity, that quasistationary states might still persist after the collapse
of a SMS star. Actual numerical simulations are however necessary to validate the viability of this
scenario and the survival of quasi-bound states of scalar fields around SMBH for cosmological timescales.
In addition to such simulations it is worth to
study the evolution of scalar fields around accreting black holes during significantly longer evolutionary times than those attempted in~\cite{Sanchis-Gual:2015}, a task we defer for the near future.

Finally, we note that the dynamical scenario presented here can also be generalized to 
rotating stars where the final state may be a rotating Kerr black hole surrounded by a scalar 
field cloud yielding a hairy black hole solution \cite{Herdeiro:2014goa, Herdeiro:2015gia}. It has 
been argued that these solutions may act as dynamical attractors in a highly dynamical process \cite{Benone:2014ssa} and thus simulations as the ones performed here would be able to describe their formation.

\section*{Acknowledgements}

PM thanks Ewald M{\"u}ller for comments on the manuscript. This work has been supported by the Spanish MINECO (AYA2013-40979-P), 
by the Generalitat Valenciana (PROMETEOII-2014-069), by the 
CONACyT-M\'exico, and by the Max Planck Institute for Astrophysics. The computations have been performed at the Servei 
d'Inform\`atica de la Universitat de Val\`encia.


\bibliography{num-rel}

\begin{thebibliography}{38}
\expandafter\ifx\csname natexlab\endcsname\relax\def\natexlab#1{#1}\fi
\expandafter\ifx\csname bibnamefont\endcsname\relax
  \def\bibnamefont#1{#1}\fi
\expandafter\ifx\csname bibfnamefont\endcsname\relax
  \def\bibfnamefont#1{#1}\fi
\expandafter\ifx\csname citenamefont\endcsname\relax
  \def\citenamefont#1{#1}\fi
\expandafter\ifx\csname url\endcsname\relax
  \def\url#1{\texttt{#1}}\fi
\expandafter\ifx\csname urlprefix\endcsname\relax\def\urlprefix{URL }\fi
\providecommand{\bibinfo}[2]{#2}
\providecommand{\eprint}[2][]{\url{#2}}

\bibitem[{\citenamefont{Mortlock et~al.}(2011)\citenamefont{Mortlock, Warren,
  Venemans, Patel, Hewett, McMahon, Simpson, Theuns, Gonzales-Solares, Adamson
  et~al.}}]{RIS_0}
\bibinfo{author}{\bibfnamefont{D.~J.} \bibnamefont{Mortlock}},
  \bibinfo{author}{\bibfnamefont{S.~J.} \bibnamefont{Warren}},
  \bibinfo{author}{\bibfnamefont{B.~P.} \bibnamefont{Venemans}},
  \bibinfo{author}{\bibfnamefont{M.}~\bibnamefont{Patel}},
  \bibinfo{author}{\bibfnamefont{P.~C.} \bibnamefont{Hewett}},
  \bibinfo{author}{\bibfnamefont{R.~G.} \bibnamefont{McMahon}},
  \bibinfo{author}{\bibfnamefont{C.}~\bibnamefont{Simpson}},
  \bibinfo{author}{\bibfnamefont{T.}~\bibnamefont{Theuns}},
  \bibinfo{author}{\bibfnamefont{E.~A.} \bibnamefont{Gonzales-Solares}},
  \bibinfo{author}{\bibfnamefont{A.}~\bibnamefont{Adamson}},
  \bibnamefont{et~al.}, \bibinfo{journal}{Nature}
  \textbf{\bibinfo{volume}{474}}, \bibinfo{pages}{616} (\bibinfo{year}{2011}),
  ISSN \bibinfo{issn}{0028-0836}, \bibinfo{note}{10.1038/nature10159}.

\bibitem[{\citenamefont{Venemans et~al.}(2015)\citenamefont{Venemans,
  Ba{\~n}ados, Decarli, Farina, Walter, Chambers, Fan, Rix, Schlafly, McMahon
  et~al.}}]{2041-8205-801-1-L11}
\bibinfo{author}{\bibfnamefont{B.~P.} \bibnamefont{Venemans}},
  \bibinfo{author}{\bibfnamefont{E.}~\bibnamefont{Ba{\~n}ados}},
  \bibinfo{author}{\bibfnamefont{R.}~\bibnamefont{Decarli}},
  \bibinfo{author}{\bibfnamefont{E.~P.} \bibnamefont{Farina}},
  \bibinfo{author}{\bibfnamefont{F.}~\bibnamefont{Walter}},
  \bibinfo{author}{\bibfnamefont{K.~C.} \bibnamefont{Chambers}},
  \bibinfo{author}{\bibfnamefont{X.}~\bibnamefont{Fan}},
  \bibinfo{author}{\bibfnamefont{H.-W.} \bibnamefont{Rix}},
  \bibinfo{author}{\bibfnamefont{E.}~\bibnamefont{Schlafly}},
  \bibinfo{author}{\bibfnamefont{R.~G.} \bibnamefont{McMahon}},
  \bibnamefont{et~al.}, \bibinfo{journal}{The Astrophysical Journal Letters}
  \textbf{\bibinfo{volume}{801}}, \bibinfo{pages}{L11} (\bibinfo{year}{2015}),
  \urlprefix\url{http://stacks.iop.org/2041-8205/801/i=1/a=L11}.

\bibitem[{\citenamefont{Bromm and Loeb}(2003)}]{Bromm:2002hb}
\bibinfo{author}{\bibfnamefont{V.}~\bibnamefont{Bromm}} \bibnamefont{and}
  \bibinfo{author}{\bibfnamefont{A.}~\bibnamefont{Loeb}},
  \bibinfo{journal}{Astrophys. J.} \textbf{\bibinfo{volume}{596}},
  \bibinfo{pages}{34} (\bibinfo{year}{2003}), \eprint{astro-ph/0212400}.

\bibitem[{\citenamefont{Haiman and Loeb}(2001)}]{Haiman01}
\bibinfo{author}{\bibfnamefont{Z.}~\bibnamefont{Haiman}} \bibnamefont{and}
  \bibinfo{author}{\bibfnamefont{A.}~\bibnamefont{Loeb}},
  \bibinfo{journal}{Astrophys. J.} \textbf{\bibinfo{volume}{552}},
  \bibinfo{pages}{459} (\bibinfo{year}{2001}).

\bibitem[{\citenamefont{Begelman et~al.}(2006)\citenamefont{Begelman,
  Volonteri, and Rees}}]{Begelman:2006db}
\bibinfo{author}{\bibfnamefont{M.~C.} \bibnamefont{Begelman}},
  \bibinfo{author}{\bibfnamefont{M.}~\bibnamefont{Volonteri}},
  \bibnamefont{and} \bibinfo{author}{\bibfnamefont{M.~J.} \bibnamefont{Rees}},
  \bibinfo{journal}{Mon.Not.Roy.Astron.Soc.} \textbf{\bibinfo{volume}{370}},
  \bibinfo{pages}{289} (\bibinfo{year}{2006}), \eprint{astro-ph/0602363}.

\bibitem[{\citenamefont{{Begelman}}(2010)}]{Begelman10}
\bibinfo{author}{\bibfnamefont{M.~C.} \bibnamefont{{Begelman}}},
  \bibinfo{journal}{Mon.Not.Roy.Astron.Soc.} \textbf{\bibinfo{volume}{402}},
  \bibinfo{pages}{673} (\bibinfo{year}{2010}), \eprint{0910.4398}.

\bibitem[{\citenamefont{{Mayer} et~al.}(2010)\citenamefont{{Mayer},
  {Kazantzidis}, {Escala}, and {Callegari}}}]{Mayer10}
\bibinfo{author}{\bibfnamefont{L.}~\bibnamefont{{Mayer}}},
  \bibinfo{author}{\bibfnamefont{S.}~\bibnamefont{{Kazantzidis}}},
  \bibinfo{author}{\bibfnamefont{A.}~\bibnamefont{{Escala}}}, \bibnamefont{and}
  \bibinfo{author}{\bibfnamefont{S.}~\bibnamefont{{Callegari}}},
  \bibinfo{journal}{\nat} \textbf{\bibinfo{volume}{466}}, \bibinfo{pages}{1082}
  (\bibinfo{year}{2010}), \eprint{0912.4262}.

\bibitem[{\citenamefont{{Shibata} and {Shapiro}}(2002)}]{Shibata02}
\bibinfo{author}{\bibfnamefont{M.}~\bibnamefont{{Shibata}}} \bibnamefont{and}
  \bibinfo{author}{\bibfnamefont{S.~L.} \bibnamefont{{Shapiro}}},
  \bibinfo{journal}{Astrophys. J.} \textbf{\bibinfo{volume}{572}},
  \bibinfo{pages}{L39} (\bibinfo{year}{2002}), \eprint{astro-ph/0205091}.

\bibitem[{\citenamefont{{Montero} et~al.}(2012)\citenamefont{{Montero},
  {Janka}, and {M{\"u}ller}}}]{MonteroJM12}
\bibinfo{author}{\bibfnamefont{P.~J.} \bibnamefont{{Montero}}},
  \bibinfo{author}{\bibfnamefont{H.-T.} \bibnamefont{{Janka}}},
  \bibnamefont{and}
  \bibinfo{author}{\bibfnamefont{E.}~\bibnamefont{{M{\"u}ller}}},
  \bibinfo{journal}{Astrophys. J.} \textbf{\bibinfo{volume}{749}},
  \bibinfo{eid}{37} (\bibinfo{year}{2012}), \eprint{1108.3090}.

\bibitem[{\citenamefont{Lidsey et~al.}(1997)\citenamefont{Lidsey, Liddle, Kolb,
  Copeland, Barreiro et~al.}}]{Lidsey:1995np}
\bibinfo{author}{\bibfnamefont{J.~E.} \bibnamefont{Lidsey}},
  \bibinfo{author}{\bibfnamefont{A.~R.} \bibnamefont{Liddle}},
  \bibinfo{author}{\bibfnamefont{E.~W.} \bibnamefont{Kolb}},
  \bibinfo{author}{\bibfnamefont{E.~J.} \bibnamefont{Copeland}},
  \bibinfo{author}{\bibfnamefont{T.}~\bibnamefont{Barreiro}},
  \bibnamefont{et~al.}, \bibinfo{journal}{Rev.Mod.Phys.}
  \textbf{\bibinfo{volume}{69}}, \bibinfo{pages}{373} (\bibinfo{year}{1997}),
  \eprint{astro-ph/9508078}.

\bibitem[{\citenamefont{Padmanabhan}(2003)}]{Padmanabhan:2002ji}
\bibinfo{author}{\bibfnamefont{T.}~\bibnamefont{Padmanabhan}},
  \bibinfo{journal}{Phys.Rept.} \textbf{\bibinfo{volume}{380}},
  \bibinfo{pages}{235} (\bibinfo{year}{2003}), \eprint{hep-th/0212290}.

\bibitem[{\citenamefont{Sahni and Wang}(2000)}]{Sahni:1999qe}
\bibinfo{author}{\bibfnamefont{V.}~\bibnamefont{Sahni}} \bibnamefont{and}
  \bibinfo{author}{\bibfnamefont{L.-M.} \bibnamefont{Wang}},
  \bibinfo{journal}{Phys.Rev.} \textbf{\bibinfo{volume}{D62}},
  \bibinfo{pages}{103517} (\bibinfo{year}{2000}), \eprint{astro-ph/9910097}.

\bibitem[{\citenamefont{Hu et~al.}(2000)\citenamefont{Hu, Barkana, and
  Gruzinov}}]{Hu:2000ke}
\bibinfo{author}{\bibfnamefont{W.}~\bibnamefont{Hu}},
  \bibinfo{author}{\bibfnamefont{R.}~\bibnamefont{Barkana}}, \bibnamefont{and}
  \bibinfo{author}{\bibfnamefont{A.}~\bibnamefont{Gruzinov}},
  \bibinfo{journal}{Phys. Rev. Lett.} \textbf{\bibinfo{volume}{85}},
  \bibinfo{pages}{1158} (\bibinfo{year}{2000}), \eprint{astro-ph/0003365}.

\bibitem[{\citenamefont{Matos and Urena-Lopez}(2000)}]{Matos:2000ng}
\bibinfo{author}{\bibfnamefont{T.}~\bibnamefont{Matos}} \bibnamefont{and}
  \bibinfo{author}{\bibfnamefont{L.~A.} \bibnamefont{Urena-Lopez}},
  \bibinfo{journal}{Class. Quant. Grav.} \textbf{\bibinfo{volume}{17}},
  \bibinfo{pages}{L75} (\bibinfo{year}{2000}), \eprint{astro-ph/0004332}.

\bibitem[{\citenamefont{Matos and Urena-Lopez}(2001)}]{Matos:2000ss}
\bibinfo{author}{\bibfnamefont{T.}~\bibnamefont{Matos}} \bibnamefont{and}
  \bibinfo{author}{\bibfnamefont{L.~A.} \bibnamefont{Urena-Lopez}},
  \bibinfo{journal}{Phys.Rev.} \textbf{\bibinfo{volume}{D63}},
  \bibinfo{pages}{063506} (\bibinfo{year}{2001}), \eprint{astro-ph/0006024}.

\bibitem[{\citenamefont{Barranco et~al.}(2011)\citenamefont{Barranco, Bernal,
  Degollado, Diez-Tejedor, Megevand et~al.}}]{Burt:2011pv}
\bibinfo{author}{\bibfnamefont{J.}~\bibnamefont{Barranco}},
  \bibinfo{author}{\bibfnamefont{A.}~\bibnamefont{Bernal}},
  \bibinfo{author}{\bibfnamefont{J.~C.} \bibnamefont{Degollado}},
  \bibinfo{author}{\bibfnamefont{A.}~\bibnamefont{Diez-Tejedor}},
  \bibinfo{author}{\bibfnamefont{M.}~\bibnamefont{Megevand}},
  \bibnamefont{et~al.}, \bibinfo{journal}{Phys.Rev.}
  \textbf{\bibinfo{volume}{D84}}, \bibinfo{pages}{083008}
  (\bibinfo{year}{2011}), \eprint{1108.0931}.

\bibitem[{\citenamefont{Sanchis-Gual et~al.}(2015)\citenamefont{Sanchis-Gual,
  Degollado, Montero, and Font}}]{Sanchis-Gual:2015}
\bibinfo{author}{\bibfnamefont{N.}~\bibnamefont{Sanchis-Gual}},
  \bibinfo{author}{\bibfnamefont{J.~C.} \bibnamefont{Degollado}},
  \bibinfo{author}{\bibfnamefont{P.~J.} \bibnamefont{Montero}},
  \bibnamefont{and} \bibinfo{author}{\bibfnamefont{J.~A.} \bibnamefont{Font}},
  \bibinfo{journal}{Phys. Rev. D} \textbf{\bibinfo{volume}{91}},
  \bibinfo{pages}{043005} (\bibinfo{year}{2015}),
  \urlprefix\url{http://link.aps.org/doi/10.1103/PhysRevD.91.043005}.

\bibitem[{\citenamefont{Witek et~al.}(2013)\citenamefont{Witek, Cardoso,
  Ishibashi, and Sperhake}}]{Witek:2012tr}
\bibinfo{author}{\bibfnamefont{H.}~\bibnamefont{Witek}},
  \bibinfo{author}{\bibfnamefont{V.}~\bibnamefont{Cardoso}},
  \bibinfo{author}{\bibfnamefont{A.}~\bibnamefont{Ishibashi}},
  \bibnamefont{and} \bibinfo{author}{\bibfnamefont{U.}~\bibnamefont{Sperhake}},
  \bibinfo{journal}{Phys.Rev.} \textbf{\bibinfo{volume}{D87}},
  \bibinfo{pages}{043513} (\bibinfo{year}{2013}), \eprint{1212.0551}.

\bibitem[{\citenamefont{Barranco et~al.}(In preparation)\citenamefont{Barranco,
  Bernal, Degollado, Diez-Tejedor, Megevand et~al.}}]{Barranco:2015}
\bibinfo{author}{\bibfnamefont{J.}~\bibnamefont{Barranco}},
  \bibinfo{author}{\bibfnamefont{A.}~\bibnamefont{Bernal}},
  \bibinfo{author}{\bibfnamefont{J.~C.} \bibnamefont{Degollado}},
  \bibinfo{author}{\bibfnamefont{A.}~\bibnamefont{Diez-Tejedor}},
  \bibinfo{author}{\bibfnamefont{M.}~\bibnamefont{Megevand}},
  \bibnamefont{et~al.} (\bibinfo{year}{In preparation}).

\bibitem[{\citenamefont{Barranco et~al.}(2014)\citenamefont{Barranco, Bernal,
  Degollado, Diez-Tejedor, Megevand et~al.}}]{Barranco:2013rua}
\bibinfo{author}{\bibfnamefont{J.}~\bibnamefont{Barranco}},
  \bibinfo{author}{\bibfnamefont{A.}~\bibnamefont{Bernal}},
  \bibinfo{author}{\bibfnamefont{J.~C.} \bibnamefont{Degollado}},
  \bibinfo{author}{\bibfnamefont{A.}~\bibnamefont{Diez-Tejedor}},
  \bibinfo{author}{\bibfnamefont{M.}~\bibnamefont{Megevand}},
  \bibnamefont{et~al.}, \bibinfo{journal}{Phys.Rev.}
  \textbf{\bibinfo{volume}{D89}}, \bibinfo{pages}{083006}
  (\bibinfo{year}{2014}), \eprint{1312.5808}.

\bibitem[{\citenamefont{Montero and Cordero-Carrion}(2012)}]{Montero:2012yr}
\bibinfo{author}{\bibfnamefont{P.~J.} \bibnamefont{Montero}} \bibnamefont{and}
  \bibinfo{author}{\bibfnamefont{I.}~\bibnamefont{Cordero-Carrion}},
  \bibinfo{journal}{Phys.Rev.} \textbf{\bibinfo{volume}{D85}},
  \bibinfo{pages}{124037} (\bibinfo{year}{2012}), \eprint{1204.5377}.

\bibitem[{\citenamefont{Sanchis-Gual et~al.}(2014)\citenamefont{Sanchis-Gual,
  Montero, Font, M\"uller, and Baumgarte}}]{Sanchis-Gual:2014}
\bibinfo{author}{\bibfnamefont{N.}~\bibnamefont{Sanchis-Gual}},
  \bibinfo{author}{\bibfnamefont{P.~J.} \bibnamefont{Montero}},
  \bibinfo{author}{\bibfnamefont{J.~A.} \bibnamefont{Font}},
  \bibinfo{author}{\bibfnamefont{E.}~\bibnamefont{M\"uller}}, \bibnamefont{and}
  \bibinfo{author}{\bibfnamefont{T.~W.} \bibnamefont{Baumgarte}},
  \bibinfo{journal}{Phys. Rev. D} \textbf{\bibinfo{volume}{89}},
  \bibinfo{pages}{104033} (\bibinfo{year}{2014}),
  \urlprefix\url{http://link.aps.org/doi/10.1103/PhysRevD.89.104033}.

\bibitem[{\citenamefont{Baumgarte and Shapiro}(1998)}]{Baumgarte98}
\bibinfo{author}{\bibfnamefont{T.~W.} \bibnamefont{Baumgarte}}
  \bibnamefont{and} \bibinfo{author}{\bibfnamefont{S.~L.}
  \bibnamefont{Shapiro}}, \bibinfo{journal}{Phys. Rev. D}
  \textbf{\bibinfo{volume}{59}}, \bibinfo{pages}{024007}
  (\bibinfo{year}{1998}).

\bibitem[{\citenamefont{Shibata and Nakamura}(1995)}]{Shibata95}
\bibinfo{author}{\bibfnamefont{M.}~\bibnamefont{Shibata}} \bibnamefont{and}
  \bibinfo{author}{\bibfnamefont{T.}~\bibnamefont{Nakamura}},
  \bibinfo{journal}{Phys. Rev. D} \textbf{\bibinfo{volume}{52}},
  \bibinfo{pages}{5428} (\bibinfo{year}{1995}).

\bibitem[{\citenamefont{Alcubierre and Mendez}(2011)}]{Alcubierre:2010is}
\bibinfo{author}{\bibfnamefont{M.}~\bibnamefont{Alcubierre}} \bibnamefont{and}
  \bibinfo{author}{\bibfnamefont{M.~D.} \bibnamefont{Mendez}},
  \bibinfo{journal}{Gen.Rel.Grav.} \textbf{\bibinfo{volume}{43}},
  \bibinfo{pages}{2769} (\bibinfo{year}{2011}), \eprint{1010.4013}.

\bibitem[{\citenamefont{Bona et~al.}(1997)\citenamefont{Bona, Mass\'o, Seidel,
  and Stela}}]{Bona:1997prd}
\bibinfo{author}{\bibfnamefont{C.}~\bibnamefont{Bona}},
  \bibinfo{author}{\bibfnamefont{J.}~\bibnamefont{Mass\'o}},
  \bibinfo{author}{\bibfnamefont{E.}~\bibnamefont{Seidel}}, \bibnamefont{and}
  \bibinfo{author}{\bibfnamefont{J.}~\bibnamefont{Stela}},
  \bibinfo{journal}{Phys. Rev. D} \textbf{\bibinfo{volume}{56}},
  \bibinfo{pages}{3405} (\bibinfo{year}{1997}),
  \urlprefix\url{http://link.aps.org/doi/10.1103/PhysRevD.56.3405}.

\bibitem[{\citenamefont{Alcubierre et~al.}(2003)\citenamefont{Alcubierre,
  Br\"ugmann, Diener, Koppitz, Pollney, Seidel, and
  Takahashi}}]{Alcubierre:2003ab}
\bibinfo{author}{\bibfnamefont{M.}~\bibnamefont{Alcubierre}},
  \bibinfo{author}{\bibfnamefont{B.}~\bibnamefont{Br\"ugmann}},
  \bibinfo{author}{\bibfnamefont{P.}~\bibnamefont{Diener}},
  \bibinfo{author}{\bibfnamefont{M.}~\bibnamefont{Koppitz}},
  \bibinfo{author}{\bibfnamefont{D.}~\bibnamefont{Pollney}},
  \bibinfo{author}{\bibfnamefont{E.}~\bibnamefont{Seidel}}, \bibnamefont{and}
  \bibinfo{author}{\bibfnamefont{R.}~\bibnamefont{Takahashi}},
  \bibinfo{journal}{Phys. Rev. D} \textbf{\bibinfo{volume}{67}},
  \bibinfo{pages}{084023} (\bibinfo{year}{2003}),
  \urlprefix\url{http://link.aps.org/doi/10.1103/PhysRevD.67.084023}.

\bibitem[{\citenamefont{Banyuls et~al.}(1997)\citenamefont{Banyuls, Font,
  Ib{\'a}nez, Mart{\'\i}, and Miralles}}]{Banyuls:97c}
\bibinfo{author}{\bibfnamefont{F.}~\bibnamefont{Banyuls}},
  \bibinfo{author}{\bibfnamefont{J.~A.} \bibnamefont{Font}},
  \bibinfo{author}{\bibfnamefont{J.~M.} \bibnamefont{Ib{\'a}nez}},
  \bibinfo{author}{\bibfnamefont{J.~M.} \bibnamefont{Mart{\'\i}}},
  \bibnamefont{and} \bibinfo{author}{\bibfnamefont{J.~A.}
  \bibnamefont{Miralles}}, \bibinfo{journal}{Astrophys. J.}
  \textbf{\bibinfo{volume}{476}}, \bibinfo{pages}{221} (\bibinfo{year}{1997}).

\bibitem[{\citenamefont{{Cordero-Carri{\'o}n} and
  {Cerd{\'a}-Dur{\'a}n}}(2012)}]{Isabel:2012arx}
\bibinfo{author}{\bibfnamefont{I.}~\bibnamefont{{Cordero-Carri{\'o}n}}}
  \bibnamefont{and}
  \bibinfo{author}{\bibfnamefont{P.}~\bibnamefont{{Cerd{\'a}-Dur{\'a}n}}},
  \bibinfo{journal}{ArXiv e-prints}  (\bibinfo{year}{2012}),
  \eprint{1211.5930}.

\bibitem[{\citenamefont{{Cordero-Carri{\'o}n} and
  {Cerd{\'a}-Dur{\'a}n}}(2014)}]{Casas:2014}
\bibinfo{author}{\bibfnamefont{I.}~\bibnamefont{{Cordero-Carri{\'o}n}}}
  \bibnamefont{and}
  \bibinfo{author}{\bibfnamefont{P.}~\bibnamefont{{Cerd{\'a}-Dur{\'a}n}}},
  \emph{\bibinfo{title}{Advances in Differential Equations and Applications}},
  SEMA SIMAI Springer Series Vol. 4 (\bibinfo{publisher}{Springer International
  Publishing Switzerland}, \bibinfo{address}{Switzerland},
  \bibinfo{year}{2014}).

\bibitem[{\citenamefont{Stergioulas and Friedmann}(1995)}]{Stergioulas:1995b}
\bibinfo{author}{\bibfnamefont{N.}~\bibnamefont{Stergioulas}} \bibnamefont{and}
  \bibinfo{author}{\bibfnamefont{J.~L.} \bibnamefont{Friedmann}},
  \bibinfo{journal}{ApJ} \textbf{\bibinfo{volume}{444}}, \bibinfo{pages}{306}
  (\bibinfo{year}{1995}).

\bibitem[{\citenamefont{Barranco et~al.}(2012)\citenamefont{Barranco, Bernal,
  Degollado, Diez-Tejedor, Megevand et~al.}}]{Barranco:2012qs}
\bibinfo{author}{\bibfnamefont{J.}~\bibnamefont{Barranco}},
  \bibinfo{author}{\bibfnamefont{A.}~\bibnamefont{Bernal}},
  \bibinfo{author}{\bibfnamefont{J.~C.} \bibnamefont{Degollado}},
  \bibinfo{author}{\bibfnamefont{A.}~\bibnamefont{Diez-Tejedor}},
  \bibinfo{author}{\bibfnamefont{M.}~\bibnamefont{Megevand}},
  \bibnamefont{et~al.}, \bibinfo{journal}{Phys.Rev.Lett.}
  \textbf{\bibinfo{volume}{109}}, \bibinfo{pages}{081102}
  (\bibinfo{year}{2012}), \eprint{1207.2153}.

\bibitem[{\citenamefont{Furuhashi and Nambu}(2004)}]{Furuhashi:2004jk}
\bibinfo{author}{\bibfnamefont{H.}~\bibnamefont{Furuhashi}} \bibnamefont{and}
  \bibinfo{author}{\bibfnamefont{Y.}~\bibnamefont{Nambu}},
  \bibinfo{journal}{Prog.Theor.Phys.} \textbf{\bibinfo{volume}{112}},
  \bibinfo{pages}{983} (\bibinfo{year}{2004}), \eprint{gr-qc/0402037}.

\bibitem[{\citenamefont{Degollado and Herdeiro}(2014)}]{Degollado:2014vsa}
\bibinfo{author}{\bibfnamefont{J.~C.} \bibnamefont{Degollado}}
  \bibnamefont{and} \bibinfo{author}{\bibfnamefont{C.~A.~R.}
  \bibnamefont{Herdeiro}}, \bibinfo{journal}{Phys.Rev.}
  \textbf{\bibinfo{volume}{D90}}, \bibinfo{pages}{065019}
  (\bibinfo{year}{2014}), \eprint{1408.2589}.

\bibitem[{\citenamefont{Maga\~na et~al.}(2012)\citenamefont{Maga\~na, Matos,
  Su\'arez, and Sanchez-Salcedo}}]{Magana:2012jc}
\bibinfo{author}{\bibfnamefont{J.}~\bibnamefont{Maga\~na}},
  \bibinfo{author}{\bibfnamefont{T.}~\bibnamefont{Matos}},
  \bibinfo{author}{\bibfnamefont{A.}~\bibnamefont{Su\'arez}}, \bibnamefont{and}
  \bibinfo{author}{\bibfnamefont{F.~J.} \bibnamefont{Sanchez-Salcedo}},
  \bibinfo{journal}{JCAP} \textbf{\bibinfo{volume}{10}} (\bibinfo{year}{2012}).

\bibitem[{\citenamefont{Herdeiro and Radu}(2014)}]{Herdeiro:2014goa}
\bibinfo{author}{\bibfnamefont{C.~A.~R.} \bibnamefont{Herdeiro}}
  \bibnamefont{and} \bibinfo{author}{\bibfnamefont{E.}~\bibnamefont{Radu}},
  \bibinfo{journal}{Phys.Rev.Lett.} \textbf{\bibinfo{volume}{112}},
  \bibinfo{pages}{221101} (\bibinfo{year}{2014}), \eprint{1403.2757}.

\bibitem[{\citenamefont{Herdeiro and Radu}(2015)}]{Herdeiro:2015gia}
\bibinfo{author}{\bibfnamefont{C.}~\bibnamefont{Herdeiro}} \bibnamefont{and}
  \bibinfo{author}{\bibfnamefont{E.}~\bibnamefont{Radu}},
  \bibinfo{journal}{Class.Quant.Grav.} \textbf{\bibinfo{volume}{32}},
  \bibinfo{pages}{144001} (\bibinfo{year}{2015}), \eprint{1501.04319}.

\bibitem[{\citenamefont{Benone et~al.}(2014)\citenamefont{Benone, Crispino,
  Herdeiro, and Radu}}]{Benone:2014ssa}
\bibinfo{author}{\bibfnamefont{C.~L.} \bibnamefont{Benone}},
  \bibinfo{author}{\bibfnamefont{L.~C.~B.} \bibnamefont{Crispino}},
  \bibinfo{author}{\bibfnamefont{C.}~\bibnamefont{Herdeiro}}, \bibnamefont{and}
  \bibinfo{author}{\bibfnamefont{E.}~\bibnamefont{Radu}},
  \bibinfo{journal}{Phys.Rev.} \textbf{\bibinfo{volume}{D90}},
  \bibinfo{pages}{104024} (\bibinfo{year}{2014}), \eprint{1409.1593}.

\end{thebibliography}

\end{document}